# Nonadiabatic Field with Triangle Window Functions on Quantum Phase Space

*Xin He, Xiangsong Cheng, Baihua Wu, and Jian Liu\**

Beijing National Laboratory for Molecular Sciences, Institute of Theoretical and Computational Chemistry, College of Chemistry and Molecular Engineering,

Peking University, Beijing 100871, China

**Corresponding Author**

\* Electronic mail: jianliupku@pku.edu.cn

**Abstract**

Recent progress on the *constraint* coordinate-momentum *phase space* (CPS) formulation of finite-state quantum systems has revealed that the triangle window function approach is an isomorphic representation of the exact population-population correlation function of the two-state system. We use the triangle window (TW) function and the CPS mapping kernel element to formulate a novel useful representation of discrete electronic degrees of freedom (DOFs). When it is employed with nonadiabatic field (NaF) dynamics, a new variant of the NaF approach (i.e., NaF-TW) is proposed. The NaF-TW expression of the population of any adiabatic state is always positive semidefinite. Extensive benchmark tests of model systems in both the condensed phase and gas phase demonstrate that the NaF-TW approach is able to faithfully capture the dynamical interplay between electronic and nuclear DOFs in a broad region, including where the states remain coupled all the time, as well as where the bifurcation characteristic of nuclear motion is important.

**TOC GRAPHICS**

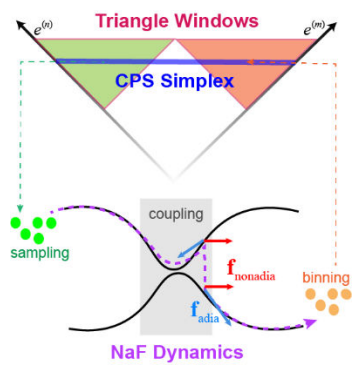

Nonadiabatic transition dynamics is crucial in understanding many important light-driven, photoemission, charge transfer, and cavity modified phenomena in natural and artificial complex molecular systems in chemistry, materials, biology, quantum information and computation, environmental science, and so forth[1-14]. In these systems, we often depict electrons by *discrete* electronic states and nuclei in continuous coordinate space. Numerical simulations of such composite systems often employ two prevailing categories of practical dynamics approaches with independent trajectories. The first category uses Born-Oppenheimer (BO) trajectories generated from different single-adiabatic-state potential energy surfaces (PESs). The surface hopping approach pioneered by Tully, Nikitin, and their co-workers[15-19] has been modified by various hopping algorithms[15-36] for connecting two independent Born-Oppenheimer trajectories on two different adiabatic PESs. The category of BO-trajectory-based dynamics often meets the challenge for nonadiabatic processes where the states remain coupled all the time, especially when the temperature is relatively low. Another category utilizes the independent mean field trajectory in the spirit of the Ehrenfest theorem[37]. In addition to the original Ehrenfest dynamics[37] for nonadiabatic transitions[38-40], a few Ehrenfest-like dynamics approaches[41-62] pioneered by Miller and co-workers[41, 42] have been developed, especially since the Meyer-Miller mapping Hamiltonian model was proposed for treating both nuclear and electronic degrees of freedom (DOFs) on the same footing[42, 63]. Among these Ehrenfest-like dynamics approaches, the symmetrical quasi-classical (SQC) method for nonadiabatic dynamics[53-55, 57, 58] is of particular interest. In the latest version of the SQC method, Cotton and Miller introduced the triangle window function (TWF) approach for discrete electronic DOFs[54, 57, 58]. The TWF approach, albeit proposed as a heuristic empirical model, is reasonably accurate for electronic dynamics (in the frozen-nuclei limit) even in the weak state-state coupling region[54]. In addition to the TWF approach offering the initial condition as well as the integral expression of the time-dependent physical property for electronic DOFs, the independent quasi-classical trajectory is generated from Ehrenfest-like dynamics of the Meyer-Miller mapping Hamiltonian[42]. The SQC with TWF for nonadiabatic dynamics[54, 55, 57] has widely been applied to various condensed phase system-bath models as well as realistic molecular systems[58, 64-72]. The category of independent mean field trajectory-based dynamics methods performs well in the nonadiabatic state-state coupling region but is more difficult in capturing the bifurcation characteristic of the nuclear motion in the asymptotic region where the state-state coupling disappears. It is shown in refs [50, 73] that the interference between different mean field trajectories (e.g., in the forward-backward or fully semiclassical initial value representation framework[74-76]) is necessary for describing the correct nuclear dynamics behavior in the asymptotic region, which requests more computational effort. In addition to the two major categories, there are some other methods using independent trajectories[77-81].

The unified phase space formulation with coordinate-momentum variables offers an exact interpretation of quantum mechanics to describe composite systems[82-90], where the *constraint* coordinate-momentum *phase space* (CPS) representation is used for *discrete* (electronic) degrees of freedom (DOFs) while the infinite (Wigner) coordinate-momentum phase space representation is used for *continuous* (nuclear) DOFs. The CPS representation (of discrete electronic DOFs) related to the quotient space $\mathrm{U}(F)/\mathrm{U}(F-1)$ was first introduced to nonadiabatic dynamics for general *F*-state systems by the sphere representation with coordinate-momentum variables in Section II of ref [84] and by the simplex representation with action-angle variables in Appendix A of ref [84]. It was shortly developed to a generalized phase space formulation, CPS with commutator variables[86, 87] in spirit of refs [82-84], which is related to the quotient space $\mathrm{U}(F)/\mathrm{U}(F-r)$ (with $1 \leq r < F$), namely, the complex Stiefel manifold[91-93]. As discussed in Appendix 3 of ref [88], the exact equations of motion (EOMs) of mapping coordinate-momentum variables of CPS for the pure finite-state quantum system are linear[82-85], which is superior to the conventional phase space approaches with angle variables[94-97] used in physics for studying dynamics of composite systems[98-102]. More importantly, the CPS formulation offers a powerful framework to derive more new (isomorphic) representations of the finite-state quantum system. This is demonstrated in ref [103] and in the present Letter.

The unified phase space formulation has recently led to a conceptually novel trajectory-based approach in the adiabatic representation of electronic states, nonadiabatic field (NaF) that is promising in faithfully describing both nuclear motion and electronic coherence/dissipation[104]. In the state-state coupling region, the nuclear EOMs of the independent trajectory of NaF involve an important nonadiabatic nuclear force term in addition to an adiabatic nuclear force term of a single electronic state (either stochastically with electronic weights or deterministically with the dominant electronic weight). This is substantially different from the two conventional categories of nonadiabatic dynamics methods that involve either BO trajectories on different PESs or mean field trajectories.

The NaF strategy[104] has been applied to the framework of Ehrenfest dynamics[37], fewest-switches surface hopping (FSSH)[15, 16], and CPS with commutator variables[84, 86-88, 105]. The investigation suggests that the nuclear EOMs of the NaF strategy should considerably improve over various surface hopping and Ehrenfest-like dynamics methods. It also indicates that the key of the most successful NaF approach includes the exact phase space representation of discrete electronic DOFs[104], which offers a consistent way in dealing with the other two critical properties of a trajectory-based quantum dynamics method[88], namely, the initial condition of the trajectory and the integral expression for evaluation of the time-dependent physical property.

In this Letter, we employ triangle window functions[54], in addition to the original mapping kernel of the $\mathrm{U}(F)/\mathrm{U}(F-1)$ CPS[84, 85], to formulate a novel representation of discrete (electronic) DOFs, which is applied with the NaF strategy to offer a more consistent trajectory-based approach for studying nonadiabatic transition dynamics.

Assume a coupled $F$-electronic-state Hamiltonian operator of the composite system

$$\hat{H} = \sum_{n,m=1}^{F} H_{nm}(\mathbf{R}, \mathbf{P}) |n\rangle\langle m|, \tag{1}$$

where $\{\mathbf{R}, \mathbf{P}\}$ are the coordinate and momentum variables for the nuclear DOFs, and $\{|n\rangle\}, n \in \{1, \cdots, F\}$ is the "complete" set of orthonormal electronic states ($F$ is in general infinite when the set of electronic states is rigorously complete). Consider the composite system in the frozen-nuclei limit of eq (1), where each element $H_{nm}(\mathbf{R}, \mathbf{P}) \equiv H_{nm}$ is constant and the Hamiltonian operator of eq (1) becomes $\hat{H} = \sum_{n,m=1}^{F} H_{nm} |n\rangle\langle m|$, which is the pure $F$-state quantum system. A generalization of the idea of the exact weighted CPS representation of ref [88] implies that the integrand function on electronic mapping CPS is not limited to the product of the element of the mapping kernel and that of the inverse mapping kernel of electronic DOFs[103]. The general phase space expression of the time correlation function between $|n\rangle\langle m|$ and $|k\rangle\langle l|$ reads

$$\begin{aligned}&\mathrm{Tr}_e\left[|n\rangle\langle m| e^{i\hat{H}t} |k\rangle\langle l| e^{-i\hat{H}t}\right] \\ &= \left(C_{nm,kl}(t)\right)^{-1} \int d\boldsymbol{\gamma}\, w(\boldsymbol{\gamma}) \int_{\mathcal{S}(\mathbf{x},\mathbf{p},\boldsymbol{\Gamma};\boldsymbol{\gamma})} F d\mathbf{x} d\mathbf{p} d\boldsymbol{\Gamma}\, \mathcal{Q}_{nm,kl}(\mathbf{x},\mathbf{p},\boldsymbol{\Gamma};\boldsymbol{\gamma};t)\end{aligned} \tag{2}$$

where $\text{Tr}_e[\ ]$ represents the trace over electronic DOFs, $\{\mathbf{x},\mathbf{p}\} = \{x^{(1)}, \cdots, x^{(F)}, p^{(1)}, \cdots, p^{(F)}\}$ are the mapping coordinate and momentum variables for discrete electronic DOFs, and $\mathbf{\Gamma}$ is the $F \times F$ commutator matrix[82, 86] that can be expressed by auxiliary coordinate-momentum variables,

$$\mathbf{\Gamma}_{mn} = \sum_{k=1}^{F} s_k \left(\xi_k^{(m)} + i\pi_k^{(m)}\right)\left(\xi_k^{(n)} - i\pi_k^{(n)}\right)/2. \tag{3}$$

In eq (2), $F\mathbf{dxdpd\Gamma}$ with $\mathrm{d}\mathbf{\Gamma} = \prod_{k,m=1}^{F} \mathrm{d}\xi_k^{(m)}\mathrm{d}\pi_k^{(m)}$ is the integral measure, $\mathcal{S}(\mathbf{x},\mathbf{p},\mathbf{\Gamma};\boldsymbol{\gamma})$ defines the phase space constraint that involves parameters $\boldsymbol{\gamma}$, $w(\boldsymbol{\gamma})$ is the quasi-distribution of parameter vector $\boldsymbol{\gamma}$, and $C_{nm,kl}(t)$ is the time-dependent normalization factor. $\mathcal{Q}_{nm,kl}(\mathbf{x},\mathbf{p},\mathbf{\Gamma};\boldsymbol{\gamma};t)$ defines the integrand function on the phase space corresponding to $|n\rangle\langle m|$ at time 0 and $e^{i\hat{H}t}|k\rangle\langle l|e^{-i\hat{H}t}$ at time $t$, which is a generalization of $K_{mn}(\mathbf{x},\mathbf{p},\mathbf{\Gamma})K_{lk}^{-1}(\mathbf{x},\mathbf{p},\mathbf{\Gamma};t)$, the product of the element of the mapping kernel and that of its inverse mapping kernel of our CPS formulations[82-89, 93, 106]. (The convention $\hbar = 1$ is used for discrete electronic DOFs throughout this Letter and its Supporting Information.)

The general expression of the time correlation function eq (2) with trajectory-based dynamics becomes

$$\text{Tr}_e\left[|n\rangle\langle m|e^{i\hat{H}t}|k\rangle\langle l|e^{-i\hat{H}t}\right] \mapsto$$
$$\left(\overline{C}_{nm,kl}(t)\right)^{-1} \int w(\boldsymbol{\gamma})\mathrm{d}\boldsymbol{\gamma}\int_{\mathcal{S}(\mathbf{x}_0,\mathbf{p}_0,\mathbf{\Gamma}_0;\boldsymbol{\gamma})} F\mathrm{d}\mathbf{x}_0\mathrm{d}\mathbf{p}_0\mathrm{d}\mathbf{\Gamma}_0\ \overline{\mathcal{Q}}_{nm,kl}\left(\{\mathbf{x},\mathbf{p},\mathbf{\Gamma}\}_{0\leq\tau\leq t};\boldsymbol{\gamma};t\right) \tag{4}$$

where $\{\mathbf{x},\mathbf{p},\mathbf{\Gamma}\}_{0\leq\tau\leq t}$ denotes the trajectory on the phase space, $\overline{\mathcal{Q}}_{nm,kl}\left(\{\mathbf{x},\mathbf{p},\mathbf{\Gamma}\}_{0\leq\tau\leq t};\boldsymbol{\gamma};t\right)$ defines the integrand function on the phase space corresponding to $|n\rangle\langle m|$ at time 0 and $e^{i\hat{H}t}|k\rangle\langle l|e^{-i\hat{H}t}$ at time $t$ for trajectory dynamics, and $\overline{C}_{nm,kl}(t)$ is the time-dependent normalization factor of the

trajectory-based dynamics approach. The expression of eq (4) is a generalization of the formalisms in the CPS formulation[82-89, 93, 106], where the trajectory is generated by the linear EOMs yielded from the symplectic structure of CPS.

Our recent work of ref [103] presents a new class of isomorphic representations of the exact population-population correlation function of the pure two-state quantum system (i.e., $F=2$). Remarkably, the TWF approach (for discrete electronic DOFs) originally proposed as a heuristic empirical model by Cotton and Miller[54], is proved as a special case of the new class of phase space representations for *exact* population dynamics[103]. The proof involves the projection of the TWF onto the $U(F)/U(F-1)$ CPS for the two-state system and integral identities for the exact population-population correlation function. Because the TWF approach is exact for the population dynamics of the two-state system[103] and reasonably accurate for that of the multistate system (i.e., $F \geq 3$)[54], it suggests that the triangle window function should be valuable for developing a novel useful representation of discrete electronic DOFs (e.g., for nonadiabatic dynamics).

Consider the pure $F$-state system of eq (1). We focus on a special class of eq (4),

$$\text{Tr}_e\left[|n\rangle\langle m|e^{i\hat{H}t}|k\rangle\langle l|e^{-i\hat{H}t}\right]$$
$$\mapsto \left(\bar{C}_{nm,kl}(t)\right)^{-1}\int d\boldsymbol{\gamma}\, w(\boldsymbol{\gamma})\int_{\mathcal{S}(\mathbf{x}_0,\mathbf{p}_0,\boldsymbol{\Gamma}_0;\boldsymbol{\gamma})} F d\mathbf{x}_0 d\mathbf{p}_0 d\boldsymbol{\Gamma}_0\, \overline{\mathcal{Q}}_{nm,kl}\left(\mathbf{x}_0,\mathbf{p}_0,\boldsymbol{\Gamma}_0;\mathbf{x}_t,\mathbf{p}_t,\boldsymbol{\Gamma}_t\right)$$

(5)

Equation (5) includes four kinds of time correlation functions of electronic DOFs, namely, population-population ($n=m$ and $k=l$), population-coherence ($n=m$ and $k \neq l$), coherence-population ($n \neq m$ and $k=l$), and coherence-coherence ($n \neq m$ and $k \neq l$) correlation functions. Commutator matrix $\boldsymbol{\Gamma}$ can be constant, that is, $\boldsymbol{\Gamma} = \gamma \mathbf{1}$, where $\gamma$ is a scalar parameter and $\mathbf{1}$ is the identity matrix. In this case, the constraint of CPS reads

$$S(\mathbf{x},\mathbf{p};\gamma) = \delta\left(\sum_{n=1}^{F} \frac{1}{2}\left(\left(x^{(n)}\right)^2 + \left(p^{(n)}\right)^2\right) - (1+F\gamma)\right) \tag{6}$$

with parameter $\gamma \in (-1/F, \infty)$, and

$$\int_{S(\mathbf{x},\mathbf{p};\gamma)} F d\mathbf{x} d\mathbf{p}\, g(\mathbf{x},\mathbf{p}) = \int F d\mathbf{x} d\mathbf{p}\, \frac{1}{\Omega(\gamma)} S(\mathbf{x},\mathbf{p};\gamma) g(\mathbf{x},\mathbf{p})\ . \tag{7}$$

where $\Omega(\gamma)$ is the normalization factor[84, 85, 88]

$$\Omega(\gamma) = \int d\mathbf{x} d\mathbf{p}\, S(\mathbf{x},\mathbf{p};\gamma) = \frac{(2\pi)^F (1+F\gamma)^{F-1}}{(F-1)!}\ . \tag{8}$$

By employing a generalization of the weighted constraint phase space[88], the expressions of time correlation functions read

$$\begin{aligned}
&\mathrm{Tr}_e\left[|n\rangle\langle m|e^{i\hat{H}t}|k\rangle\langle l|e^{-i\hat{H}t}\right] \\
&\mapsto \left(\bar{C}_{nm,kl}(t)\right)^{-1} \int_{-1/F}^{+\infty} w(\gamma) d\gamma \int_{S(\mathbf{x}_0,\mathbf{p}_0;\gamma)} F d\mathbf{x}_0 d\mathbf{p}_0\, \overline{\mathcal{Q}}_{nm,kl}(\mathbf{x}_0,\mathbf{p}_0;\mathbf{x}_t,\mathbf{p}_t)
\end{aligned} \tag{9}$$

For example, the CMM approach of ref [69] can be reinterpreted by eq (9) with $w(\gamma) = \delta(\gamma - \gamma_0)$ and $\overline{\mathcal{Q}}_{nm,kl}(\mathbf{x}_0,\mathbf{p}_0;\mathbf{x}_t,\mathbf{p}_t) = K_{mn}(\mathbf{x}_0,\mathbf{p}_0) K_{lk}^{-1}(\mathbf{x}_t,\mathbf{p}_t)$, where $K_{mn}(\mathbf{x},\mathbf{p})$ and $K_{lk}^{-1}(\mathbf{x}_t,\mathbf{p}_t)$ are matrix elements of eqs (7) and (8) of ref [69] with phase space parameter $\gamma_0$, respectively. In this Letter, we employ the TWF[54] for the integral expression of the electronic correlation function. The unified expression reads

$$\begin{aligned}
&\mathrm{Tr}_e\left[|n\rangle\langle m|e^{i\hat{H}t}|k\rangle\langle l|e^{-i\hat{H}t}\right] \\
&\mapsto \left(\bar{C}_{nm,kl}(t)\right)^{-1} \int_0^{1-1/F} w(\gamma) d\gamma \int_{S(\mathbf{x}_0,\mathbf{p}_0;\gamma)} F d\mathbf{x}_0 d\mathbf{p}_0\, \overline{\mathcal{Q}}_{nm,kl}(\mathbf{x}_0,\mathbf{p}_0;\mathbf{x}_t,\mathbf{p}_t), \\
&= \left(\bar{C}_{nm,kl}(t)\right)^{-1} \int d\mathbf{x}_0 d\mathbf{p}_0\, \frac{F \cdot F!}{(2\pi)^F (F^F - 1)} \overline{\mathcal{Q}}_{nm,kl}(\mathbf{x}_0,\mathbf{p}_0;\mathbf{x}_t,\mathbf{p}_t)
\end{aligned} \tag{10}$$

where $w(\gamma) = \dfrac{F^2}{F^F - 1}(1 + F\gamma)^{F-1}$ and $\int_0^{1-1/F} w(\gamma) d\gamma = 1$ . In the expression of $\mathrm{Tr}_e\left[|n\rangle\langle n|e^{i\hat{H}t}|m\rangle\langle m|e^{-i\hat{H}t}\right]$ of eq (10),

$$\overline{\mathcal{Q}}_{nn,mm}(\mathbf{x}_0,\mathbf{p}_0;\mathbf{x}_t,\mathbf{p}_t) = \overline{w}_n^{\mathrm{SQC}}(\mathbf{x}_0,\mathbf{p}_0) K_{nn}^{\mathrm{SQC}}(\mathbf{x}_0,\mathbf{p}_0) K_{mm}^{\mathrm{bin}}(\mathbf{x}_t,\mathbf{p}_t), \tag{11}$$

where

$$\overline{w}_n^{\mathrm{SQC}}(\mathbf{x}_0,\mathbf{p}_0) = \dfrac{2(F^F - 1)}{F \cdot F!}\left(2 - \dfrac{1}{2}\left((x_0^{(n)})^2 + (p_0^{(n)})^2\right)\right)^{2-F}. \tag{12}$$

The time-dependent normalization factor reads

$$\begin{aligned}\overline{C}_{nn,mm}(t) &= \sum_{k=1}^{F} \int_0^{1-1/F} w(\gamma) d\gamma \int_{\mathcal{S}(\mathbf{x}_0,\mathbf{p}_0;\gamma)} F d\mathbf{x}_0 d\mathbf{p}_0 \overline{\mathcal{Q}}_{nn,kk}(\mathbf{x}_0,\mathbf{p}_0;\mathbf{x}_t,\mathbf{p}_t) \\ &= \sum_{k=1}^{F} \int d\mathbf{x}_0 d\mathbf{p}_0 \dfrac{F \cdot F!}{(2\pi)^F (F^F - 1)} \overline{\mathcal{Q}}_{nn,kk}(\mathbf{x}_0,\mathbf{p}_0;\mathbf{x}_t,\mathbf{p}_t)\end{aligned}, \tag{13}$$

whose initial value $\overline{C}_{nn,mm}(0)$ is 1. For $n \neq m$ or $k \neq l$, $\overline{C}_{nm,kl}(t) \equiv 1$. The TWF for $|n\rangle\langle n|$, the population of the $n$-th state at time $0$, is an indicator function,

$$K_{nn}^{\mathrm{SQC}}(\mathbf{x}_0,\mathbf{p}_0) \equiv \begin{cases} 1 & \text{if } (\mathbf{x}_0,\mathbf{p}_0) \in \mathcal{M}_n \\ 0 & \text{otherwise} \end{cases}, \tag{14}$$

where $\mathcal{M}_n$ includes the following set of phase space points,

$$\mathcal{M}_n(\mathbf{x},\mathbf{p}) : \left\{(\mathbf{x},\mathbf{p}) \left| \begin{array}{l} 1 \leq \dfrac{1}{2}\left((x^{(n)})^2 + (p^{(n)})^2\right) \leq 2; \\ \forall k \neq n, \dfrac{1}{2}\left((x^{(k)})^2 + (p^{(k)})^2 + (x^{(n)})^2 + (p^{(n)})^2\right) \leq 2 \end{array}\right.\right\}. \tag{15}$$

For any point $(\mathbf{x},\mathbf{p})$ in $\mathcal{M}_n(\mathbf{x},\mathbf{p})$, the value of $\sum_{n=1}^{F}\dfrac{1}{2}\left((x^{(n)})^2 + (p^{(n)})^2\right)$ lies in region $[1, F]$, so that the domain of $\gamma$ is $\gamma \in [0, 1-1/F]$. The TWF for $e^{i\hat{H}t}|m\rangle\langle m|e^{-i\hat{H}t}$, the population of the $m$-th state at time $t$, is also an indicator function,

$$K_{mm}^{bin}(\mathbf{x}_t, \mathbf{p}_t) \equiv \begin{cases} 1 \text{ if } (\mathbf{x}_t, \mathbf{p}_t) \in \mathcal{M}_m^{bin} \\ 0 \text{ otherwise} \end{cases}, \quad (16)$$

where

$$\mathcal{M}_m^{bin}(\mathbf{x}, \mathbf{p}) : \left\{ (\mathbf{x}, \mathbf{p}) \left| \begin{array}{l} 1 \leq \frac{1}{2}\left(\left(x^{(m)}\right)^2 + \left(p^{(m)}\right)^2\right); \\ \forall k \neq m, \frac{1}{2}\left(\left(x^{(k)}\right)^2 + \left(p^{(k)}\right)^2\right) \leq 1 \end{array} \right. \right\}. \quad (17)$$

Because functions $K_{nn}^{SQC}(\mathbf{x}_0, \mathbf{p}_0)$ and $K_{mm}^{bin}(\mathbf{x}_t, \mathbf{p}_t)$ of eq (10) and eq (13) are always non-negative, the population-population correlation function remains positive semidefinite all the time for the choice of the electronic basis set $\{|n\rangle\}$ of eq (1). The expression of eq (10) for the population-population correlation function is exact for only the pure two-state system. We note that the phase space formalism of eqs (10) and (11) is not the only option, and an alternative formulation derived for triangle window functions is presented in Section S5 of the Supporting Information.

Unfortunately, the SQC approach with either triangle or square window functions[54, 107] does not lead to exact results for the other three kinds of electronic correlation functions even for the pure two-state case ($F=2$). It is crucial to construct the new phase space representation with the triangle window function used in eqs (14) and (15) for the initial condition such that the other three kinds of electronic correlation functions are exact. Here, we employ the element of the mapping kernel of the $U(F)/U(F-1)$ CPS[84, 85] to accomplish the task.

In the expression of eq (10) for $\text{Tr}_e\left[|n\rangle\langle n|e^{i\hat{H}t}|k\rangle\langle l|e^{-i\hat{H}t}\right]$ with $k \neq l$, the population-coherence correlation function,

$$\overline{\mathcal{Q}}_{nn,kl}(\mathbf{x}_0, \mathbf{p}_0; \mathbf{x}_t, \mathbf{p}_t) = \overline{w}_n^{SQC}(\mathbf{x}_0, \mathbf{p}_0) K_{nn}^{SQC}(\mathbf{x}_0, \mathbf{p}_0) K_{lk}^{CMM}(\mathbf{x}_t, \mathbf{p}_t), \quad (18)$$

where

$$K_{lk}^{\text{CMM}}(\mathbf{x}_t, \mathbf{p}_t) \equiv \frac{1}{2}\left(x_t^{(l)} + ip_t^{(l)}\right)\left(x_t^{(k)} - ip_t^{(k)}\right) \tag{19}$$

is the element of the mapping kernel of the $\text{U}(F)/\text{U}(F-1)$ CPS[84, 85]. The time-dependent normalization factor of eq (5) is constant, $\overline{C}_{nn,kl}(t) \equiv 1$. When the initial electronic density matrix includes the coherence term, $|n\rangle\langle m|$ (where $n \neq m$), in the expression of eq (10) for both the coherence-population (where $n \neq m$ and $k = l$) and coherence-coherence (where $n \neq m$ and $k \neq l$) correlation functions,

$$\overline{\mathcal{Q}}_{nm,kl}(\mathbf{x}_0, \mathbf{p}_0; \mathbf{x}_t, \mathbf{p}_t) = \overline{w}_{nm}^{\text{coh}}(\mathbf{x}_0, \mathbf{p}_0) K_{mn}^{\text{CMM}}(\mathbf{x}_0, \mathbf{p}_0) K_{lk}^{\text{CMM}}(\mathbf{x}_t, \mathbf{p}_t) , \tag{20}$$

where $\overline{w}_{nm}^{\text{coh}}(\mathbf{x}_0, \mathbf{p}_0) = \frac{6}{5} \sum_{i=n,m} \overline{w}_i^{\text{SQC}}(\mathbf{x}_0, \mathbf{p}_0) K_{ii}^{\text{SQC}}(\mathbf{x}_0, \mathbf{p}_0)$ involves the triangle window functions for the $n$-th and $m$-th states at time $0$. The time-dependent normalization factor $\overline{C}_{nm,kl}(t) \equiv 1$ for $n \neq m$ is also a constant.

The new electronic representation of eq (10) for evaluation of the electronic density matrix at time *t*, which is denoted as the CPS with triangle window functions (CPS-TW) approach. It has three important properties:

1) In the frozen-nuclei limit, the representation of the electronic population-population correlation is exact for the two-state system ($F=2$) and is expected to be reasonably accurate for the multistate system ($F \geq 3$). (See more discussion in ref [103].)

2) The representation of the other three kinds of correlation functions yields the exact frozen-nuclei limit for all cases ($F \geq 2$). (See more discussion in Section S1 of the Supporting Information.)

3) When the phase space expression of the total density operator of both electronic and nuclear DOFs is initially non-negative, the expression of the electronic population-population correlation (eq (10)) is guaranteed to be positive semidefinite for all cases ($F \geq 2$).

In comparison to the exact CPS representations of electronic DOFs, the third property (of the CPS-TW approach) is indispensable for solving the negative population problem of phase space mapping dynamics methods for general $F$-state systems, e.g., as shown in Figures 4 and 7 of ref [86] and in Figure S12 of the Supporting Information of ref [104]. In addition to the TWF approach, other cases of the novel class of phase space representations introduced in ref [103] can also be utilized with the CPS formulation to offer the integral expression for the four kinds of electronic correlation functions, which is expected to have the same properties of the CPS-TW approach.

By employing the CPS-TW approach instead of the exact CPS representations for electronic DOFs, in addition to using the infinite Wigner coordinate-momentum phase space for nuclear DOFs, we obtain the expression of the correlation function for both electronic and nuclear DOFs

$$\begin{aligned}
&\mathrm{Tr}_{n,e}\left[\left(|n\rangle\langle m|\otimes \hat{\rho}_{\mathrm{nuc}}\right)e^{i\hat{H}t/\hbar}\left(|k\rangle\langle l|\otimes \hat{A}_{\mathrm{nuc}}\right)e^{-i\hat{H}t/\hbar}\right] \\
&\mapsto \left(\overline{C}_{nm,kl}(t)\right)^{-1}(2\pi\hbar)^{-N}\int d\mathbf{R}_0 d\mathbf{P}_0 \int_0^{1-1/F} w(\gamma)d\gamma \int_{\mathcal{S}(\mathbf{x}_0,\mathbf{p}_0;\gamma)} F d\mathbf{x}_0 d\mathbf{p}_0 \\
&\quad \times \rho_W(\mathbf{R}_0,\mathbf{P}_0) A_W(\mathbf{R}_t,\mathbf{P}_t)\overline{\mathcal{Q}}_{nm,kl}(\mathbf{x}_0,\mathbf{p}_0;\mathbf{x}_t,\mathbf{p}_t)
\end{aligned} \quad (21)$$

Here, $\mathrm{Tr}_n[\ ]$ represents the trace over nuclear DOFs, $\rho_W(\mathbf{R},\mathbf{P})$ and $A_W(\mathbf{R},\mathbf{P})$ are the Wigner phase space functions of nuclear operators $\hat{\rho}_{\mathrm{nuc}}$ and $\hat{A}_{\mathrm{nuc}}$, respectively. For instance, $A_W(\mathbf{R},\mathbf{P}) = \int d\boldsymbol{\Delta}\langle\mathbf{R}-\boldsymbol{\Delta}/2|\hat{A}_{\mathrm{nuc}}|\mathbf{R}+\boldsymbol{\Delta}/2\rangle\exp(-i\boldsymbol{\Delta}\cdot\mathbf{P}/\hbar)$. In addition, when nuclear DOFs are also considered,

$$\begin{aligned}
\overline{C}_{nn,mm}(t) = \sum_{k=1}^{F}(2\pi\hbar)^{-N}\int d\mathbf{R}_0 d\mathbf{P}_0 \int_0^{1-1/F} w(\gamma)d\gamma \int_{\mathcal{S}(\mathbf{x}_0,\mathbf{p}_0;\gamma)} F d\mathbf{x}_0 d\mathbf{p}_0 \\
\times \rho_W(\mathbf{R}_0,\mathbf{P}_0)\overline{\mathcal{Q}}_{nn,kk}(\mathbf{x}_0,\mathbf{p}_0;\mathbf{x}_t,\mathbf{p}_t)
\end{aligned} \quad (22)$$

Provided that eq (21) is the integral expression of the time-dependent property, we propose a new variant of the NaF approach, namely, NaF with triangle window functions (NaF-TW), in the adiabatic representation (of electronic states).

Consider the full Hamiltonian of nuclei and electrons of the system

$$\hat{H} = \frac{1}{2}\hat{\mathbf{P}}^T \mathbf{M}^{-1} \hat{\mathbf{P}} + \hat{H}_{el}(\hat{\mathbf{R}}) = \frac{1}{2}\hat{\mathbf{P}}^T \mathbf{M}^{-1} \hat{\mathbf{P}} + \sum_k E_k(\mathbf{R})|\phi_k(\mathbf{R})\rangle\langle\phi_k(\mathbf{R})| \;, \qquad (23)$$

where $\mathbf{M} = \text{diag}\{m_j\}$ is the diagonal nuclear mass matrix, $\hat{H}_{el}(\hat{\mathbf{R}})$ is the electronic Hamiltonian that includes the kinetic energy of electrons and all the electrostatic potential among electrons and nuclei, and $E_k(\mathbf{R})$ denotes the adiabatic PES of the *k*-th adiabatic electronic state, $|\phi_k(\mathbf{R})\rangle$. The expression of the right-hand side (RHS) for the full Hamiltonian operator of eq (23) was first employed for phase space mapping methods for nonadiabatic dynamics in refs [88, 108]. The nonadiabatic coupling vector in the adiabatic representation is $\mathbf{d}_{mn}(\mathbf{R}) = \left\langle \phi_m(\mathbf{R}) \middle| \frac{\partial \phi_n(\mathbf{R})}{\partial \mathbf{R}} \right\rangle$, of which the *J*-th component is $d_{mn}^{(J)}(\mathbf{R})$. Note that $-i\mathbf{d}^{(J)}(\mathbf{R})$ is a Hermitian matrix of electronic DOFs and that vector $-i\mathbf{d}(\mathbf{R})$ implies a nonabelian gauge field[88, 109]. Assume that $\{|\phi_n\rangle\}, n \in \{1,\cdots,F\}$ are effectively complete to describe the process where the gauge field tensor, $\frac{\partial(-i\mathbf{d}^{(J)})}{\partial R_I} - \frac{\partial(-i\mathbf{d}^{(I)})}{\partial R_J} + i[-i\mathbf{d}^{(I)}, -i\mathbf{d}^{(J)}]_{\text{ele}}$, is close to zero such that we can construct the so-called quasi-diabatic basis[110]. The null gauge field tensor indicates that the diabatic basis is strictly defined[88, 110]. E.g., it occurs when all adiabatic electronic states are involved, or when the adiabatic basis is parametrized along only a single DOF of the nuclear path[88, 110]. Because the diabatic basis is, in general, not well-defined, the gauge field tensor should be ignored with caution. Following the EOMs of NaF proposed in ref [104], the EOMs of NaF-TW similarly reads

$$\dot{\tilde{\mathbf{g}}}(\mathbf{R}) = -i\mathbf{V}^{(\text{eff})}(\mathbf{R},\mathbf{P})\tilde{\mathbf{g}}(\mathbf{R})$$
$$\dot{\mathbf{R}} = \mathbf{M}^{-1}\mathbf{P} \qquad , \qquad (24)$$
$$\dot{\mathbf{P}} = \mathbf{f}_{\text{nonadia}}(\mathbf{R}) - \nabla_{\mathbf{R}}E_{j_{\text{occ}}}(\mathbf{R})$$

where $\tilde{\mathbf{g}}(\mathbf{R}) = \tilde{\mathbf{x}}(\mathbf{R}) + i\tilde{\mathbf{p}}(\mathbf{R})$, $(\tilde{\mathbf{x}},\tilde{\mathbf{p}})$ are electronic mapping variables in the adiabatic representation, $\mathbf{P}$ is the kinematic nuclear momentum of the adiabatic representation[55, 86, 88, 104] (equivalently, the mapping diabatic momentum[88, 104]), the elements of the effective potential matrix, $\mathbf{V}^{(\text{eff})}$, are functions of nuclear phase space variables,

$$V_{nk}^{(\text{eff})}(\mathbf{R},\mathbf{P}) = E_n(\mathbf{R})\delta_{nk} - i\mathbf{M}^{-1}\mathbf{P}\cdot\mathbf{d}_{nk}(\mathbf{R}) \quad , \qquad (25)$$

and the nonadiabatic nuclear force reads

$$\mathbf{f}_{\text{nonadia}}(\mathbf{R}) = -\sum_{k\neq l}\left[(E_k(\mathbf{R}) - E_l(\mathbf{R}))\mathbf{d}_{lk}(\mathbf{R})\right]\tilde{\rho}_{kl}(\mathbf{R}) \quad . \qquad (26)$$

In eq (26), $\tilde{\rho}_{kl}(\mathbf{R})$ is the element in the $k$-th row and $l$-th column of the matrix,

$$\tilde{\boldsymbol{\rho}}(\mathbf{R}) = \frac{(1+F/3)}{\text{Tr}_e\left[\tilde{\mathbf{g}}\tilde{\mathbf{g}}^{\dagger}\right]}\tilde{\mathbf{g}}\tilde{\mathbf{g}}^{\dagger} - \mathbf{1}/3 \quad . \qquad (27)$$

The $1+F/3$ term of eq (27) corresponds to the value of the so-called "zero-point energy parameter", $1/3$, suggested in ref [54]. (Please see more discussion in Section S4 of the Supporting Information.) The nonadiabatic nuclear force, eq (26), intrinsically accounts for nonadiabatic transition processes in the state-state coupling region and disappears in the region where the state-state coupling vanishes. Its importance in the nuclear EOMs has been demonstrated by the applications to a few benchmark condensed phase model systems[104]. Following ref [104], the adiabatic nuclear force $-\nabla_{\mathbf{R}}E_{j_{\text{occ}}}(\mathbf{R})$ of eq (24) is contributed from the single-state adiabatic nuclear force that has the largest weight, i.e., $-\sum_k \nabla_{\mathbf{R}}E_k(\mathbf{R})\left(\prod_{j\neq k}h(\tilde{\rho}_{kk} - \tilde{\rho}_{jj})\right)$ with the Heaviside function $h(y) = \{1 \text{ if } y \geq 0 \text{ else } 0\}$. That is, the contribution of adiabatic force ingredients with smaller weights is neglected. We focus on this approach of the single-state adiabatic nuclear force from the dominant weight, although other approaches are also possible[104]. The NaF mapping energy in the adiabatic representation, $H_{\text{NaF}}(\mathbf{R}, \mathbf{P}, \tilde{\mathbf{x}}(\mathbf{R}), \tilde{\mathbf{p}}(\mathbf{R})) \equiv \frac{1}{2} \mathbf{P}^T \mathbf{M}^{-1} \mathbf{P} + E_{J_{\text{occ}}}(\mathbf{R})$, is conserved by rescaling the adiabatic kinematic nuclear momentum $\mathbf{P}$ (that is equivalent to the mapping diabatic momentum[88, 104]) along the momentum vector after the integration of the EOMs of eq (24) as well as when the largest weight is switched[104]. If it is impossible to conserve the NaF energy when the largest weight is switched, the switching is prohibited, with no change of $\mathbf{P}$, and the single-state adiabatic nuclear force is still contributed by the gradient of the previously occupied adiabatic PES. The algorithm of NaF-TW is described in detail in Section S2 of the Supporting Information.

We consider a series of typical benchmark condensed phase and gas phase nonadiabatic model systems where numerically exact results are available, which have been used for testing the numerical performance of approximate dynamics methods in refs. [86, 88, 104]. The latest SQC approach with triangle window functions of ref [57] (which is denoted as SQC-TW), the conventional Ehrenfest dynamics[37], the fewest-switches surface hopping (FSSH)[16] algorithm described in ref [28], and the original NaF (with $\gamma = 1/2$) of ref [104] are also tested for comparison. The initial condition for nuclear DOFs is sampled from the Wigner distribution on nuclear coordinate-momentum phase space, which takes care of nuclear quantum effects in all the trajectory-based nonadiabatic dynamics methods for fair comparison. All simulations are performed in the adiabatic representation. When exact results are available in only the diabatic representation, the diabatic initial condition is transformed to its adiabatic counterpart, and dynamics results in the adiabatic representation are transformed back to the corresponding diabatic results.

We first consider standard system-bath models, where the system is bilinearly coupled with harmonic bath DOFs of a dissipative environment in the condensed phase. The coupling imparts a substantial influence from the bath environment and yields the reduced dynamics of the system across a broad spectrum ranging from coherent to dissipative regimes. Such models serve as pivotal tools for understanding important processes governing electron/exciton energy transfer dynamics in the realm of chemical and biological reactions. Methodologies that yield numerically exact results for condensed phase system-bath models, most in the diabatic representation, include quasi-adiabatic propagator path integral (QuAPI)[111-113] and more efficient small matrix PI (SMatPI)[114, 115], hierarchy equations of motion (HEOM)[116-120], (multilayer) multiconfiguration time-dependent Hartree [(ML-)MCTDH][121-123], time-dependent density matrix renormalization group (TD-DMRG)[124], and so forth. We use the two-site spin-boson model and the seven-site Fenna–Matthews–Olson (FMO) monomer model for testing trajectory-based nonadiabatic dynamics methods.

Figure 1 investigates four typical spin-boson models at low temperature, which range from weak to strong system-bath coupling. Three hundred discrete bath modes are utilized for the Ohmic spectral density. Initially, the (nuclear) bath DOFs are at thermal equilibrium and the system is in the diabatic excited state. (Please see more numerical details in Section S3-A of the Supporting Information.) In comparison to numerically exact data, while Ehrenfest dynamics produces the worst results, FSSH performs slightly better but does not capture the correct asymptotic behavior for a relatively long time. In contrast, SQC-TW, NaF, and NaF-TW yield results that are in overall good agreement with exact data. In addition, Section S3-F of the Supporting Information tests the two-state system-bath model for electron transfer reactions. Figure S4 (of the Supporting Information) illustrates that SQC-TW, NaF, and NaF-TW, using the CPS formulation or the CPS-TW representation that yields *exact* electronic dynamics of the pure two-state quantum system, noticeably outperform the original SQC method using square window functions of ref [125].

Figure 2 considers the seven-site FMO monomer model related to the photosynthetic organism of green sulfur bacteria. One hundred discrete bath modes per site are employed for the Debye spectral density. At time $t=0$, the (nuclear) bath DOFs are at thermal equilibrium at 77 K and the first site is occupied. Both Ehrenfest dynamics and FSSH perform poorly even for relatively short time and fail to even qualitatively capture the steady-state behavior in the long-time limit. In comparison, SQC-TW, NaF, and NaF-TW show much better performance and are capable of reasonably describing the evolution of both electronic population and "coherence", from the fast relaxation behavior at short time to the asymptotic behavior at long time.

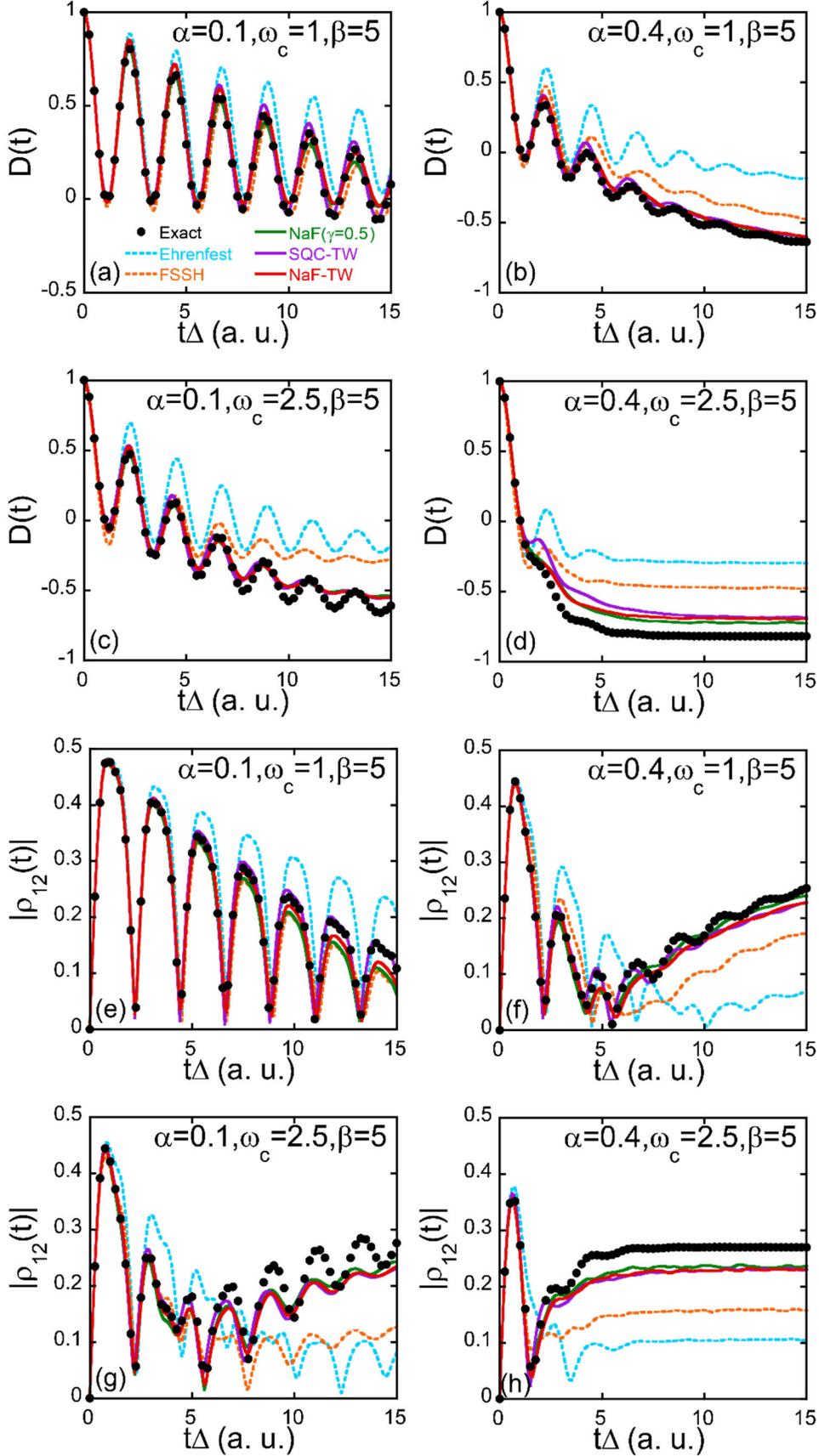

**Figure 1.** Panels (a-d): The population difference between State 1 and State 2, $D(t)$, as a function of time for the four spin-boson models with the Ohmic spectral density at the inverse temperature $\beta = 5$. Panels (e-h): The modulus of the off-diagonal (coherence) element of the reduced electronic density matrix, $|\rho_{12}(t)|$, as a function of time. Panels (a-d) or (e-h) represent the results of the spin-boson models with parameters $\{\alpha = 0.1, \omega_c = 1\}$, $\{\alpha = 0.4, \omega_c = 1\}$, $\{\alpha = 0.1, \omega_c = 2.5\}$, and $\{\alpha = 0.4, \omega_c = 2.5\}$, respectively. Black points: (exact results produced by) eHEOM. Cyan long-dashed lines: Ehrenfest dynamics. Orange short-dashed lines: FSSH. Green solid lines: NaF. Purple solid lines: SQC-TW. Red solid lines: NaF-TW. Converged results are obtained using three hundred discrete bath modes. For SQC-TW, the expression of the population-population correlation function is equivalent to that of ref [57] of Cotton and Miller, while eqs (10), (18) and (19) are used for the population-coherence correlation function because its SQC expression with triangle window functions of ref [57] is not exact for even the pure two-state system. More details of the parameters of the spin-boson models and the simulations are described in Section S3-A of the Supporting Information.

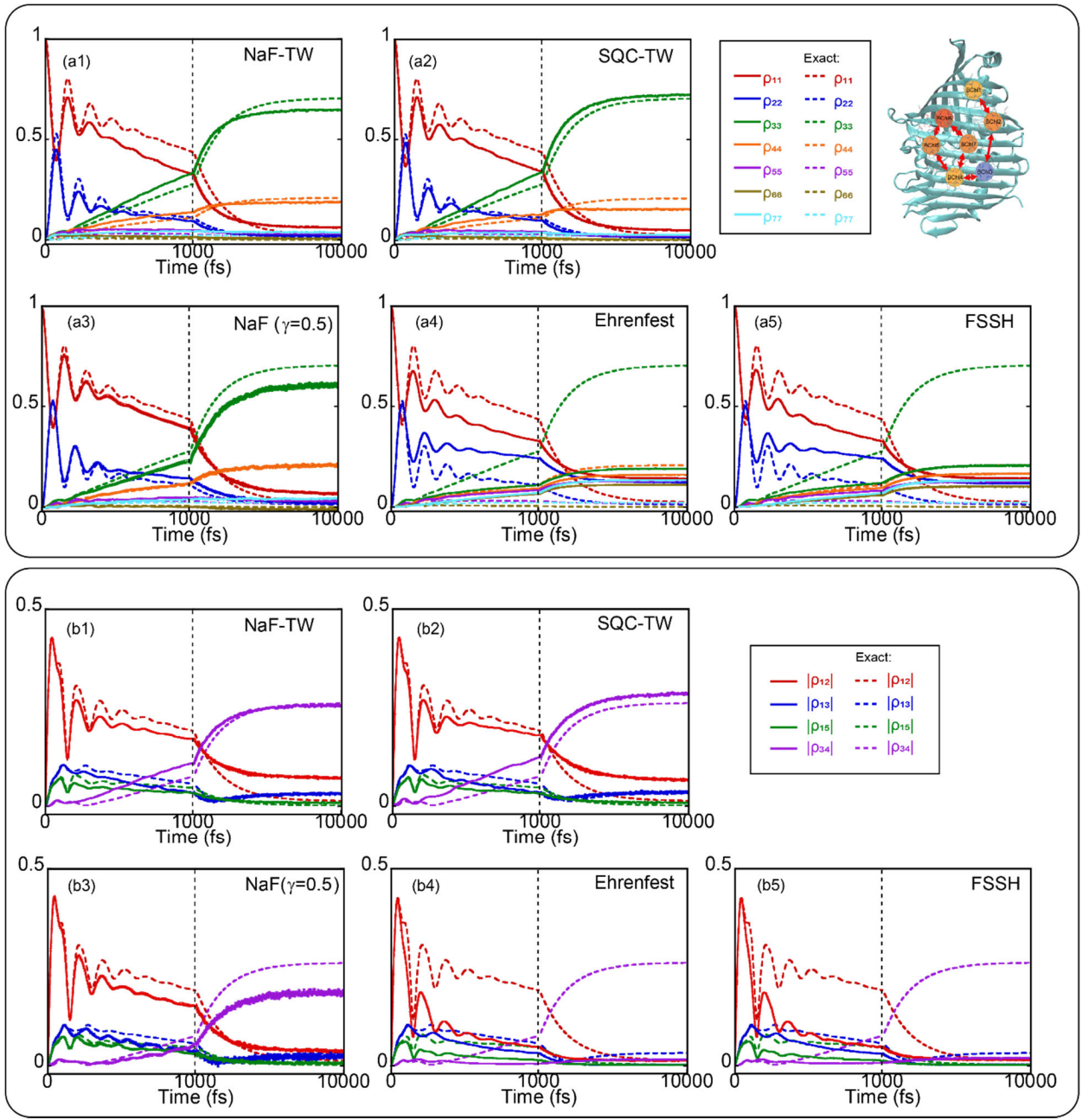

**Figure 2.** Initially occupied site is Site 1. Panel (a): The population dynamics of the seven-site FMO monomer model at temperature 77 K. The red, blue, green, orange, purple, brown, and cyan solid lines represent the population of Sites 1-7, respectively. Panel (b): Dynamics of the off-diagonal (coherence) terms of the reduced electronic density matrix of the same model. The red, blue, green, and purple solid lines illustrate $|\rho_{12}(t)|$, $|\rho_{13}(t)|$, $|\rho_{15}(t)|$, and $|\rho_{34}(t)|$, respectively.

The exact results produced by HEOM are presented by dashed lines in corresponding colors. Subpanels (a1-a5) or Subpanels (b1-b5) denote the results of NaF-TW, SQC-TW, NaF ($\gamma = 0.5$), Ehrenfest dynamics, and FSSH, respectively. One hundred discrete bath modes for each site are employed to obtain converged results. For SQC-TW, eqs (10), (18) and (19) are used for the population-coherence correlation function because its SQC expression with triangle window functions for the multistate system is not available in the literature. More details of the FMO model and the simulations are depicted in Section S3-A of the Supporting Information.

We then consider two typical models of cavity quantum electrodynamics (cQED), where the matter system is tightly coupled to the vacuum field in a confined optical cavity[10, 126-128]. The first atom-in-cavity model involves two atomic energy levels, and the second one includes three energy levels. The highest energy level is initially occupied. More details of the models and initial conditions are described in Supporting Information S2-B. Figure 3 shows that both Ehrenfest dynamics and FSSH lead to significant deviation since a relatively short time and meet challenges in capturing the recoherence around $t=1800$ au. In contrast, SQC-TW, NaF, and NaF-TW yield much more accurate data for the population dynamics of all energy levels and are capable of semiquantitatively describing the short time behavior, as well as the reabsorption and re-emission processes around $t=1800$ au. The comparison between NaF and NaF-TW demonstrates that NaF dynamics performs robustly well, regardless of whether CPS or CPS-TW is used for the electronic representation. The results yielded by NaF-TW are close to those produced by the SQC-TW approach where independent mean-field trajectories are employed.

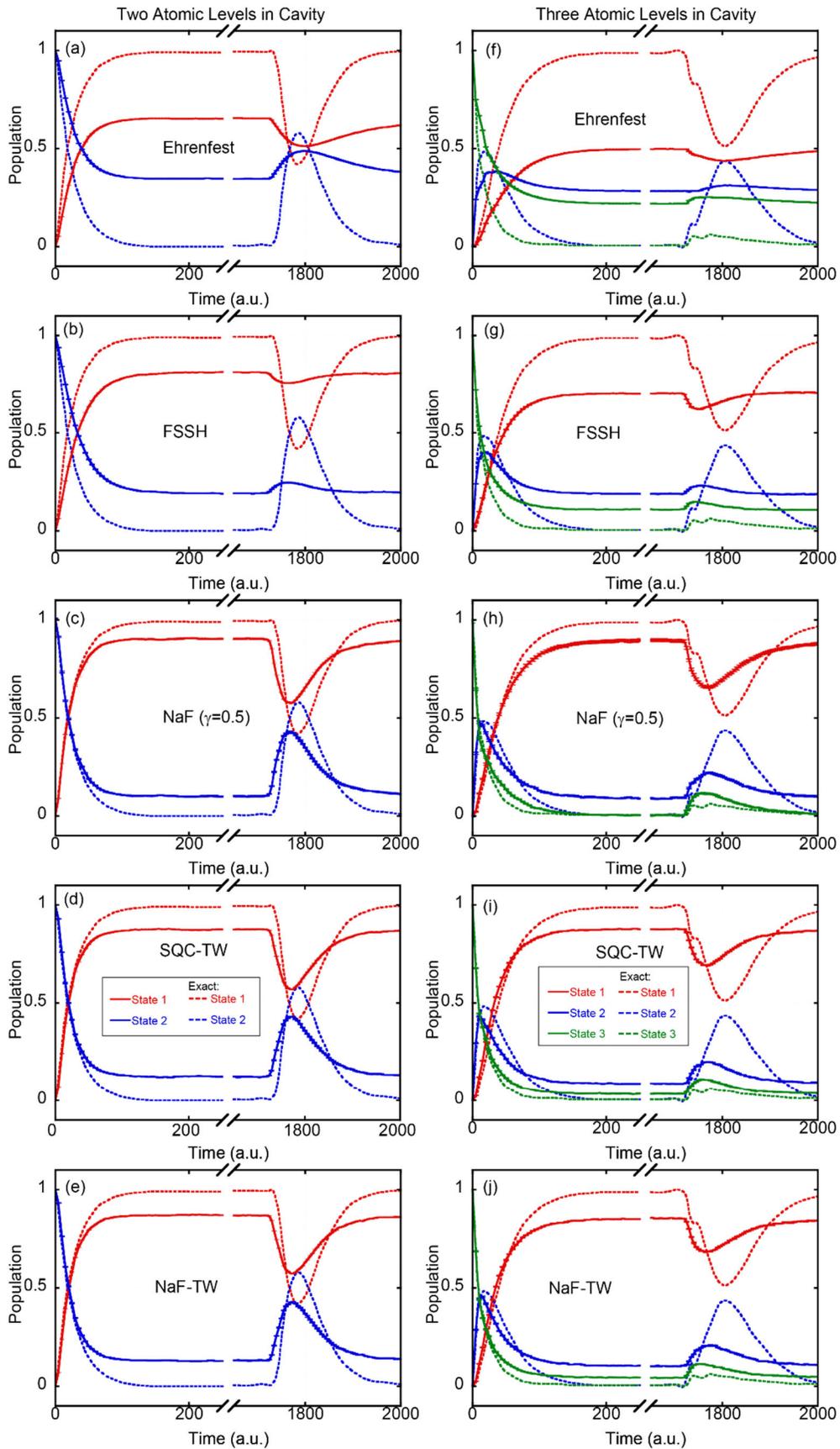

**Figure 3.** Panels (a-e): Results for the population dynamics of the two-level atom-in-cavity model. The red and blue solid lines represent the population of State 1 and that of State 2, respectively, while the dashed lines in corresponding colors demonstrate the exact results. Panels (f-j): Results of the population dynamics of the three-level atom-in-cavity model. The red, blue, and green solid lines represent the population of State 1, State 2, and State 3, respectively, while the dashed lines in corresponding colors demonstrate the exact results. Panels (a-e) or (f-j) present the results of Ehrenfest, FSSH, NaF ($\gamma = 0.5$), SQC-TW, and NaF-TW, respectively. The exact results are obtained from refs [129, 130]. Four hundred standing-wave modes for the optical field are used to obtain converged data. More details of the models and simulations are presented in Section S3-B of the Supporting Information.

In all the three types of models above, the performance of NaF-TW is comparable to that of SQC-TW. Both NaF-TW and NaF outperform FSSH, which suggests that the NaF strategy is superior to conventional SH approaches in studying systems where the electronic states remain coupled all the time.

The third set of tests focuses on the linear vibronic coupling model (LVCM) that captures the characteristic of the pivotal conical intersection (CI) region of the molecular system in various photodriven phenomena. We use two LVCMs where MCTDH results in the diabatic representation are available. The first test case involves the two-electronic-state LVCM with three nuclear modes and that with 24 nuclear modes of refs. [131, 132] which mimic the S1/S2 conical intersection of the pyrazine molecule. The initial condition is set as the cross-product of the vibronic ground state and the excited electronically diabatic state (S2) [131, 132]. The second test case employs a 3-electronic-state 2-nuclear-mode LVCM of the Cr(CO)$_5$ molecule, where the initial condition is the cross-product of a Gaussian nuclear wave packet and the first excited electronically diabatic state as described in ref [133]. More details of the LVCMs are demonstrated in Section S3-C of the Supporting Information. Figure 4 and Figure 5 show the results of the population dynamics in all these LVCM cases. FSSH, NaF, and NaF-TW significantly outperform Ehrenfest dynamics. While FSSH performs slightly better for the 2-state 24-mode case of pyrazine in Figure 4(b), NaF and NaF-TW are overall superior for the 2-state 3-mode case of pyrazine in Figure 4(a) as well as for the 3-state 2-mode LVCM of the $Cr(CO)_5$ molecule. NaF-TW considerably improves over SQC-TW for the peaks of long time dynamics of the pyrazine molecule in Figure 4(a,b) and for the population oscillation behavior when the evolution crosses or recrosses the CI region of the realistic gas phase $Cr(CO)_5$ molecular system, as shown in Figure 5(a-c). NaF and NaF-TW also outperform SQC-TW and Ehrenfest dynamics in reproducing the expectation values of coordinate and momentum variables of the nuclear modes in LVCMs of the pyrazine molecule and the $Cr(CO)_5$ molecule, as presented in Figure 4(c-f) and Figure 5(d), respectively. The results of nuclear dynamics generated by FSSH are comparable to those yielded by NaF and NaF-TW.

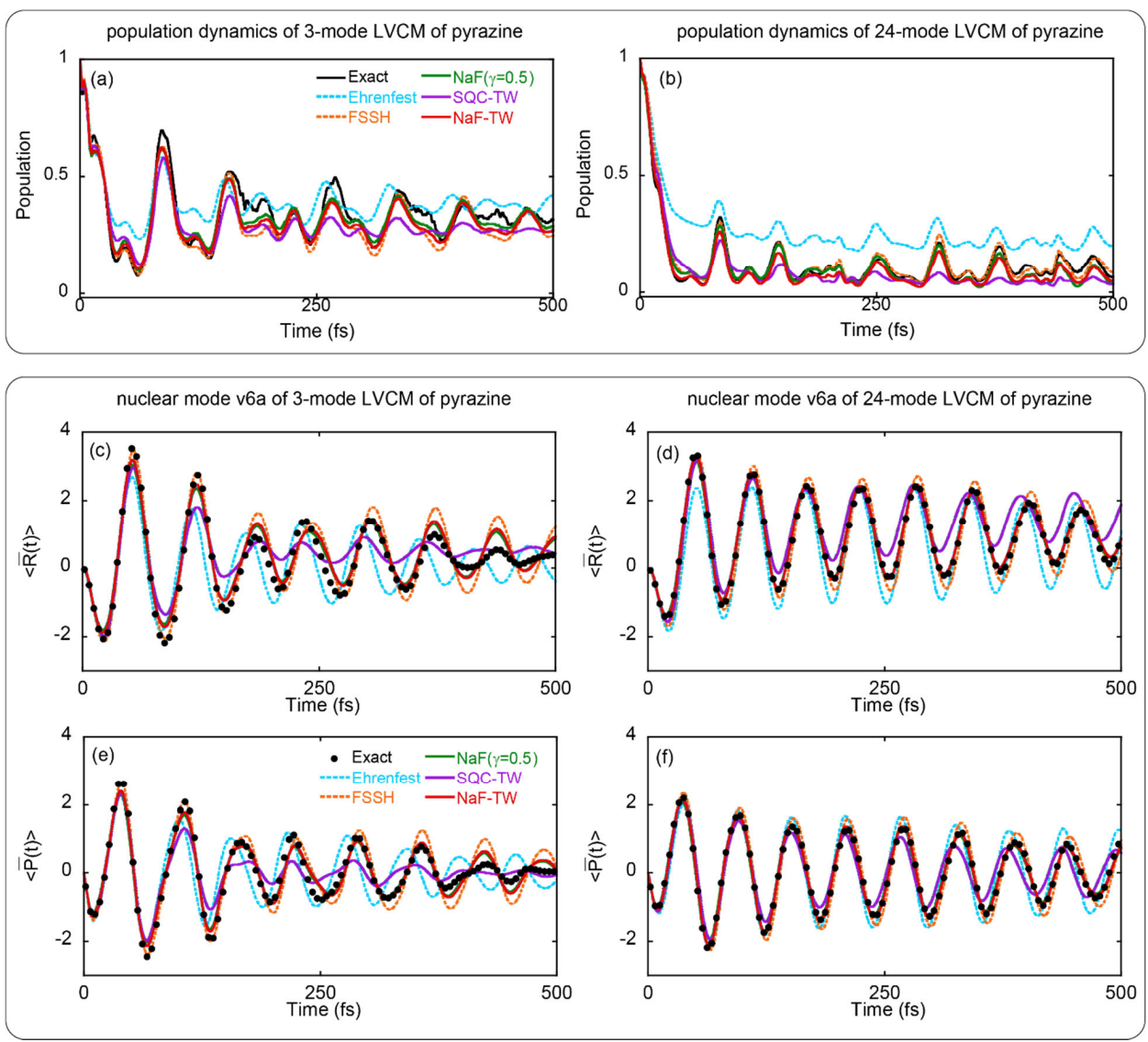

**Figure 4.** Panels (a,b) denote the population dynamics of the second state of the 2-state LVCM with 3 modes for the pyrazine molecule[131] and that with 24 modes for the same molecule[132], respectively. Panels (c-f) denote the average coordinate and momentum of nuclear mode $v_{6a}$ of the 3-mode and 24-mode LVCMs (for the pyrazine molecule), respectively. Cyan dashed lines: Ehrenfest dynamics. Orange dashed lines: FSSH. Green solid lines: NaF ($\gamma = 0.5$). Purple solid lines: SQC-TW. Red solid lines: NaF-TW. Black solid lines and black points represent exact results produced by the MCTDH package[134].

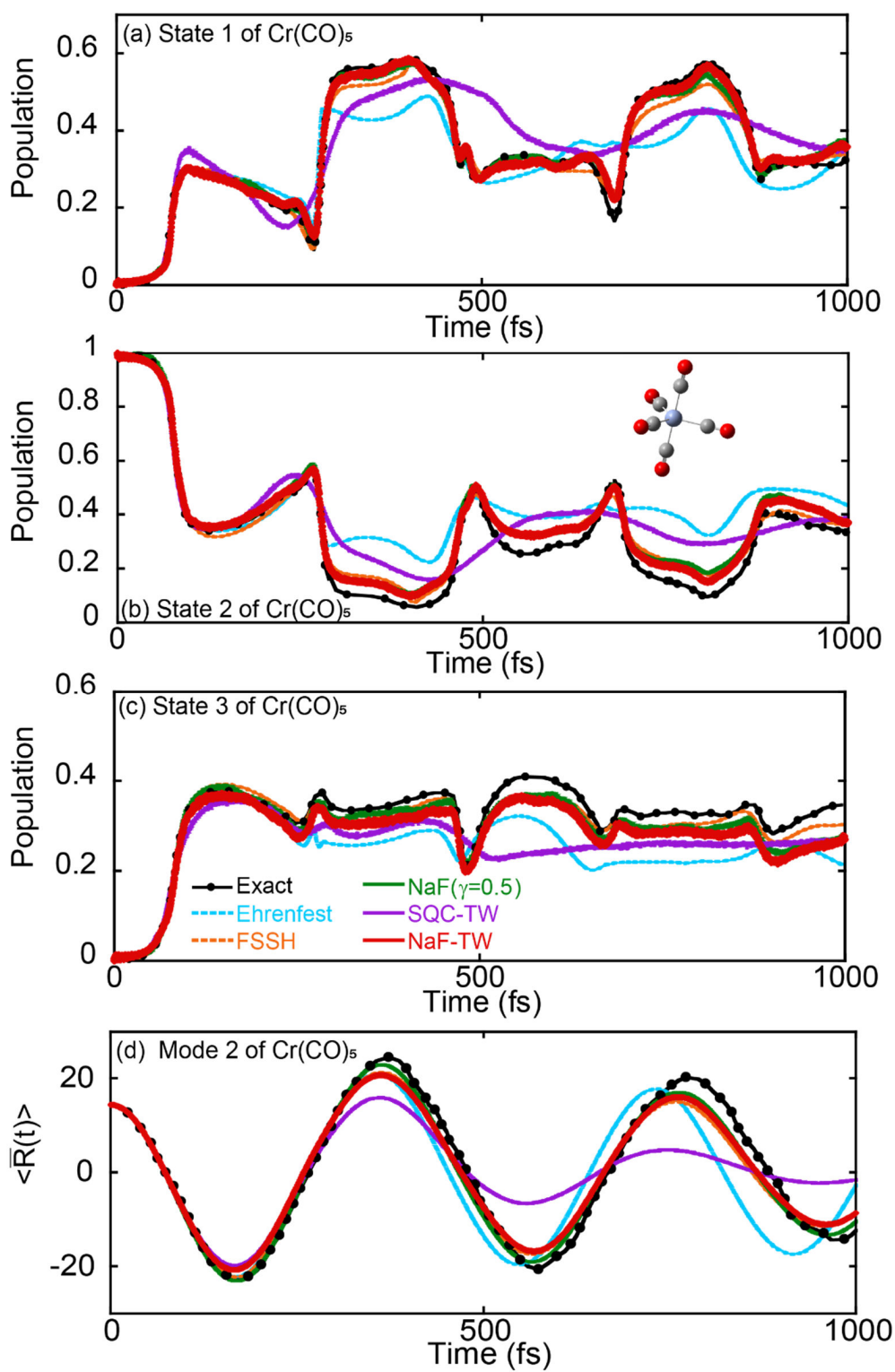

**Figure 5:** Panels (a-c) denote the population dynamics of States 1-3 of the 3-state 2-mode LVCM for the Cr(CO)$_5$ molecule[133], respectively. Panel (d) presents the expectation value of the coordinate variable of the second nuclear normal mode as a function of time. Cyan dashed lines: Ehrenfest dynamics. Orange dashed lines: FSSH. Green solid lines: NaF ($\gamma = 0.5$). Purple solid lines: SQC-TW. Red solid lines: NaF-TW. In Panels (a-d), black solid lines with black points denote exact data (obtained by MCTDH) from ref [133].

Finally, we test typical gas phase models with one anharmonic nuclear DOF where asymptotic regions are involved. The first case includes the coupled three-electronic-state photodissociation models of Miller and co-workers[46]. Numerical details are presented in Section S3-D of the Supporting Information. We focus on Model 3, which is the most challenging. Its nuclear momentum distribution in the long time limit produced by the numerically exact discrete variable representation (DVR) method[135] includes one peak in the higher momentum region and another one in the lower momentum region, while that yielded by Ehrenfest dynamics has only one peak and entirely misses the two-peak characteristic in the asymptotic region. Figure 6 demonstrates the results of Model 3. Although SQC-TW noticeably outperforms Ehrenfest dynamics for the electronic population dynamics, it leads to a broader asymptotic nuclear momentum distribution with only one peak. This indicates that SQC-TW with mean field trajectories is not capable of semiquantitatively capturing the two peaks in the nuclear momentum distribution in the long time limit, which is a consequence of the bifurcation nature of the nuclear motion in the asymptotic region where the nonadiabatic coupling vanishes. As shown in Figure 6, NaF-TW is superior to SQC-TW and yields results similar to those of FSSH and NaF, which are close to exact data by DVR. The second case involves Tully's standard scattering models[16], among which the extended coupling region (ECR) model is the most challenging one. More details are presented in Section S3-E of the Supporting Information. Quantum dynamics of the ECR model involves both the nuclear wavepacket that transmits forwardly and the one that reflects backwardly in asymptotic regions. The dramatic bifurcation characteristic has considerable influence on both electronic and nuclear dynamics. The performance of Ehrenfest dynamics is poor for the ECR model. Figure 7 shows that SQC-TW improves over Ehrenfest dynamics but is unable to reproduce the sharp step-like change in the transmission/reflection probability on State 2 as a function of the momentum of the center of the initial nuclear Gaussian wavepacket. In comparison, NaF-TW, as well as NaF and FSSH, leads to reasonably accurate electronic dynamics for the ECR model. Figure 7(e) and Figure 7(f) demonstrate that, while SQC-TW does not perform well in describing nuclear dynamics for the ECR model in comparison to the exact DVR data, NaF-TW yields a much more accurate nuclear momentum distribution in the asymptotic region. Although NaF-TW and SQC-TW share the same CPS-TW expression of eq (21) for the electronic correlation function, the comparison in Figures 6 and 7 suggests that NaF dynamics (of NaF-TW) is more consistent than Ehrenfest-like dynamics (of SQC-TW) in describing the correct correlation between electronic and nuclear dynamics.

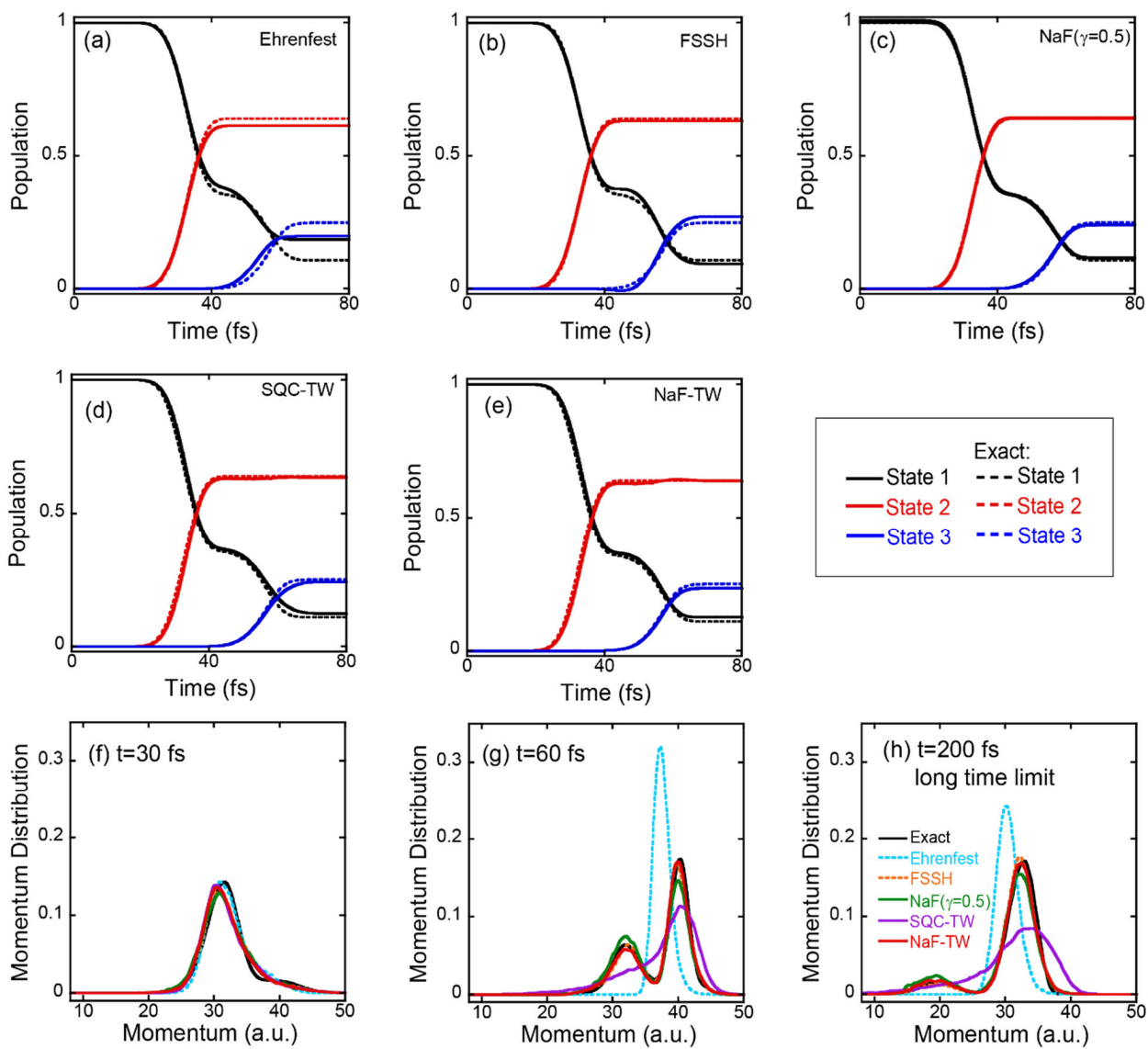

**Figure 6.** Panels (a-e) denote the population dynamics of the third photodissociation model of ref [46], where the black, red, and blue solid lines represent the population of States 1-3, respectively, and the exact results produced by DVR are presented by the dashed lines in corresponding colors. Panels (f-h) present the nuclear momentum distribution at $t = 30$, $60$, and $200$ fs [which is in the asymptotic (long time) limit], respectively. The cyan dashed, orange dashed, green solid, purple solid, and red solid lines represent the results of Ehrenfest dynamics, FSSH, NaF ($\gamma = 0.5$), SQC-

TW, and NaF-TW, respectively. The exact nuclear momentum distribution obtained by DVR is presented by the black solid line.

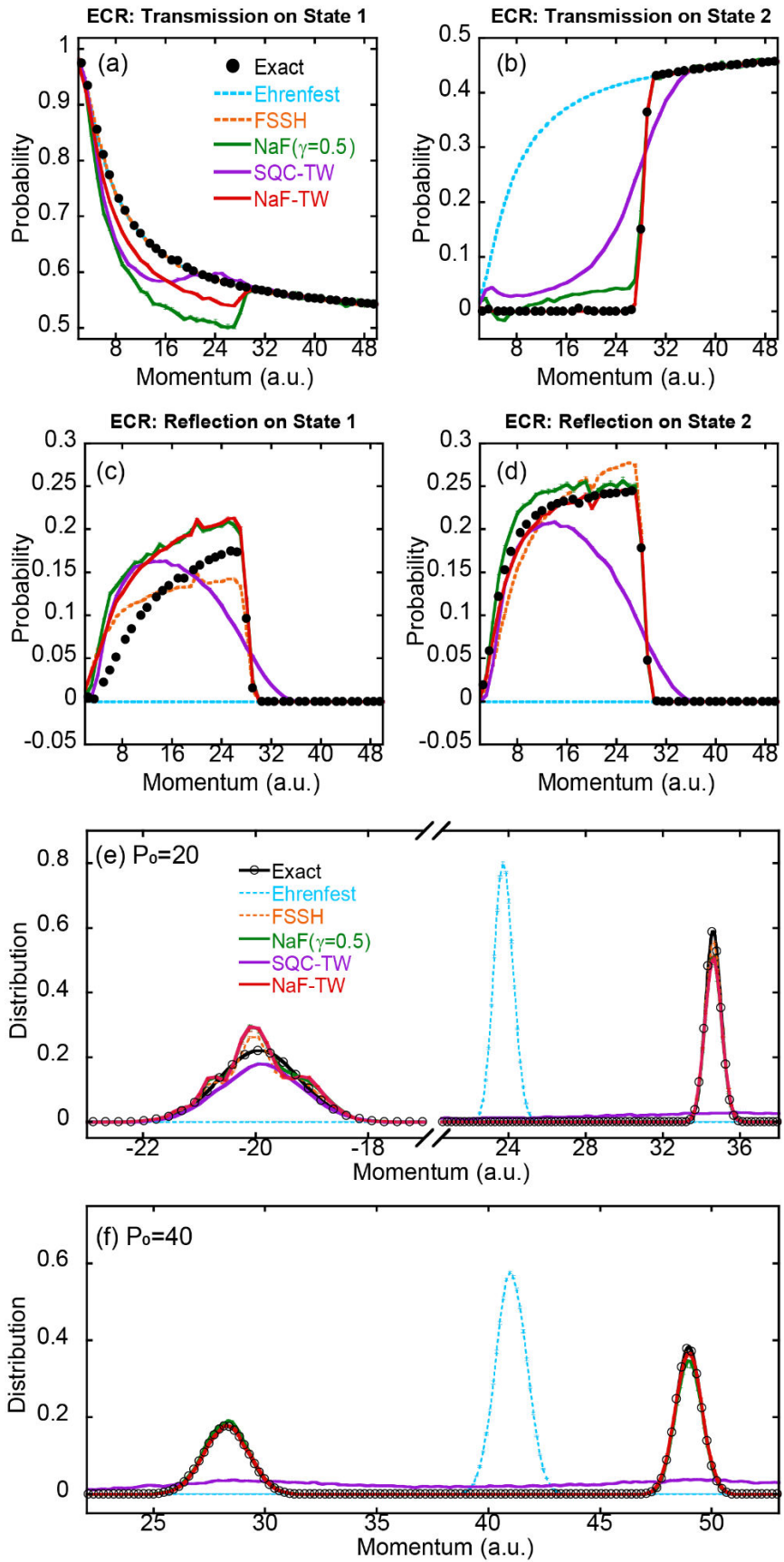

**Figure 7**. Panels (a,b) denote the transmission probability on (adiabatic) State 1 and that on State 2 for Tully's ECR model. Panels (c,d) denote the reflection probability on (adiabatic) State 1 and that on State 2 for the same model. Panels (e) and (f) present the asymptotic nuclear momentum distribution for the initial momentum $P_0 = 20$ of the center of the nuclear Gaussian wavepacket and that for $P_0 = 40$, respectively. Cyan dashed lines: Ehrenfest dynamics. Orange dashed lines: FSSH. Green solid lines: NaF ($\gamma = 0.5$). Purple solid lines: SQC-TW. Red solid lines: NaF-TW. Black points: Exact results by DVR.

In comparison to the exact CPS formulation for discrete (electronic) DOFs, although the CPS-TW representation of the population-population correlation function (i.e., the population dynamics) is exact for only the pure two-state system[103], its applications to three-state or multistate nonadiabatic systems (e.g., in Figures 2, 5, and 6) are also reasonably accurate in practice. As long as the phase space function of the initial total density operator (of both nuclear and electronic DOFs) is non-negative, the CPS-TW expression of the population dynamics (of electronic states) is guaranteed to be positive semidefinite, irrespective of the number of electronic states and the approximation of nuclear dynamics. The advantage of the CPS-TW representation helps NaF-TW outperform NaF in the cases of Figure 6(f-h) and Figure 7.

When the CPS-TW representation of electronic DOFs is used, in addition to NaF-TW, we show another variant that employs the commutator matrix as used in the original NaF approach[104], which is denoted as NaF-TW2. As demonstrated by the extensive numerical tests in Section S4 of the Supporting Information, the numerical performance of NaF-TW2 is very similar to that of NaF-TW.[101]

In summary, since the unified phase space formulation with coordinate-momentum variables offers a powerful tool for studying composite systems, we use the triangle window functions[54, 57, 58] and the $\mathrm{U}(F)/\mathrm{U}(F-1)$ CPS[82, 84, 85] to construct the CPS-TW representation for discrete (electronic) DOFs and employ it with the recently developed NaF strategy[104]. It leads to the NaF-TW approach for nonadiabatic transition dynamics. We test the performance of NaF-TW extensively for a series of standard benchmark condensed phase and gas phase nonadiabatic systems where numerically exact data are feasible for comparison. NaF-TW is able to capture the dynamical correlation between electronic and nuclear DOFs in a reasonably accurate manner. The performance of NaF-TW is similar to that of the SQC approach[57] in (molecular) systems where the electronic states remain coupled all the time. In addition, NaF-TW significantly outperforms the SQC approach in (molecular) systems where the evolution involves the asymptotic region where the state-state coupling disappears. While NaF-TW captures the correct correlation of electronic and nuclear dynamics in the asymptotic region, more advanced semiclassical methods[50, 73] beyond the SQC approach are necessary to accomplish the same task when independent mean field trajectories are used.

The comprehensive benchmark numerical tests in the main text as well as in the Supporting Information suggest that NaF dynamics is overall superior to conventional surface hopping dynamics and Ehrenfest-like dynamics in a broad region. Although the CPS formulation or CPS-TW representation may also be used with various surface hopping dynamics or Ehrenfest-like dynamics methods, NaF dynamics is highly recommended for its superiority. We also note that, since NaF or NaF-TW only involves the independent trajectory without the phase, it is rather difficult to perform well in the deep quantum tunneling or quantum coherence region. The semiclassical-like approach or time-dependent multiconfiguration approach, which includes the NaF trajectory with its corresponding phase, will be a potential tool for studying the much more challenging quantum mechanical region.

Because the CPS formulation and CPS-TW representation can also be used for the interpretation of discrete quantum states of light atoms or those of high-frequency vibrational DOFs, the NaF strategy can be employed to study nuclear quantum effects in proton/hydrogen transfer processes[21, 56, 136]. It is expected that further development of NaF-TW and the CPS formulations "in ever-increasing levels of abstraction"[137] will lead to a promising and robust trajectory-based approach for investigating electronically (and/or vibrationally) nonadiabatic transition phenomena (internal conversion, intersystem crossing, electron/hole/proton/hydrogen transfer, etc.) and dynamic processes with important quantum effects in complex/large composite systems in chemistry, biology, materials, environmental science, quantum information, quantum computation, and so forth.

# ■ ASSOCIATED CONTENT

**Supporting Information**.

The Supporting Information is available free of charge at https://pubs.acs.org/doi/10.1021/acs.jpclett.4c00793.

> Five sections: proof that the CPS-TW expression of the time correlation function involving coherence terms is exact for electronic dynamics of the pure $F$-state quantum system; numerical details in evaluation of the time correlation function in NaF-TW, NaF, and SQC-TW; simulation details for models in the main text; comparison of different strategies employing triangle window functions; mapping triangle window functions onto constraint phase space leads to novel phase space formulations for the $F$-state quantum systems (PDF)

■ **AUTHOR INFORMATION**

**Corresponding Author**

*E-mail: jianliupku@pku.edu.cn

**ORCID**

Xin He: 0000-0002-5189-7204

Xiangsong Cheng: 0000-0001-8793-5092

Baihua Wu: 0000-0002-1256-6859

Jian Liu: 0000-0002-2906-5858

**Notes**

The authors declare no competing financial interest.

■ **ACKNOWLEDGMENT**

This work was supported by the National Science Fund for Distinguished Young Scholars Grant No. 22225304. We acknowledge the High-performance Computing Platform of Peking University, Beijing PARATERA Tech Co., Ltd., and Guangzhou supercomputer center for providing computational resources. We thank Youhao Shang, Haocheng Lu, and Bingqi Li for useful discussions. We also thank Bill Miller for having encouraged us to investigate the window function approach.

# Supporting Information:

# Nonadiabatic Field with Triangle Window Functions on Quantum Phase Space

*Xin He, Xiangsong Cheng, Baihua Wu, and Jian Liu\**

Beijing National Laboratory for Molecular Sciences, Institute of Theoretical and Computational Chemistry, College of Chemistry and Molecular Engineering,

Peking University, Beijing 100871, China

AUTHOR INFORMATION

**Corresponding Author**

\* Electronic mail: jianliupku@pku.edu.cn

## S1: Proof that the CPS-TW Expression of the Time Correlation Function Involving Coherence Terms is Exact for Electronic Dynamics of the Pure *F*-State Quantum System

In the literature, when $n \neq m$, $|n\rangle\langle m|$ is also denoted as a coherence term, and $|n\rangle\langle n|$ represents a population term. In the four kinds of time correlation functions of electronic degrees of freedom (DOFs), while only the population-population correlation function does not involve any coherence terms, all of the other three kinds include coherence terms. As proved in ref [1], the CPS-TW expression for the population-population correlation function, which includes eqs (11)-(17) of the main text (of the present letter), is exact for the pure two-state quantum system. Although it is not exact for the population dynamics of the pure *F*-state quantum system when $F \geq 3$, the numerical performance is reasonably good. Below we focus on the CPS-TW expressions for the three kinds of time correlation functions involving coherence terms, which are eqs (10), (12), (14)-(15) and (18)-(21) of the main text.

Consider the pure *F*-electronic-state system, i.e., the frozen nuclei-limit of the Hamiltonian of eq (1) of the main text, which reads $\hat{H} = \sum_{n,m=1}^{F} H_{nm} |n\rangle\langle m|$. Here, $H_{nm}$ is set as the constant element in the *n*-th row and *m*-th column of the $F \times F$ Hermitian matrix $\mathbf{H}$. (When only electronic DOFs are involved, the convention $\hbar = 1$ is adopted.) For simplicity, following ref [2] of Cotton and Miller, we define the action variables $\{\mathbf{e}\}$ and angle variables $\{\mathbf{\theta}\}$ as

$$e^{(k)} = \frac{1}{2}\left(\left(x^{(k)}\right)^2 + \left(p^{(k)}\right)^2\right)$$
$$\theta^{(k)} = \arctan\left(p^{(k)}/x^{(k)}\right)$$
(S1)

We utilize the transformation between coordinate-momentum variables and action-angle variables, eq (S1), to obtain the expression of the population-coherence correlation function (where $k \neq l$):

$$\operatorname{Tr}_e\left[|n\rangle\langle n|e^{i\hat{H}t}|k\rangle\langle l|e^{-i\hat{H}t}\right]$$

$$\mapsto \int \frac{2\mathrm{d}\mathbf{x}_0 \mathrm{d}\mathbf{p}_0}{(2\pi)^F \left(2 - \frac{1}{2}\left((x_0^{(n)})^2 + (p_0^{(n)})^2\right)\right)^{F-2}} K_{nn}^{\mathrm{SQC}}(\mathbf{x}_0, \mathbf{p}_0) K_{lk}^{\mathrm{CMM}}(\mathbf{x}_t, \mathbf{p}_t) , \tag{S2}$$

$$= \int \frac{2\mathrm{d}\mathbf{e}_0 \mathrm{d}\boldsymbol{\theta}_0}{(2\pi)^F (2 - e_0^{(n)})^{F-2}} K_{nn}^{\mathrm{SQC}}(\mathbf{e}_0, \boldsymbol{\theta}_0) K_{lk}^{\mathrm{CMM}}(\mathbf{e}_t, \boldsymbol{\theta}_t)$$

where

$$K_{nn}^{\mathrm{SQC}}(\mathbf{e},\boldsymbol{\theta}) \equiv K_{nn}^{\mathrm{SQC}}(\mathbf{e}) = h(e^{(n)} - 1) \prod_{n' \neq n} h(2 - e^{(n)} - e^{(n')}) \tag{S3}$$

and

$$K_{lk}^{\mathrm{CMM}}(\mathbf{e},\boldsymbol{\theta}) = \sqrt{e^{(l)} e^{(k)}} e^{i(\theta^{(l)} - \theta^{(k)})} \tag{S4}$$

are derived from substitution of eq (S1) into eqs (14)-(15) and eq (19) of the main text.

For $n \neq m$, the expression of the coherence-population or coherence-coherence correlation function reads

$$\operatorname{Tr}_e\left[|n\rangle\langle m|e^{i\hat{H}t}|k\rangle\langle l|e^{-i\hat{H}t}\right]$$

$$\mapsto \int \frac{12\mathrm{d}\mathbf{x}_0 \mathrm{d}\mathbf{p}_0}{5(2\pi)^F} \sum_{i=n,m} \frac{K_{ii}^{\mathrm{SQC}}(\mathbf{x}_0, \mathbf{p}_0)}{\left(2 - \frac{1}{2}\left((x_0^{(i)})^2 + (p_0^{(i)})^2\right)\right)^{F-2}} K_{mn}^{\mathrm{CMM}}(\mathbf{x}_0, \mathbf{p}_0) K_{lk}^{\mathrm{CMM}}(\mathbf{x}_t, \mathbf{p}_t) . \tag{S5}$$

$$= \int \frac{12\mathrm{d}\mathbf{e}_0 \mathrm{d}\boldsymbol{\theta}_0}{5(2\pi)^F} \sum_{i=n,m} \frac{K_{ii}^{\mathrm{SQC}}(\mathbf{e}_0, \boldsymbol{\theta}_0)}{(2 - e_0^{(i)})^{F-2}} K_{mn}^{\mathrm{CMM}}(\mathbf{e}_0, \boldsymbol{\theta}_0) K_{lk}^{\mathrm{CMM}}(\mathbf{e}_t, \boldsymbol{\theta}_t)$$

In comparison, for general $n, m, k, l \in \{1, \cdots, F\}$, the exact time correlation function reads

$$\operatorname{Tr}_e\left[|n\rangle\langle m|e^{i\hat{H}t}|k\rangle\langle l|e^{-i\hat{H}t}\right] = U_{ln}(t) U_{km}^*(t) , \tag{S6}$$

where the time evolution operator $e^{-i\hat{H}t}$ is represented by a $F \times F$ unitary matrix $\mathbf{U}(t) = e^{-i\mathbf{H}t}$.

We prove that the correlation function defined by eq (S2) or eq (S5) for electronic DOFs is exact, when the equations of motion (EOMs) of mapping electronic variables are

$$\dot{\mathbf{g}} = -i\mathbf{H}\mathbf{g}, \tag{S7}$$

which leads to the solution of $\mathbf{g}_t$,

$$\mathbf{g}_t = \mathbf{U}(t)\mathbf{g}_0 \equiv e^{-i\mathbf{H}t}\mathbf{g}_0. \tag{S8}$$

Equation (S7) is isomorphic to the time-dependent Schrödinger equation. Following eq (S8), the time evolution of off-diagonal (i.e., $l \ne k$) elements of the CMM kernel reads

$$\begin{aligned} K_{lk}^{\text{CMM}}(\mathbf{e}_t, \boldsymbol{\theta}_t) &= K_{lk}^{\text{CMM}}(\mathbf{x}_t, \mathbf{p}_t) \\ &= \frac{1}{2}\left(x_t^{(l)} + ip_t^{(l)}\right)\left(x_t^{(k)} - ip_t^{(k)}\right) \\ &= \frac{1}{2}(\mathbf{U}(t)\mathbf{g}_0)_l \left(\mathbf{g}_0^\dagger \mathbf{U}^\dagger(t)\right)_k \\ &= \sum_{r,s=1}^{F} U_{lr}(t) U_{ks}^*(t) \sqrt{e_0^{(r)} e_0^{(s)}} e^{i\left(\theta_0^{(r)} - \theta_0^{(s)}\right)} \end{aligned} \tag{S9}$$

**1) The Proof of the Exact *Population-Coherence* Correlation Function for Electronic Dynamics of the Pure *F*-State Quantum System**

When only electronic DOFs are involved, by using eq (S9), and also utilizing that for any $\theta_0^{(k)}, k = 1, \ldots, F$, and integer $m$,

$$\int_0^{2\pi} e^{im\theta_0^{(k)}} \, \mathrm{d}\theta_0^{(k)} = 2\pi\delta_{m0} = \delta_{m0}\int_0^{2\pi} \mathrm{d}\theta_0^{(k)}, \tag{S10}$$

it is straight-forward to show the population-coherence correlation function eq (S2) leads to

$$\begin{aligned} &\int \frac{2\mathrm{d}\mathbf{e}_0 \mathrm{d}\boldsymbol{\theta}_0}{(2\pi)^F \left(2 - e_0^{(n)}\right)^{F-2}} K_{nn}^{\text{SQC}}(\mathbf{e}_0, \boldsymbol{\theta}_0) K_{lk}^{\text{CMM}}(\mathbf{e}_t, \boldsymbol{\theta}_t) \\ &= \int \frac{2\mathrm{d}\mathbf{e}_0 \mathrm{d}\boldsymbol{\theta}_0}{(2\pi)^F \left(2 - e_0^{(n)}\right)^{F-2}} K_{nn}^{\text{SQC}}(\mathbf{e}_0) \sum_{r,s=1}^{F} U_{lr}(t) U_{ks}^*(t) \sqrt{e_0^{(r)} e_0^{(s)}} e^{i\left(\theta_0^{(r)} - \theta_0^{(s)}\right)} \\ &= \int \frac{2\mathrm{d}\mathbf{e}_0}{\left(2 - e_0^{(n)}\right)^{F-2}} K_{nn}^{\text{SQC}}(\mathbf{e}_0) \sum_{r=1}^{F} U_{lr}(t) U_{kr}^*(t) e_0^{(r)} \end{aligned} \tag{S11}$$

Then, using the properties that for $l \neq k$, $\sum_{r=1}^{F} U_{lr}(t)U_{kr}^*(t) = 0$ and that for any $r \neq s \neq n$,

$$\int \frac{2d\mathbf{e}}{(2-e^{(n)})^{F-2}} K_{nn}^{SQC}(\mathbf{e}) e^{(r)} = \int \frac{2d\mathbf{e}}{(2-e^{(n)})^{F-2}} K_{nn}^{SQC}(\mathbf{e}) e^{(s)}, \quad (S12)$$

eq (S11) is transformed to

$$\int \frac{2d\mathbf{e}_0}{(2-e_0^{(n)})^{F-2}} K_{nn}^{SQC}(\mathbf{e}_0) \sum_{r=1}^{F} U_{lr}(t) U_{kr}^*(t) e_0^{(r)}$$
$$= U_{ln}(t) U_{kn}^*(t) \int \frac{2d\mathbf{e}_0}{(2-e_0^{(n)})^{F-2}} K_{nn}^{SQC}(\mathbf{e}_0) \left( e_0^{(n)} - e_0^{(r \neq n)} \right) \quad (S13)$$

In ref [2] Cotton and Miller obtained the following integrals,

$$\int \frac{2d\mathbf{e}}{(2-e^{(n)})^{F-2}} K_{nn}^{SQC}(\mathbf{e}) e^{(n)}$$
$$= \int_1^2 2de^{(n)} e^{(n)} (2-e^{(n)})^{2-F} \prod_{n' \neq n} \int_0^{2-e^{(n)}} de^{(n')}, \quad (S14)$$
$$= \frac{4}{3}$$

and

$$\int \frac{2d\mathbf{e}}{(2-e^{(n)})^{F-2}} K_{nn}^{SQC}(\mathbf{e}) e^{(r \neq n)}$$
$$= \int_1^2 2de^{(n)} (2-e^{(n)})^{2-F} \int_0^{2-e^{(n)}} de^{(r)} e^{(r)} \prod_{n' \neq \{n,r\}} \int_0^{2-e^{(n)}} de^{(n')}. \quad (S15)$$
$$= \frac{1}{3}$$

Substitution of eqs (S14) and (S15) into eq (S13) yields

S4

$$\int \frac{2\mathrm{d}\mathbf{e}_0 \mathrm{d}\boldsymbol{\theta}_0}{(2\pi)^F \left(2-e_0^{(n)}\right)^{F-2}} K_{nn}^{\mathrm{SQC}}(\mathbf{e}_0,\boldsymbol{\theta}_0) K_{lk}^{\mathrm{CMM}}(\mathbf{e}_t,\boldsymbol{\theta}_t)$$

$$= U_{ln}(t) U_{kn}^*(t) \left(\frac{4}{3}-\frac{1}{3}\right) \tag{S16}$$

$$= \mathrm{Tr}\left[|n\rangle\langle n|e^{i\hat{H}t}|k\rangle\langle l|e^{-i\hat{H}t}\right]$$

Thus, eq (S2), the expression of the population-coherence correlation function, reproduces the exact result of the pure *F*-electronic-state system.

**2) The Proof of the Exact *Coherence-Population* Correlation Function and *Coherence-Coherence* Correlation Function for Electronic Dynamics of the Pure *F*-State Quantum System**

Utilizing eqs (S9)-(S10), for eq (S5) with $n \neq m$, we obtain

$$\begin{aligned}
&\int \frac{12\mathrm{d}\mathbf{e}_0 \mathrm{d}\boldsymbol{\theta}_0}{5(2\pi)^F} \sum_{i=n,m} \frac{K_{ii}^{\mathrm{SQC}}(\mathbf{e}_0,\boldsymbol{\theta}_0)}{\left(2-e_0^{(i)}\right)^{F-2}} K_{mn}^{\mathrm{CMM}}(\mathbf{e}_0,\boldsymbol{\theta}_0) K_{lk}^{\mathrm{CMM}}(\mathbf{e}_t,\boldsymbol{\theta}_t) \\
&= \int \frac{12\mathrm{d}\mathbf{e}_0 \mathrm{d}\boldsymbol{\theta}_0}{5(2\pi)^F} \sum_{i=n,m} \frac{K_{ii}^{\mathrm{SQC}}(\mathbf{e}_0,\boldsymbol{\theta}_0)}{\left(2-e_0^{(i)}\right)^{F-2}} \sqrt{e_0^{(m)} e_0^{(n)}}\, e^{i\left(\theta_0^{(m)}-\theta_0^{(n)}\right)} \\
&\quad \times \sum_{r,s=1}^{F} U_{lr}(t) U_{ks}^*(t) \sqrt{e_0^{(r)} e_0^{(s)}}\, e^{i\left(\theta_0^{(r)}-\theta_0^{(s)}\right)} \\
&= U_{ln}(t) U_{km}^*(t) \int \frac{12\mathrm{d}\mathbf{e}_0}{5} \sum_{i=n,m} \frac{K_{ii}^{\mathrm{SQC}}(\mathbf{e}_0)}{\left(2-e_0^{(i)}\right)^{F-2}} e_0^{(m)} e_0^{(n)}
\end{aligned} \tag{S17}$$

we can derive that

$$\begin{aligned}
&\int 2\mathrm{d}\mathbf{e}\, \frac{K_{kk}^{\mathrm{SQC}}(\mathbf{e})}{\left(2-e^{(k)}\right)^{F-2}} e^{(k)} e^{(j\neq k)} \\
&= \int_1^2 2\mathrm{d}e^{(k)} e^{(k)} \left(2-e^{(k)}\right)^{2-F} \int_0^{2-e^{(k)}} \mathrm{d}e^{(j)} e^{(j)} \prod_{k'\neq\{j,k\}} \int_0^{2-e^{(k)}} \mathrm{d}e^{(k')} \\
&= \frac{5}{12}
\end{aligned} \tag{S18}$$

S5

By substituting eq (S18) into eq (S17), we obtain

$$\int \frac{12 d\mathbf{e}_0 d\boldsymbol{\theta}_0}{5(2\pi)^F} \sum_{i=n,m} \frac{K_{ii}^{SQC}(\mathbf{e}_0, \boldsymbol{\theta}_0)}{(2-e_0^{(i)})^{F-2}} K_{mn}^{CMM}(\mathbf{e}_0, \boldsymbol{\theta}_0) K_{lk}^{CMM}(\mathbf{e}_t, \boldsymbol{\theta}_t)$$

$$= U_{ln}(t) U_{km}^*(t) \left( \frac{6}{5} \times 2 \times \frac{5}{12} \right) \qquad , \qquad (S19)$$

$$= \text{Tr}\left[ |n\rangle\langle m| e^{i\hat{H}t} |k\rangle\langle l| e^{-i\hat{H}t} \right]$$

so that eq (S5), the expression of the coherence-population correlation function as well as the coherence-coherence correlation function, is also exact for electronic dynamics of the pure $F$-state quantum system.

**S2: Numerical Details in Evaluation of the Time Correlation Function in NaF-TW, NaF and SQC-TW**

As discussed in ref [3], in evaluation of the time correlation function of eq (21) of the main text, i.e.,

$$\text{Tr}_{n,e}\left[ \left( |n\rangle\langle m| \otimes \hat{\rho}_{nuc} \right) e^{i\hat{H}t/\hbar} \left( |k\rangle\langle l| \otimes \hat{A}_{nuc} \right) e^{-i\hat{H}t/\hbar} \right], \qquad (S20)$$

trajectory-based quantum dynamics methods involve three key elements: the initial condition, the integral expression of the time correlation function, and the EOMs for trajectories. The main text introduces NaF-TW, which employs the triangle window (TW) functions for electronic DOFs for the time correlation function. In this section, we present numerical details of NaF-TW, NaF and SQC-TW for the three key elements in the phase space formulation.

**S2-A: Initial Conditions for NaF-TW, NaF and SQC-TW**

The integral expression of the time correlation function of eq (21) in the main text is interpreted as the summation over trajectories with proper initial conditions. The initial value of phase space variables of nuclear DOFs is sampled from the Wigner distribution

S6

$\rho_W(\mathbf{R},\mathbf{P}) = \int d\Delta \langle \mathbf{R}-\Delta/2|\hat{\rho}_{\text{nuc}}|\mathbf{R}+\Delta/2\rangle \exp(-i\Delta\cdot\mathbf{P}/\hbar)$. The sampling procedures of the initial value of phase space variables of electronic DOFs, including coordinate-momentum variables $(\mathbf{x},\mathbf{p})$ and commutator matrix variables $\Gamma$ in the diabatic representation (or $(\tilde{\mathbf{x}},\tilde{\mathbf{p}})$ and $\tilde{\Gamma}$ in the adiabatic representation), are separately described for the NaF-TW, NaF and SQC-TW methods.

1) **Initial Value of Coordinate-Momentum Variables $(\mathbf{x}_0,\mathbf{p}_0)$ or $(\tilde{\mathbf{x}}_0,\tilde{\mathbf{p}}_0)$ for Electronic DOFs in NaF**

In NaF, the initial value of $(\mathbf{x}_0,\mathbf{p}_0)$ in the diabatic representation or $(\tilde{\mathbf{x}}_0,\tilde{\mathbf{p}}_0)$ in the adiabatic representation, is uniformly sampled on the corresponding CPS, $\mathcal{S}(\mathbf{x},\mathbf{p};\gamma)$ or $\mathcal{S}(\tilde{\mathbf{x}},\tilde{\mathbf{p}};\gamma)$. This process includes the following steps:

**Step 1**: Generate a $2F$-dimensional random vector $\boldsymbol{\xi}$ whose components are independently sampled from the standard normal (Gaussian) distribution, i.e., $\xi^{(k)} \sim \mathcal{N}(0,1)$ with $k=1,\cdots,2F$.

**Step 2**: Normalize the vector $\boldsymbol{\xi}$: for $k=1,\cdots,2F$, $\xi_{\text{norm}}^{(k)} = \xi^{(k)}/\sqrt{\sum_{i=1}^{2F}(\xi^{(i)})^2}$.

**Step 3:** The initial value of coordinate variables is proportional to the value of the first $F$ components of $\boldsymbol{\xi}_{\text{norm}}$, while the initial value of momentum variables is proportional to that of the remaining $F$ components, i.e.,

$$\begin{aligned} x_0^{(k)} &= \sqrt{2(1+F\gamma)}\xi_{\text{norm}}^{(k)} \\ p_0^{(k)} &= \sqrt{2(1+F\gamma)}\xi_{\text{norm}}^{(F+k)} \end{aligned}, \quad k=1,\cdots,F \qquad (S21)$$

in the diabatic representation, or similarly

$$\begin{aligned} \tilde{x}_0^{(k)} &= \sqrt{2(1+F\gamma)}\xi_{\text{norm}}^{(k)} \\ \tilde{p}_0^{(k)} &= \sqrt{2(1+F\gamma)}\xi_{\text{norm}}^{(F+k)} \end{aligned}, \quad k=1,\cdots,F \qquad (S22)$$

in the adiabatic representation.

## 2) Initial Value of Coordinate-Momentum Variables $(\mathbf{x}_0, \mathbf{p}_0)$ or $(\tilde{\mathbf{x}}_0, \tilde{\mathbf{p}}_0)$ for Electronic DOFs in NaF-TW or SQC-TW

In either NaF-TW or SQC-TW, the initial sampling procedure of electronic phase space variables $(\mathbf{x}_0, \mathbf{p}_0)$ in the diabatic representation reads:

**Step 1**: Choose the initially occupied state. When we consider the population-population ($n = m$ and $k = l$) or population-coherence ($n = m$ and $k \neq l$) correlation function, set the occupied state $j_{\text{occ}}$ as $n$. When we calculate the coherence-population ($n \neq m$ and $k = l$) or coherence-coherence ($n \neq m$ and $k \neq l$) correlation function, $j_{\text{occ}}$ is randomly assigned to be either of $n$ and $m$ with the equal probability.

**Step 2**: When the occupied state is $j_{\text{occ}}$, generate the initial value of action-angle variables $\{\mathbf{e}, \boldsymbol{\theta}\}$ by using the algorithm of Cotton and Miller described in ref [2]. First, iteratively generate two real random numbers $\zeta$ and $\zeta_2$ from a uniform distribution on domain $[0,1]$ until their sum $\zeta + \zeta_2 < 1$, and record the value of $\zeta$. The probability density function of $\zeta$ is proportional to $1 - \zeta$ for $\zeta \in [0,1]$ according to the algorithm. Subsequently, sample a set of random numbers $\{\eta^{(k)}, k = 1, \ldots, F \text{ and } k \neq j_{\text{occ}}\}$ from a uniform distribution on domain $[0, 1-\zeta]$. The initial value of action variables is assigned as

$$e_0^{(k)} = \begin{cases} 1 + \zeta, & \text{if } k = j_{\text{occ}} \\ \eta^{(k)}, & \text{otherwise} \end{cases}. \tag{S23}$$

The initial value of each angle variable, $\theta_0^{(k)}$ ($k = 1, \ldots, F$), is independently sampled from a uniform distribution on domain $[0, 2\pi)$.

**Step 3:** The initial value of coordinate-momentum variables for electronic DOFs is obtained from that of action-angle variables by the transformation

$$x_0^{(k)} = \sqrt{2e_0^{(k)}} \cos\theta_0^{(k)}$$
$$p_0^{(k)} = \sqrt{2e_0^{(k)}} \sin\theta_0^{(k)} \quad \text{(S24)}$$

Similarly, in the adiabatic representation, we generate the initial value of action-angle variables $(\tilde{\mathbf{e}}_0, \tilde{\boldsymbol{\theta}}_0)$ from Step 2 and subsequently obtain $(\tilde{\mathbf{x}}_0, \tilde{\mathbf{p}}_0)$ by Step 3.

### 3) Initial Value of Commutator Matrix Variables $\Gamma_0$ or $\tilde{\Gamma}_0$ for Electronic DOFs in NaF, NaF-TW, or SQC-TW

NaF-TW, NaF, and SQC-TW differ in the treatment of the initial condition of commutator matrix $\Gamma_0$ in the diabatic representation (or $\tilde{\Gamma}_0$ in the adiabatic representation). When we use NaF-TW in the main text, $\Gamma_0$ (or $\tilde{\Gamma}_0$) is set to the constant matrix $\gamma\mathbf{I}$. As comparison, when we employ NaF or SQC-TW, the initial commutator matrix is expressed as

$$\Gamma_0^{(nm)} = \left(\frac{1}{2}\left(\left(x_0^{(n)}\right)^2 + \left(p_0^{(n)}\right)^2\right) - \delta_{n,j_{\text{occ}}}\right)\delta_{nm} \quad \text{(S25)}$$

in the diabatic representation, or

$$\tilde{\Gamma}_0^{(nm)} = \left(\frac{1}{2}\left(\left(\tilde{x}_0^{(n)}\right)^2 + \left(\tilde{p}_0^{(n)}\right)^2\right) - \delta_{n,j_{\text{occ}}}\right)\delta_{nm} \quad \text{(S26)}$$

in the adiabatic representation.

As mentioned in the main text, when the initial condition is defined and the initial sampling is prepared in the diabatic representation, electronic phase space variables $(\mathbf{x}_0, \mathbf{p}_0)$ as well as commutator matrix $\Gamma_0$ are transformed to their counterparts in the adiabatic representation before performing NaF dynamics.

## S2-B: Numerical Integrators for the Equations of Motion for Trajectories

### 1) Numerical Integrator for NaF-TW or NaF

There are six steps of the numerical integrator for one time interval $\Delta t$ for updating nuclear coordinate $\mathbf{R}$, nuclear kinematic momentum $\mathbf{P}$, electronic phase variables $\tilde{\mathbf{g}} = \tilde{\mathbf{x}} + i\tilde{\mathbf{p}}$, and commutator matrix $\tilde{\mathbf{\Gamma}}$ in the adiabatic representation. As first discussed in ref [3], nuclear kinematic momentum $\mathbf{P}$ in the adiabatic representation is equivalent to the mapping nuclear momentum in the diabatic representation.

**Step 1**: Propagate the nuclear kinematic momentum within half a time step $\Delta t/2$

$$\mathbf{P}_{t+\Delta t/2} \leftarrow \mathbf{P}_t + \left(-\nabla_{\mathbf{R}} E_{j_{\text{occ}}(t)}(\mathbf{R}_t) + \mathbf{f}_{\text{nonadia}}(t)\right)\frac{\Delta t}{2}, \tag{S27}$$

where $-\nabla_{\mathbf{R}} E_{j_{\text{occ}}}(\mathbf{R}_t)$ represents the state-specific adiabatic nuclear force evolving on the $j_{\text{occ}}(t)$-th adiabatic state, and $\mathbf{f}_{\text{nonadia}}(t) = -\sum_{k \neq l}\left[(E_k(\mathbf{R}_t) - E_l(\mathbf{R}_t))\mathbf{d}_{lk}(\mathbf{R}_t)\right]\tilde{\rho}_{kl}(t)$ stands for the nonadiabatic nuclear force. Here, $\tilde{\rho}(t)$ denotes the electronic quasi-density matrix, which is

$$\tilde{\rho}(t) = \frac{1}{2}\tilde{\mathbf{g}}_t\tilde{\mathbf{g}}_t^\dagger - \tilde{\mathbf{\Gamma}}_t \tag{S28}$$

for NaF, or

$$\tilde{\rho}(t) = (1 + F/3)\tilde{\mathbf{g}}_t\tilde{\mathbf{g}}_t^\dagger / \text{Tr}_e\left[\tilde{\mathbf{g}}_t\tilde{\mathbf{g}}_t^\dagger\right] - 1/3 \tag{S29}$$

for NaF-TW.

**Step 2**: Propagate the nuclear coordinate within a full time step $\Delta t$

$$\mathbf{R}_{t+\Delta t} \leftarrow \mathbf{R}_t + \mathbf{M}^{-1}\mathbf{P}_{t+\Delta t/2}\Delta t. \tag{S30}$$

**Step 3**: Propagate electronic phase variables within a full time step $\Delta t$ according to

$$\tilde{\mathbf{g}}_{t+\Delta t} \leftarrow \tilde{\mathbf{U}}(\mathbf{R}_{t+\Delta t}, \mathbf{P}_{t+\Delta t/2}; \Delta t)\tilde{\mathbf{g}}_t. \tag{S31}$$

Here, $\tilde{\mathbf{U}}(\mathbf{R}_t, \mathbf{P}_t; \Delta t) = \exp\left[-i\Delta t \mathbf{V}^{(\text{eff})}(\mathbf{R}_t, \mathbf{P}_t)\right]$ denotes the unitary short-time propagator within a full time step $\Delta t$ in the adiabatic representation. Moreover, in the case of NaF, the EOM for commutator matrix $\tilde{\boldsymbol{\Gamma}}$ reads $\dot{\tilde{\boldsymbol{\Gamma}}} = i\left[\tilde{\boldsymbol{\Gamma}} \mathbf{V}^{(\text{eff})}(\mathbf{R}, \mathbf{P}) - \mathbf{V}^{(\text{eff})}(\mathbf{R}, \mathbf{P})\tilde{\boldsymbol{\Gamma}}\right]$, leading to the propagator within a full time step $\Delta t$,

$$\tilde{\boldsymbol{\Gamma}}_{t+\Delta t} \leftarrow \tilde{\mathbf{U}}(\mathbf{R}_{t+\Delta t}, \mathbf{P}_{t+\Delta t/2}; \Delta t)\tilde{\boldsymbol{\Gamma}}_t \tilde{\mathbf{U}}^\dagger(\mathbf{R}_{t+\Delta t}, \mathbf{P}_{t+\Delta t/2}; \Delta t). \tag{S32}$$

When it is more convenient to utilize potential energy surfaces in the diabatic representation, the short-time propagator in the adiabatic representation can be replaced with $\mathbf{T}^\dagger(\mathbf{R}_{t+\Delta t})\exp\left[-i\Delta t\mathbf{V}(\mathbf{R}_{t+\Delta t})\right]\mathbf{T}(\mathbf{R}_t)$, where $\mathbf{T}(\mathbf{R})$ represents the diabatic-to-adiabatic transformation matrix with elements $T_{nm}(\mathbf{R}) = \langle n|\phi_m(\mathbf{R})\rangle$ and $\mathbf{V}(\mathbf{R})$ denotes the diabatic potential energy matrix.

**Step 4**: Update $j_{\text{occ}}(t)$ to a new occupied state $j_{\text{occ}}(t+\Delta t)$ based on the switching strategy, i.e., select the state with the largest weight. We then rescale nuclear kinematic momentum $\mathbf{P}$ if $j_{\text{occ}}(t+\Delta t) \neq j_{\text{occ}}(t)$,

$$\mathbf{P}_{t+\Delta t/2} \leftarrow \mathbf{P}_{t+\Delta t/2}\sqrt{\frac{H_{\text{NaF}}(\mathbf{R}_{t+\Delta t}, \mathbf{P}_{t+\Delta t/2}, \tilde{\mathbf{x}}_{t+\Delta t}, \tilde{\mathbf{p}}_{t+\Delta t}) - E_{j_{\text{occ}}(t+\Delta t)}(\mathbf{R}_{t+\Delta t})}{\mathbf{P}^T_{t+\Delta t/2}\mathbf{M}^{-1}\mathbf{P}_{t+\Delta t/2}/2}}. \tag{S33}$$

If $H_{\text{NaF}}(\mathbf{R}_{t+\Delta t}, \mathbf{P}_{t+\Delta t/2}, \tilde{\mathbf{x}}_{t+\Delta t}, \tilde{\mathbf{p}}_{t+\Delta t}) < E_{j_{\text{occ}}(t+\Delta t)}(\mathbf{R}_{t+\Delta t})$, the switching of the adiabatic nuclear force component is frustrated. In such a case we keep $j_{\text{occ}}(t+\Delta t) = j_{\text{occ}}(t)$ and the rescaling step of eq (S33) is skipped.

**Step 5**: Propagate the nuclear kinematic momentum within the other half time step $\Delta t/2$

$$\mathbf{P}_{t+\Delta t} \leftarrow \mathbf{P}_{t+\Delta t/2} + \left(-\nabla_\mathbf{R} E_{j_{occ}(t+\Delta t)}(\mathbf{R}_{t+\Delta t}) + \mathbf{f}_{nonadia}(t+\Delta t)\right)\frac{\Delta t}{2}. \quad (S34)$$

**Step 6**: Finally, rescale the nuclear kinematic momentum $\mathbf{P}$ again to ensure energy conservation in the mapping variables:

$$\mathbf{P}_{t+\Delta t} \leftarrow \mathbf{P}_{t+\Delta t}\sqrt{\frac{H_{\text{NaF}}(\mathbf{R}_0,\mathbf{P}_0,\tilde{\mathbf{x}}_0,\tilde{\mathbf{p}}_0) - E_{j_{occ}(t+\Delta t)}(\mathbf{R}_{t+\Delta t})}{\mathbf{P}_{t+\Delta t}^T \mathbf{M}^{-1}\mathbf{P}_{t+\Delta t}/2}}. \quad (S35)$$

If $H_{\text{NaF}}(\mathbf{R}_0,\mathbf{P}_0,\tilde{\mathbf{x}}_0,\tilde{\mathbf{p}}_0) < E_{j_{occ}(t+\Delta t)}(\mathbf{R}_{t+\Delta t})$, i.e., the total mapping energy on phase space is smaller than the pure adiabatic potential energy, it indicates that the time step size $\Delta t$ is relatively large for the integrator from time $t$ to time $t+\Delta t$. In such a case, one should then choose a smaller time step size $\Delta t$ and repeat Steps 1-6 for the update of $(\mathbf{R}_{t+\Delta t},\mathbf{P}_{t+\Delta t},\tilde{\mathbf{x}}_{t+\Delta t},\tilde{\mathbf{p}}_{t+\Delta t})$ from $(\mathbf{R}_t,\mathbf{P}_t,\tilde{\mathbf{x}}_t,\tilde{\mathbf{p}}_t)$. The time step size $\Delta t$ should be adjusted in the region where the sum of adiabatic and nonadiabatic nuclear force terms is relatively large.

The rescaling of $\mathbf{P}$ along its direction is used in the numerical propagator in order to ensure the energy conservation. Alternatively, in Steps 1 and 5, $\mathbf{f}_{nonadia}$ of eq (S27) and eq (S34) can also be replaced by its projected component $\mathbf{f}_{nonadia}^\perp$ perpendicular to the velocity vector, $\dot{\mathbf{R}} = \mathbf{M}^{-1}\mathbf{P}$. That is,

$$\mathbf{f}_{nonadia}^\perp = \mathbf{f}_{nonadia} - \frac{\mathbf{f}_{nonadia} \cdot \dot{\mathbf{R}}}{\dot{\mathbf{R}} \cdot \dot{\mathbf{R}}}\dot{\mathbf{R}}. \quad (S36)$$

**2) Numerical Integrator for SQC-TW or Ehrenfest Dynamics**

In contrast to the aforementioned NaF integrator, both SQC-TW and Ehrenfest dynamics employ the mean field picture instead of the nonadiabatic field picture. In the mean field picture, the integrator also includes Steps 1, 2, 3, and 5 of the NaF/NaF-TW integrator. The difference is that, in the integrator for the EOMs for trajectories of either SQC-TW or Ehrenfest dynamics, the full nuclear force (i.e., the time derivative of $\mathbf{P}$) in Step 1 and Step 5 is defined by

$$\dot{\mathbf{P}} = \mathbf{f}_{\mathrm{mf}}(t) = -\sum_{k,l}\left[\nabla_{\mathbf{R}}E_k(\mathbf{R}_t)\delta_{kl} + \left(E_k(\mathbf{R}_t) - E_l(\mathbf{R}_t)\right)\mathbf{d}_{lk}(\mathbf{R}_t)\right]\tilde{\rho}_{kl}(t) \ . \tag{S37}$$

When SQC-TW is used, $\tilde{\rho}(t) = \frac{1}{2}\tilde{\mathbf{g}}_t\tilde{\mathbf{g}}_t^\dagger - \tilde{\Gamma}_0$, where commutator matrix $\tilde{\Gamma}$ does *not* evolve with time. This corresponds to the "trajectory-adjusted zero-point energy" treatment of the SQC approach with triangle window functions of Cotton and Miller in ref[4]. (We note that, in the SQC-TW approach, only the expression of the electronic population-population correlation function is the same as that of ref[4], while those of the other three kinds of time correlation functions for electronic DOFs are not from ref[4] but constructed in the main text of the present letter.)

When Ehrenfest dynamics is employed, we have $\tilde{\rho}(t) = \frac{1}{2}\tilde{\mathbf{g}}_t\tilde{\mathbf{g}}_t^\dagger$ instead.

3) **Numerical Integrator for FSSH**

FSSH employs the adiabatic nuclear force picture rather than the nonadiabatic field picture. In the FSSH algorithm[5,6], the full nuclear force in Step 1 and Step 5 (of the numerical integrator of NaF/NaF-TW) is $\dot{\mathbf{P}} = \mathbf{f}_{\mathrm{adia}}(t) = -\nabla_{\mathbf{R}}E_{j_{\mathrm{occ}}(t)}(\mathbf{R}_t)$ instead, and Step 4 (of the numerical integrator of NaF/NaF-TW) is replaced with the FSSH strategy (i.e., eq (S51) of the Supporting Information of ref[7]) to determine a new occupied state $j_{\mathrm{occ}}(t + \Delta t)$. The detailed description of the numerical integrator for FSSH is described in Section S7 of the Supporting Information of ref[7].

## S2-C: Numerical Evaluation of the Time Correlation Function for NaF, NaF-TW and SQC-TW

### 1) Evaluation of the Time Correlation Function for NaF

The sampling of the initial condition of phase space variables of NaF, as described by the instruction in Sub-section S2-A, is performed multiple times for generating a collection of $N_{\text{samp}}$ numbers of initial conditions for trajectories, i.e., $\left\{ \left( \mathbf{x}_{0;[s]}, \mathbf{p}_{0;[s]}, \mathbf{\Gamma}_{0;[s]}, \mathbf{R}_{0;[s]}, \mathbf{P}_{0;[s]} \right), s = 1, \ldots, N_{\text{samp}} \right\}$. Each initial condition is then used for evolving a trajectory by the numerical integrator of NaF described in Sub-section S2-B, i.e., $\left( \mathbf{x}_{0;[s]}, \mathbf{p}_{0;[s]}, \mathbf{\Gamma}_{0;[s]}, \mathbf{R}_{0;[s]}, \mathbf{P}_{0;[s]} \right)$ evolves to $\left( \mathbf{x}_{t;[s]}, \mathbf{p}_{t;[s]}, \mathbf{\Gamma}_{t;[s]}, \mathbf{R}_{t;[s]}, \mathbf{P}_{t;[s]} \right)$ at time $t$. The time correlation function is evaluated by the following summation,

$$\text{Tr}_{n,e} \left[ \left( |n\rangle\langle m| \otimes \hat{\rho}_{\text{nuc}} \right) e^{i\hat{H}t/\hbar} \left( |k\rangle\langle l| \otimes \hat{A}_{\text{nuc}} \right) e^{-i\hat{H}t/\hbar} \right]$$
$$\mapsto \frac{1}{N_{\text{samp}}} \sum_{s=1}^{N_{\text{samp}}} \left[ \frac{1}{2} \left( x_{0;[s]}^{(m)} + i p_{0;[s]}^{(m)} \right) \left( x_{0;[s]}^{(n)} - i p_{0;[s]}^{(n)} \right) - \gamma \delta_{mn} \right] \quad \text{(S38)}$$
$$\times \left[ \frac{1+F}{2(1+F\gamma)^2} \left( x_{t;[s]}^{(l)} + i p_{t;[s]}^{(l)} \right) \left( x_{t;[s]}^{(k)} - i p_{t;[s]}^{(k)} \right) - \frac{1-\gamma}{1+F\gamma} \delta_{lk} \right] A_W \left( \mathbf{R}_{t;[s]}, \mathbf{P}_{t;[s]} \right)$$

### 2) Evaluation of the Time Correlation Function for NaF-TW and SQC-TW

According to the numerical procedures of NaF-TW or those of SQC-TW in Sub-sections S2-A and S2-B, $N_{\text{samp}}$ numbers of initial conditions $\left\{ \left( \mathbf{x}_{0;[s]}, \mathbf{p}_{0;[s]}, \mathbf{R}_{0;[s]}, \mathbf{P}_{0;[s]} \right), s = 1, \ldots, N_{\text{samp}} \right\}$ are generated and propagated to $\left\{ \left( \mathbf{x}_{t;[s]}, \mathbf{p}_{t;[s]}, \mathbf{R}_{t;[s]}, \mathbf{P}_{t;[s]} \right), s = 1, \ldots, N_{\text{samp}} \right\}$ for NaF-TW; or similarly, $N_{\text{samp}}$ numbers of initial conditions $\left\{ \left( \mathbf{x}_{0;[s]}, \mathbf{p}_{0;[s]}, \mathbf{\Gamma}_{0;[s]}, \mathbf{R}_{0;[s]}, \mathbf{P}_{0;[s]} \right), s = 1, \ldots, N_{\text{samp}} \right\}$ are sampled and evolved to $\left\{ \left( \mathbf{x}_{t;[s]}, \mathbf{p}_{t;[s]}, \mathbf{R}_{t;[s]}, \mathbf{P}_{t;[s]} \right), s = 1, \ldots, N_{\text{samp}} \right\}$ for time $t$ for SQC-TW. The population-population ($n = m$ and $k = l$) or population-coherence ($n = m$ and $k \neq l$) correlation function is evaluated by the following summation over trajectories,

$$\mathrm{Tr}_{n,e}\left[\left(|n\rangle\langle n| \otimes \hat{\rho}_{\mathrm{nuc}}\right)e^{i\hat{H}t/\hbar}\left(|k\rangle\langle l| \otimes \hat{A}_{\mathrm{nuc}}\right)e^{-i\hat{H}t/\hbar}\right]$$

$$\mapsto \begin{cases} \dfrac{\sum_{s=1}^{N_{\mathrm{samp}}} K_{kk}^{\mathrm{bin}}\left(\mathbf{x}_{t;[s]}, \mathbf{p}_{t;[s]}\right) A_W\left(\mathbf{R}_{t;[s]}, \mathbf{P}_{t;[s]}\right)}{\sum_{s=1}^{N_{\mathrm{samp}}} \sum_{i=1}^{F} K_{ii}^{\mathrm{bin}}\left(\mathbf{x}_{t;[s]}, \mathbf{p}_{t;[s]}\right)}, & \text{for } k = l \\[2ex] \dfrac{1}{N_{\mathrm{samp}}} \sum_{s=1}^{N_{\mathrm{samp}}} K_{kl}^{\mathrm{CMM}}\left(\mathbf{x}_{t;[s]}, \mathbf{p}_{t;[s]}\right) A_W\left(\mathbf{R}_{t;[s]}, \mathbf{P}_{t;[s]}\right), & \text{for } k \neq l \end{cases}, \quad \text{(S39)}$$

where $K_{kk}^{\mathrm{bin}}(\mathbf{x}, \mathbf{p})$ and $K_{kl}^{\mathrm{CMM}}(\mathbf{x}, \mathbf{p})$ are defined as eqs (16) and (19) of the main text (of the present letter). The coherence-population ($n \neq m$ and $k = l$) or coherence-coherence ($n \neq m$ and $k \neq l$) correlation function is estimated by

$$\mathrm{Tr}_{n,e}\left[\left(|n\rangle\langle m| \otimes \hat{\rho}_{\mathrm{nuc}}\right)e^{i\hat{H}t/\hbar}\left(|k\rangle\langle l| \otimes \hat{A}_{\mathrm{nuc}}\right)e^{-i\hat{H}t/\hbar}\right]$$
$$\mapsto \frac{12}{5N_{\mathrm{samp}}} \sum_{s=1}^{N_{\mathrm{samp}}} K_{mn}^{\mathrm{CMM}}\left(\mathbf{x}_{0;[s]}, \mathbf{p}_{0;[s]}\right) K_{lk}^{\mathrm{CMM}}\left(\mathbf{x}_{t;[s]}, \mathbf{p}_{t;[s]}\right) A_W\left(\mathbf{R}_{t;[s]}, \mathbf{P}_{t;[s]}\right). \quad \text{(S40)}$$

## S3: Simulation Details for Models

### S3-A: Reduced Electronic Dynamics for System-Bath Models.

For system-bath models, the system is always bilinearly coupled with harmonic bath DOFs of a dissipative environment in the condensed phase. The system-bath coupling, representing the substantial influence from the bath environment, governs the dynamics of the system across a spectrum ranging from coherent to dissipative regimes. These models serve as pivotal tools for understanding electron/exciton dynamics in chemical and biological reactions. Numerically exact results of the spin-boson model can be achieved by quasi-adiabatic propagator path integral (QuAPI)[8-10] and more efficient small matrix PI (SMatPI)[11, 12], hierarchy equations of motion (HEOM)[13-17], (multi-layer) multi-configuration time-dependent Hartree [(ML-)MCTDH] [18-20], and time-dependent density matrix renormalization group (TD-DMRG)[21].

In the present paper, we test two distinctive models: the two-site spin-boson models and the seven-site Fenna–Matthews–Olson (FMO) Monomer model.

*Spin-Boson Model*: The two-site spin-boson models stand as a fundamental prototype for comprehension of electron transfer and energy transport phenomena. The Hamiltonian for spin-boson models reads $\hat{H} = \hat{H}_B + \hat{H}_{SB} + \hat{H}_S$ with employing the environment bath part $\hat{H}_B$, the linear coupling term $\hat{H}_{S-B}$ and the system part $\hat{H}_S$ as

$$\hat{H}_B = \sum_{j=1}^{N_b} \frac{1}{2}\left(\hat{P}_j^2 + \omega_j^2 \hat{R}_j^2\right)$$
$$\hat{H}_{SB} = \sum_{j=1}^{N_b} c_j \hat{R}_j \left(|1\rangle\langle 1| - |2\rangle\langle 2|\right) \quad \text{(S41)}$$
$$\hat{H}_S = \varepsilon\left(|1\rangle\langle 1| - |2\rangle\langle 2|\right) + \Delta\left(|1\rangle\langle 2| + |2\rangle\langle 1|\right)$$

Here $\varepsilon$ denotes the energy bias while $\Delta$ signifies the tunneling between states $|1\rangle$ and $|2\rangle$. The bath is discretized into a series of quantum harmonic oscillators, with $\{\hat{P}_j\}$, $\{\hat{R}_j\}$, $\{\omega_j\}$ and $\{c_j\}$ representing the mass-weighted momentum, coordinate, frequencies and the coupling coefficients of the $j$-th oscillator, respectively. For spin-boson models, we adopt the discretization scheme proposed in refs [22, 23] for the Ohmic spectral density $J(\omega) = \frac{\pi}{2}\alpha\omega\exp(-\omega/\omega_c)$ with the Kondo parameter $\alpha$ and the cut-off frequency $\omega_c$, as depicted below:

$$\begin{cases} \omega_j = -\omega_c \ln\left(1 - \frac{j}{1+N_b}\right) \\ c_j = \omega_j \sqrt{\frac{\alpha\omega_c}{N_b+1}} \end{cases}, \quad j = 1, \cdots, N_b. \quad \text{(S42)}$$

In our simulation, we investigate four specific spin-boson models outlined in ref [24]. These models span a range of system-bath coupling strengths from weak to strong (small to large $\alpha$), and the cut-off frequency $\omega_c$ from low to high. In addition, all simulations are performed at low temperatures ($\beta = 5$) and utilize three hundred discrete bath modes to characterize the Ohmic spectral density within the spin-boson models.

*FMO Monomer Model*: The FMO complex derived from green sulfur bacteria serves as a prototype system crucial for investigating photosynthetic organisms[25-32]. Specifically referencing ref [29], the FMO monomer employs a site-exciton model, encompassing a seven-site structure coupled with a harmonic bath, whose three parts of Hamiltonian alternatively reads,

$$\hat{H}_B = \sum_{n=1}^{F}\sum_{j=1}^{N_b}\left(\hat{P}_{nj}^2 + \omega_j^2 \hat{R}_{nj}^2\right)/2$$

$$\hat{H}_{SB} = \sum_{n=1}^{F}\sum_{j=1}^{N_b} c_j \hat{R}_{nj} |n\rangle\langle n|$$

$$\hat{H}_S = \sum_{n,m=1}^{F} H_{S,nm} |n\rangle\langle m|$$

$$= \begin{pmatrix} 12410 & -87.7 & 5.5 & -5.9 & 6.7 & -13.7 & -9.9 \\ -87.7 & 12530 & 30.8 & 8.2 & 0.7 & 11.8 & 4.3 \\ 5.5 & 30.8 & 12210 & -53.5 & -2.2 & -9.6 & 6.0 \\ -5.9 & 8.2 & -53.5 & 12320 & -70.7 & -17.0 & -63.3 \\ 6.7 & 0.7 & -2.2 & -70.7 & 12480 & 81.1 & -1.3 \\ -13.7 & 11.8 & -9.6 & -17.0 & 81.1 & 12630 & 39.7 \\ -9.9 & 4.3 & 6.0 & -63.3 & -1.3 & 39.7 & 12440 \end{pmatrix} \text{cm}^{-1} . \quad (S43)$$

Here the variables $\{\hat{P}_{nj}, \hat{R}_{nj}\}$ denote the mass-weighted momentum and coordinate of the *j*-th quantum harmonic oscillator for the bath on *n*-th site, and the bath frequencies $\{\omega_j\}$ and system-bath coupling coefficients $\{c_j\}$ are determined by discretizing the spectral density. Specifically, we utilize the Debye spectral density $J(\omega) = 2\lambda\omega_c\omega/(\omega^2 + \omega_c^2)$ for each site, with parameters $\lambda = 35 \text{ cm}^{-1}$ for the bath reorganization energy and $\omega_c = 106.14 \text{ cm}^{-1}$ for the characteristic frequency. We use the discretization scheme of refs [19, 33, 34],

$$\omega_j = \omega_c \tan\left(\frac{\pi}{2} - \frac{\pi j}{2(N_b+1)}\right), \quad j=1,\cdots,N_b$$
$$c_j = \omega_j \sqrt{\frac{2\lambda}{N_b+1}}, \quad j=1,\cdots,N_b$$
(S44)

for the Debye spectral density. (In comparison, eq (S42) is the discretization scheme for the Ohmic spectral density by following refs [22, 23].) A challenging temperature $T = 77\text{K}$ is investigated as studied in our previous work[35]. The first site of the system is initially occupied, and the bath DOFs are sampled from the Wigner distributions of the corresponding harmonic oscillators. In the FMO monomer model, one hundred discrete bath modes per site (totaling $N_b = 100$) are used to represent the continuous Debye spectral density.

We consider the decoupled initial condition for both spin-boson models and the FMO monomer model, where the system is in the excited state and the bath modes are at the thermal equilibrium (i.e., the quantum Boltzmann distribution for the pure bath Hamiltonian operator). Specially, the bath modes are sampled from the corresponding Wigner distribution

$$\rho_W(\mathbf{R},\mathbf{P}) \propto \exp\left[-\sum_{j=1}^{N_b} \frac{\beta}{2Q(\omega_j)}\left(P_j^2 + \omega_j^2 R_j^2\right)\right]$$
(S45)

with $Q(\omega) = \frac{\beta\hbar\omega/2}{\tanh(\beta\hbar\omega/2)}$ as the quantum corrector[36]. For both the spin-boson models and the FMO monomer model, we use the exact results obtained from our previous work[24] using HEOM/extended HEOM (eHEOM)[37, 38] for comparison.

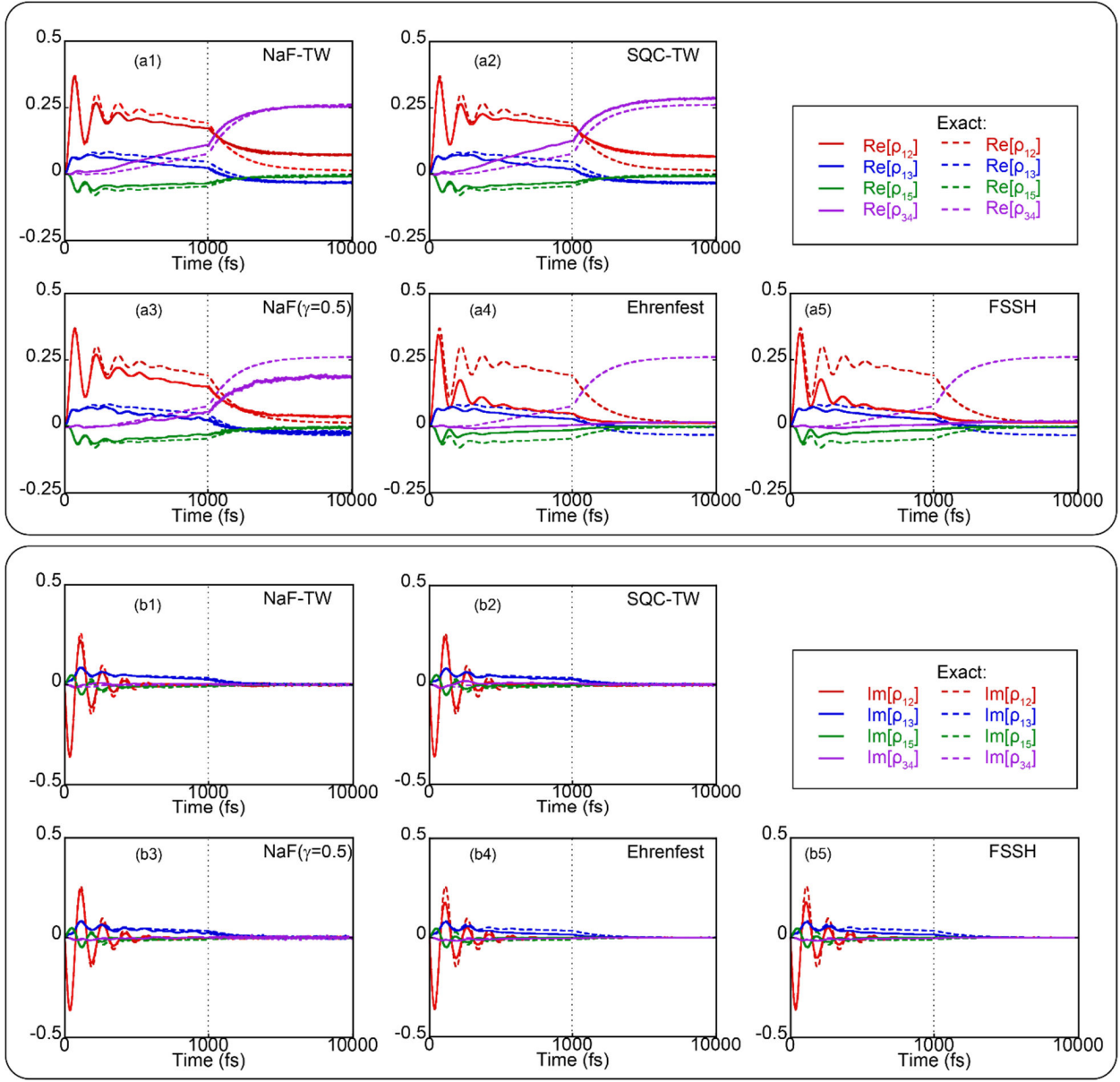

**Figure S1**. Dynamics of the off-diagonal (coherence) terms of the reduced electronic density matrix of the seven-site FMO monomer model at temperature 77 K. The real part and imaginary parts are presented in panel (a) and panel (b), respectively. Panel (a): The red, blue, green, and purple solid lines illustrate the real part of $\rho_{12}(t)$, $\rho_{13}(t)$, $\rho_{15}(t)$ and $\rho_{34}(t)$, respectively. Panel (b): Similar to panel (a), but the red, blue, green, and purple solid lines depict the imaginary part of $\rho_{12}(t)$, $\rho_{13}(t)$, $\rho_{15}(t)$ and $\rho_{34}(t)$, respectively. The exact results produced by HEOM are presented by dashed lines in corresponding colors. Sub-panels (a1-a5) or sub-panels (b1-b5) denote the results of NaF-TW, SQC-TW, NaF($\gamma=0.5$), Ehrenfest dynamics, and FSSH, respectively. One hundred discrete bath modes for each site are employed to obtain converged results.

**S3-B: Cavity Quantum Electrodynamic for Atom-in-Cavity Models.**

It has been observed that several significant phenomena manifest in cavity quantum electrodynamics (cQED), particularly under conditions where matter is tightly coupled to the vacuum field within a confined optical cavity[39-42]. In the main text, we examine benchmark cQED models featuring a multi-level hydrogen atom confined within a one-dimensional lossless cavity[43-46]. The atomic system is coupled to multi-cavity-modes, whose Hamiltonian is described by $F$ atomic energy levels:

$$\hat{H} = \sum_{n=1}^{F} \varepsilon_n |n\rangle\langle n| + \sum_{j=1}^{N} \frac{1}{2}\left(\hat{P}_j^2 + \omega_j^2 \hat{R}_j^2\right) + \sum_{j=1}^{N} \omega_j \lambda_j(r_0) \hat{R}_j \sum_{n \neq m}^{F} \mu_{nm} |n\rangle\langle m|, \quad \text{(S46)}$$

where $\varepsilon_n$ is the atomic energy level of the $n$-th atomic state, and we employ a three-state model with the energy levels $\varepsilon_1 = -0.6738$, $\varepsilon_2 = -0.2798$, $\varepsilon_3 = -0.1547$, and $\mu_{nm}$ denotes the transitional dipole moment with nonzero values $\mu_{12} = -1.034$, $\mu_{23} = -2.536$ (all in atomic units). The first model involves full three atomic levels, and the second one is a reduced two-level model where only the two lowest levels are considered. The variables, $\hat{R}_j, \hat{P}_j, \omega_j$ denote the canonical coordinate, canonical momentum, and frequency of the $j$-th optical field mode, respectively, and the atom-optical field interaction reads

$$\lambda_j(r_0) = \sqrt{\frac{2}{\varepsilon_0 L}} \sin\left(\frac{j\pi r_0}{L}\right), \quad j = 1, \cdots, N, \quad \text{(S47)}$$

where $L$, $\varepsilon_0$ and $r_0$ denote the volume length of cavity, the vacuum permittivity, and the location of the atom, respectively. We fix $L = 236200$ a.u. and $r_0 = L/2$, and employ four hundred standing-wave modes for the optical field, with the frequency $\omega_j$ of the $j$-th mode set to $j\pi c/L$ (Here $c = 137.036$ a.u. denotes the light speed in vacuum). Initially, the highest atomic level is excited, while all cavity modes remain in the vacuum state. Subsequently, spontaneous emission occurs, releasing a photon that traverses the cavity and reflects back to interact with the atom. The re-absorption and re-emission processes then follow. We compare our results with benchmark data obtained from truncated configuration interaction calculations, as reported in refs [43, 44].

## S3-C: Dynamics around Conical Intersection for Linear Vibronic Coupling Models.

The linear vibronic coupling model (LVCM) is a straightforward yet powerful model that simulates molecular systems, particularly those where the conical intersection (CI) region is pivotal, such as in light-induced processes. In the diabatic representation, the Hamiltonian of LVCM is expressed as

$$\hat{H} = \sum_{k=1}^{N} \frac{\omega_k}{2}\left(\hat{P}_k^2 + \hat{R}_k^2\right) + \sum_{n=1}^{F}\left(E_n + \sum_{k=1}^{N} \kappa_k^{(n)} \hat{R}_k\right)|n\rangle\langle n| + \sum_{n\neq m}^{F}\left(\sum_{k=1}^{N} \lambda_k^{(nm)} \hat{R}_k\right)|n\rangle\langle m|, \quad (S48)$$

where $E_n$ $(n = 1,...,F)$ is the vertical excitation energy of the $n$-th state, while $\hat{P}_k$ and $\hat{R}_k$ $(k = 1,...,N)$ denote dimensionless weighted normal-mode momentum and coordinate of the $k$-th nuclear DOF, respectively, with the corresponding frequency $\omega_k$. In addition, parameters $\kappa_k^{(n)}$ and $\lambda_k^{(nm)}$ represent the linear coupling coefficients of the $k$-th nuclear vibronic DOF with the diagonal and off-diagonal elements of the electronic density, respectively.

In our first case study, we explore two versions of the linear vibronic coupling model (LVCM) applied to the S1/S2 conical intersection of the pyrazine molecule. One LVCM variant includes three nuclear modes, while the other incorporates 24 nuclear modes. Detailed parameters for these models can be found in refs [47,48]. The initial state consists of the cross-product between the vibronic ground state and the electronically excited diabatic state (S2). Furthermore, we investigate a typical three-electronic-state LVCM with two nuclear modes for the Cr(CO)$_5$ molecule, as detailed in ref [49]. Here, the initial condition comprises the cross-product of the first electronically excited diabatic state and a Gaussian nuclear wave-packet. The Gaussian wave-packet is centered at the minimum point of the ground state of the Cr(CO)$_6$ molecule, where a carbonyl group dissociates. The width of each mode is determined by the corresponding vibrational frequencies. While we employ the diabatic representation for initializing and evaluating dynamical properties, consistent with the approach in MCTDH, we switch to the adiabatic representation for real-time dynamics to ensure fair comparison among different non-adiabatic dynamics methods. The parameter lists for the LVCMs applied to both the pyrazine and Cr(CO)$_5$ molecules can be found in the Supporting Information, specifically in Tables S2-S4 of ref [7].

**S3-D: Photodissociation Dynamics of Gas Phase Models with One Nuclear Degree of Freedom.**

We further test gas phase models with asymptotic regions. We consider the coupled three-electronic-state photodissociation models of Miller and coworkers [50]. Each PES is described by a Morse oscillator and the coupling terms are depicted by Gaussian functions:

$$V_{ii}(R) = D_i \left[1 - e^{-\beta_i(R-R_i)}\right]^2 + C_i, \quad i = 1,2,3.$$
$$V_{ij}(R) = V_{ji}(R) = A_{ij} e^{-\alpha_{ij}(R-R_{ij})^2}, \quad i,j = 1,2,3; \text{ and } i \neq j.$$

(S49)

where the detailed parameters match those of Model III as described in ref [50]. The Wigner distribution for nuclear DOF is $\rho_W(R,P) \propto \exp(-m\omega(R-R_e)^2 - P^2/m\omega)$, where $m = 20000$ a.u. is the mass of the nuclear DOF, $\omega = 0.005$ a.u. is the vibrational frequency of ground state, and $R_e$ denotes the center of wavepacket. The first electronic diabatic state is initially occupied. Numerically exact results for the models can be obtained by the discrete variable representation (DVR) approach of ref [51].

### S3-E: Nonadiabatic Scattering Dynamics of Tully's Models.

We investigate Tully's three models[6] for nonadiabatic scattering dynamics, namely, the single avoided crossing (SAC) model, the dual avoided crossing (DAC) model and the extended coupling region (ECR) model. The results of the ECR model are demonstrated in the main text. The ECR model poses a formidable challenge for mapping-based methods, involving some trajectories that transmit forward while others reflect—a complex scenario inadequately described within mean-field approximations. The Hamiltonian of ECR model reads

$$\begin{aligned} V_{11}(R) &= +E_0 \\ V_{22}(R) &= -E_0 \\ V_{12}(R) &= V_{21}(R) = C\left[e^{BR}h(-R) + (2-e^{-BR})h(R)\right] \end{aligned} \quad , \quad (S50)$$

where $B = 0.9$, $C = 0.1$, $E_0 = -0.0006$, and $h(R)$ represents the Heaviside function. Initially, the system with mass $m = 2000$ a.u. occupies in the electronic adiabatic ground state, and the nuclear wavefunction $\psi(R) \propto e^{-\alpha(R-R_0)^2/2 + iP_0(R-R_0)}$ leads to the corresponding Wigner distribution $\rho_W(R,P) \propto e^{-\alpha(R-R_0)^2 + (P-P_0)^2/\alpha}$. The center of the initial wavefunction, denoted by $R_0$, is positioned at $-13$, with a width parameter of $\alpha = 1$. The initial momentum is set to $P_0$ varying from 2 to 50.

In addition, we also provide supplemental results of trajectory-based dynamics for the SAC model and the DAC model[6]. The SAC model reads

$$V_{11}(R) = A(1-e^{-B|R|})\text{sgn}(R)$$
$$V_{22}(R) = -V_{11}(R) \qquad . \qquad (S51)$$
$$V_{12}(R) = V_{21}(R) = Ce^{-DR^2}$$

with $A = 0.01$, $B = 1.6$, $C = 0.005$, $D = 1.0$, and initial $R_0 = -3.8$; and the DAC model reads

$$V_{11}(R) = 0$$
$$V_{22}(R) = -Ae^{-BR^2} + E_0 \qquad . \qquad (S52)$$
$$V_{12}(R) = V_{21}(R) = Ce^{-DR^2}$$

with $A = 0.1$, $B = 0.28$, $C = 0.015$, $D = 0.06$, $E_0 = 0.05$ and initial $R_0 = -10$. Other parameters keep the same as those used in the ECR model. Figure S2 presents the transmission probabilities of the SAC and the DAC models. For the SAC model, the results of all methods are close to each other. For the DAC model, NaF, SQC-TW and NaF-TW slightly outperform Ehrenfest dynamics and FSSH, especially in the high kinematic energy region where $P_0 > 20$. Figure S3 depicts the nuclear distribution of the SAC and DAC models in the long time limit, when the center of the momentum of the initial nuclear Gaussian wavepacket is $P_0 = 10$ or $P_0 = 20$. NaF-TW, NAF, and FSSH outperform SQC-TW and Ehrenfest dynamics in the description of the nuclear dynamics.

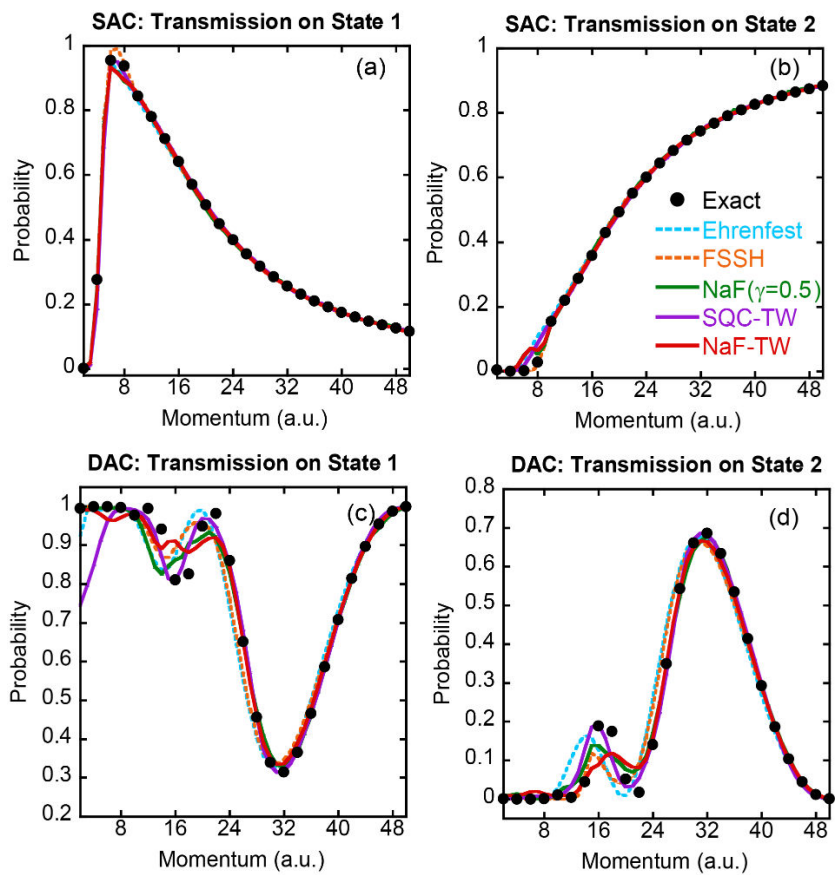

**Figure S2.** Panels (a-b) describe the transmission probability on (adiabatic) State 1 and that on State 2 for Tully's SAC model. Panels (c-d) present the transmission probability on (adiabatic) State 1 and that on State 2 for Tully's DAC model. Black points: Exact results produced by DVR. Cyan dashed lines: Ehrenfest dynamics. Orange dashed lines: FSSH. Purple solid lines: SQC-TW. Green solid lines: NaF. Red solid lines: NaF-TW.

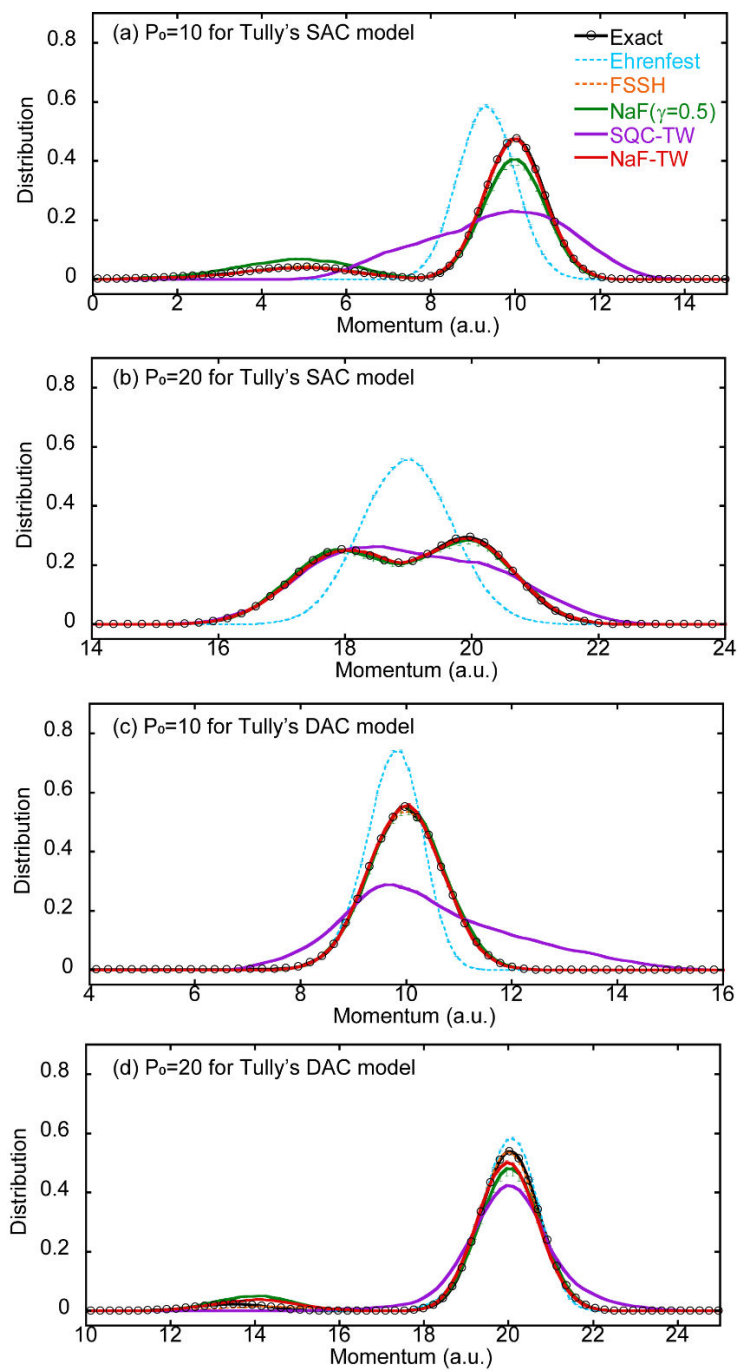

**Figure S3.** Panels (a-b) describe the asymptotic nuclear momentum distribution for Tully's SAC model, with the initial momentum of the incoming nuclear Gaussian wavepacket centered at $P_0 = 10$ and $P_0 = 20$, respectively. Panels (c-d): same as panels (a-b) but for Tully's DAC model.

Black circles: Exact results produced by DVR. Cyan dashed lines: Ehrenfest dynamics. Purple solid lines: SQC-TW. Green solid lines: NaF. Red solid lines: NaF-TW.

### S3-F: Electron Transfer Rate for the Two-Level Model in the Condensed Phase

We apply NaF-TW to calculate the electron transfer rate in the condensed phase. We adopt the model employed in refs [52, 53], which elucidates a two-level system coupled with a solvent bath:

$$\hat{H} = \hat{H}_s(\hat{R}_s, \hat{P}_s) + \hat{H}_b(\hat{\mathbf{R}}, \hat{\mathbf{P}}, \hat{R}_s). \tag{S53}$$

Here,

$$\begin{aligned}\hat{H}_s(\hat{R}_s, \hat{P}_s) = \frac{1}{2}\hat{P}_s^2 &+ \left(\frac{1}{2}\varepsilon + \frac{1}{2}\Omega^2 \hat{R}_s^2 + c_s \hat{R}_s\right)|1\rangle\langle 1| \\ &+ \left(-\frac{1}{2}\varepsilon + \frac{1}{2}\Omega^2 \hat{R}_s^2 - c_s \hat{R}_s\right)|2\rangle\langle 2| \\ &+ \Delta(|1\rangle\langle 2| + |2\rangle\langle 1|)\end{aligned} \tag{S54}$$

$$\hat{H}_b(\hat{\mathbf{R}}, \hat{\mathbf{P}}, \hat{R}_s) = \sum_{n=1}^{N_b}\left[\frac{1}{2}\hat{P}_n^2 + \frac{1}{2}\omega_n^2\left(\hat{R}_n + \frac{c_n}{\omega_n^2}\hat{R}_s\right)^2\right], \tag{S55}$$

where $\{\hat{P}_s, \hat{R}_s\}$ denote the nuclear momentum and coordinate of the reaction DOF, respectively, and $\{\hat{P}_n, \hat{R}_n\}$ $(n=1,\cdots,N_b)$ represent those of solvent bath DOFs. The frequencies $\{\omega_n\}$ and the coefficients $\{c_n\}$ $(n=1,\cdots,N_b)$ for solvent bath are obtained by discretizing the Ohmic spectral density with the Kondo parameter $\alpha = 9.49\times 10^{-6}$ and the cut-off frequency $\omega_c = \Omega = 3.5\times 10^{-4}$ a.u. The reorganization energy $\lambda = 2.39\times 10^{-2}$ a.u., the coupling $\Delta = 5\times 10^{-5}$ a.u., and the parameter $c_s = \Omega\sqrt{\lambda/2}$. Parameter $\varepsilon$ represents the energy bias. One hundred discrete bath DOFs are utilized in the calculations.

The electron transfer rate is obtained from[54]

S27

$$k = \int_0^\infty dt \, \mathrm{Re} \, C_{FF}(t), \tag{S56}$$

where the flux-flux correlation function is given by

$$C_{FF}(t) = \mathrm{Tr}\left[\hat{\rho}_{\mathrm{nuc}} \hat{F} e^{i\hat{H}t} \hat{F} e^{-i\hat{H}t}\right], \tag{S57}$$

with the flux operator defined as $\hat{F} = i\Delta(|1\rangle\langle 2| - |2\rangle\langle 1|)$. The initial Wigner density distribution for nuclear DOFs reads

$$\rho_{\mathrm{nuc}}(\mathbf{R}, \mathbf{P}, R_s, P_s) \propto \exp\left[-\sum_{n=1}^{N_b} \frac{\beta(P_n^2 + \omega_n^2 R_n^2)}{2Q(\omega_n)}\right] \exp\left[-\frac{\beta(P_s^2 + \Omega^2 R_s^2 + 2c_s R_s)}{2Q(\Omega)}\right], \tag{S58}$$

with the temperature set at 300 K.

In the region of weak coupling and relative high temperature, the Marcus electron transfer rate theory[55] is anticipated to perform reasonably well and is utilized as a reference for comparison. The results of electron transfer rate produced by NaF-TW, NaF ($\gamma = 0.5$), SQC-TW, SQC-TW2 and NaF-TW2 are compared with those predicted by the Marcus theory in Figure S4. (The details of SQC-TW2 and NaF-TW2 are described in Section S4 of this Supporting Information.) We also present the results (from ref [53]) obtained by an earlier version of SQC where square window functions were utilized. Figure S4 depicts the relationship between the electronic transfer rate $k$ and the bias $\varepsilon$. The results obtained from NaF-TW, NaF ($\gamma = 0.5$), SQC-TW, SQC-TW2 and NaF-TW2 exhibit good agreement with the Marcus theory in both the normal and inverted regimes. The three approaches overall outperform the SQC approach with square window functions in ref [53].

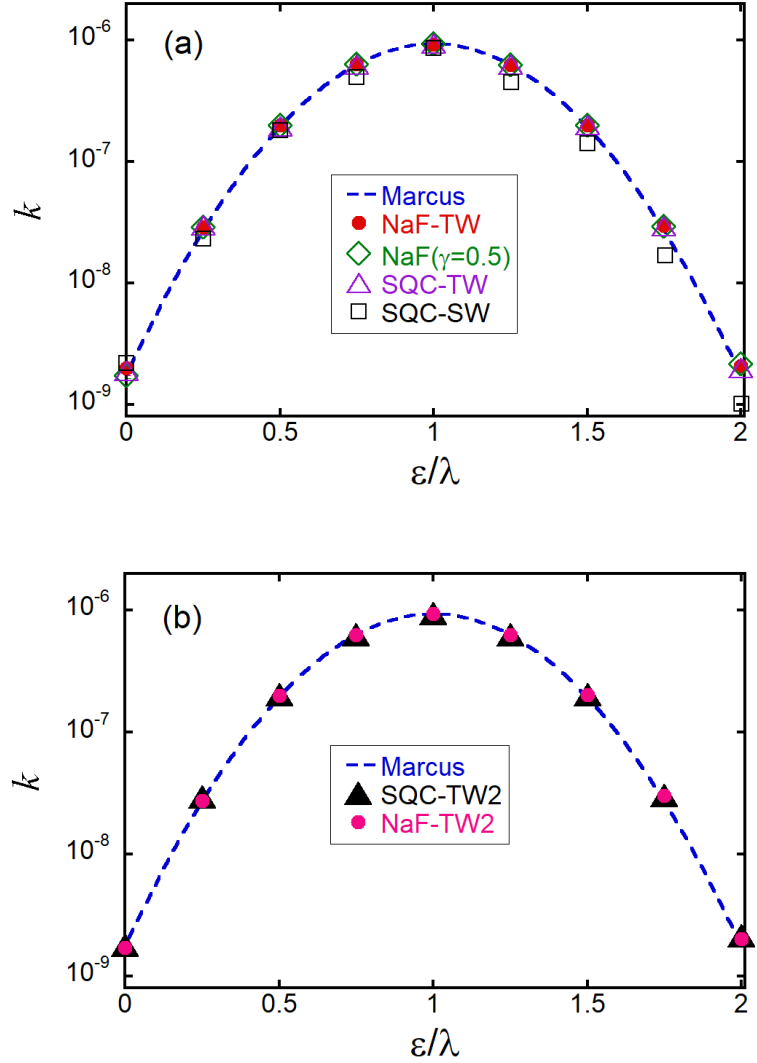

**Figure S4.** The electronic transfer rate $k$ against $\varepsilon/\lambda$. In panel (a), blue dashed line, red points, green hollow diamonds and purple hollow triangles represent the results obtained from the Marcus theory, NaF-TW, NaF ($\gamma=0.5$) and SQC-TW, respectively. Black hollow squares demonstrate the results of SQC with square window functions (denoted as SQC-SW in the figure) obtained from ref [53]. In panel (b), blue dashed line, black solid triangles and magenta points represent the results yielded from the Marcus theory, SQC-TW2 and NaF-TW2, respectively.

**S4: Comparison of Different Strategies Employing Triangle Window Functions**

For the SQC with triangle window functions, we compare two strategies. The first SQC-TW version as presented in the main text, involves the "trajectory-adjusted zero-point energy" treatment of the SQC-TW approach of Cotton and Miller in ref [4]. In this approach, we choose $\tilde{\rho}(t) = \frac{1}{2}\tilde{\mathbf{g}}_t\tilde{\mathbf{g}}_t^\dagger - \tilde{\Gamma}_0$ where (time-independent) commutator matrix $\tilde{\Gamma}$ is fixed during real-time evolution. In addition, we propose another version denoted as SQC-TW2, where we let $\tilde{\rho}(t) = \frac{1}{2}\tilde{\mathbf{g}}_t\tilde{\mathbf{g}}_t^\dagger - \tilde{\Gamma}_t$ and (time-dependent) commutator matrix $\tilde{\Gamma}$ evolves according to eq (S32). In Figures S5-S7, we present the comparison between SQC-TW and SQC-TW2 for three different types of models: the spin-boson model, the LVCM of the pyrazine molecule, and the FMO monomer model. The numerical performance of SQC-TW and that of SQC-TW2 do not differ much from each other in the model tests.

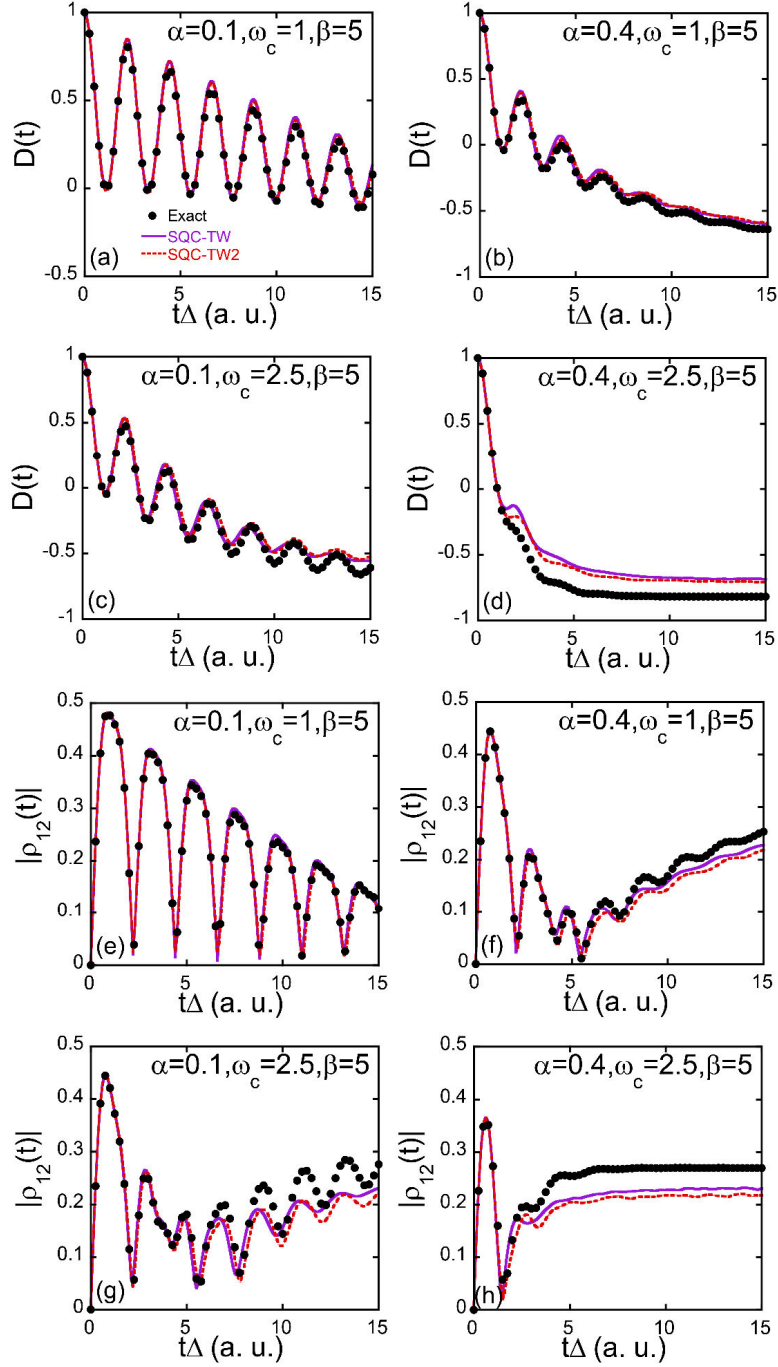

**Figure S5.** Comparison of the population dynamics (panels (a-d)) and coherence dynamics (panels (e-h)) of SQC-TW and those of SQC-TW2 for four spin-boson models discussed in the main text. The parameters for the models in panels (a-h) are consistent with those detailed in Figure 1 of the

S31

main text. Purple lines: SQC-TW. Red dashed lines: SQC-TW2. Numerically exact results are denoted by black points.

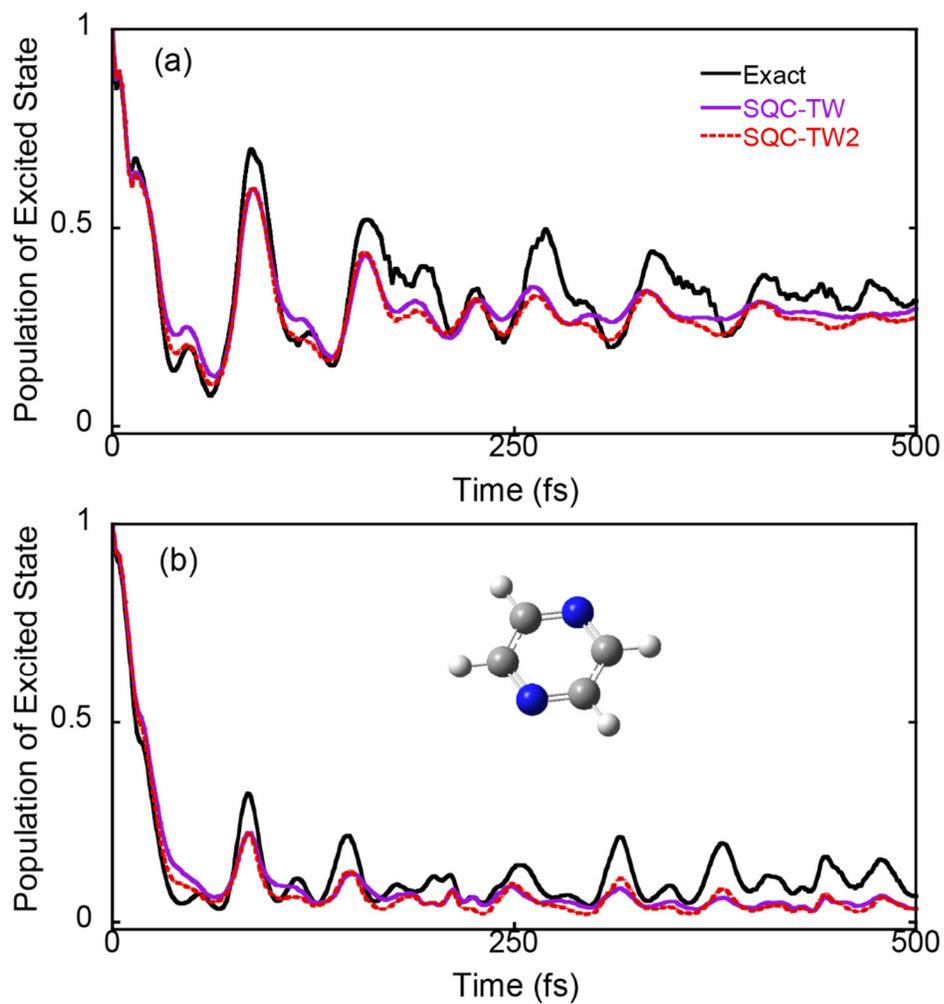

**Figure S6.** Comparisons of SQC-TW and SQC-TW2 results. Panels (a-b) depict the population dynamics of the second state of the 2-state LVCM with 3 modes for pyrazine[47] and that with 24 modes for the same molecule[48], respectively. Purple lines: SQC-TW. Red dashed lines: SQC-TW2. Black lines: Numerically exact results produced by MCTDH [56].

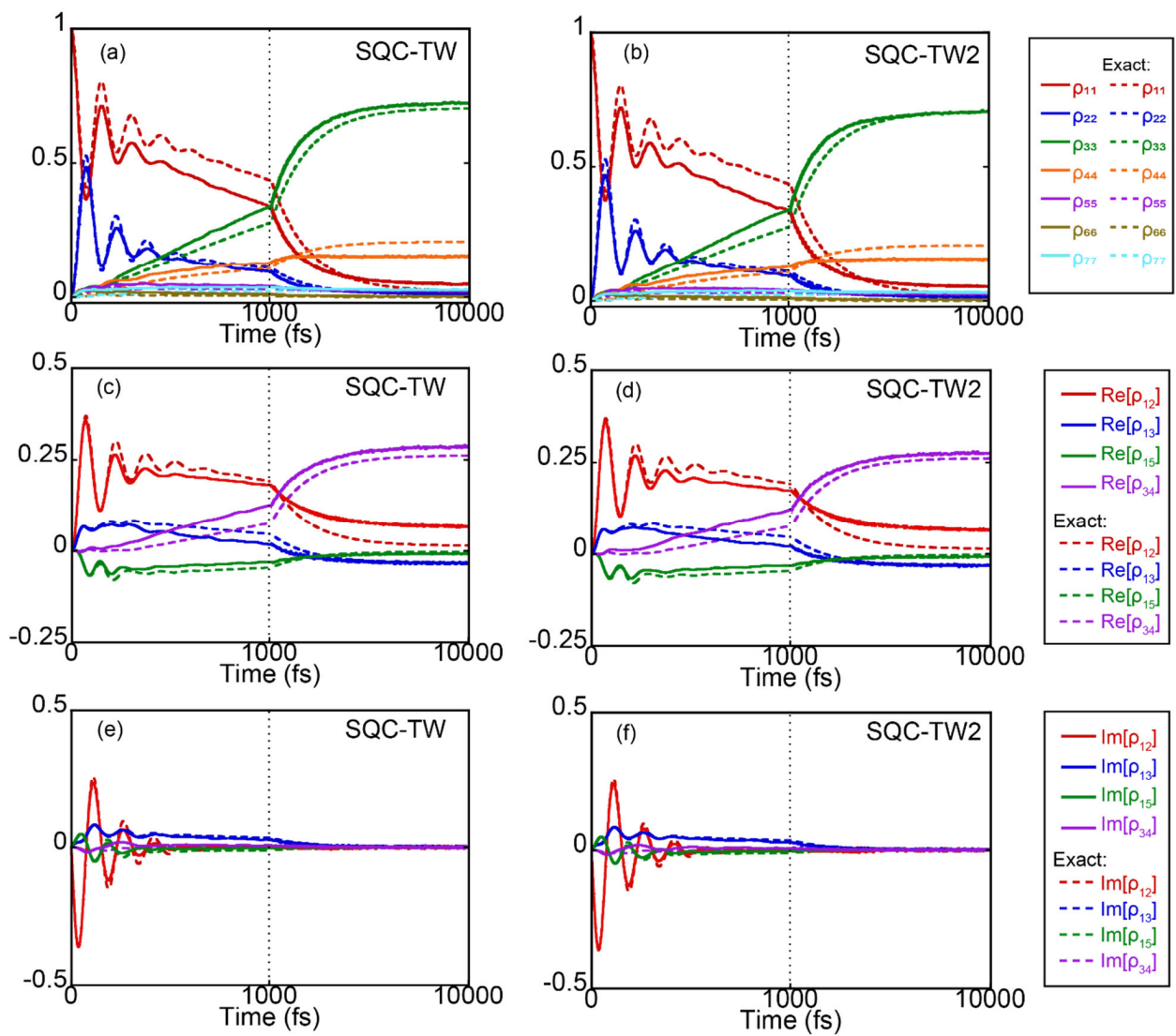

**Figure S7.** Comparisons of SQC-TW and SQC-TW2. Panels (a-b) depict the population dynamics of the FMO monomer reproduced by the SQC-TW and SQC-TW2, respectively. Panels (c-f) demonstrate the corresponding coherence dynamics of the FMO monomer for SQC-TW and SQC-TW2, respectively. The line colors in all panels match those used in Figure 2 of the main text.

We also compare two NaF-TW strategies. The first, as presented in the main text of the present paper, employs $\tilde{\rho}(t) = (1+F/3)\tilde{\mathbf{g}}_t\tilde{\mathbf{g}}_t^\dagger / \text{Tr}_e[\tilde{\mathbf{g}}_t\tilde{\mathbf{g}}_t^\dagger] - \mathbf{1}/3$. In the second approach, denoted as NaF-TW2, we alternatively employ $\tilde{\rho}(t) = \frac{1}{2}\tilde{\mathbf{g}}_t\tilde{\mathbf{g}}_t^\dagger - \tilde{\Gamma}_t$, where commutator matrix $\tilde{\Gamma}_t$ varies with time, and the sampling of initial value of $\tilde{\Gamma}_0$ is according to the details described in Sub-section S2-A. NaF-TW and NaF-TW2 exhibit nearly comparable performance across all benchmark models. In Figures S8-S12, we compare their performances in the spin-boson model, the LVCM of pyrazine molecule, the FMO monomer and Tully's models, respectively.

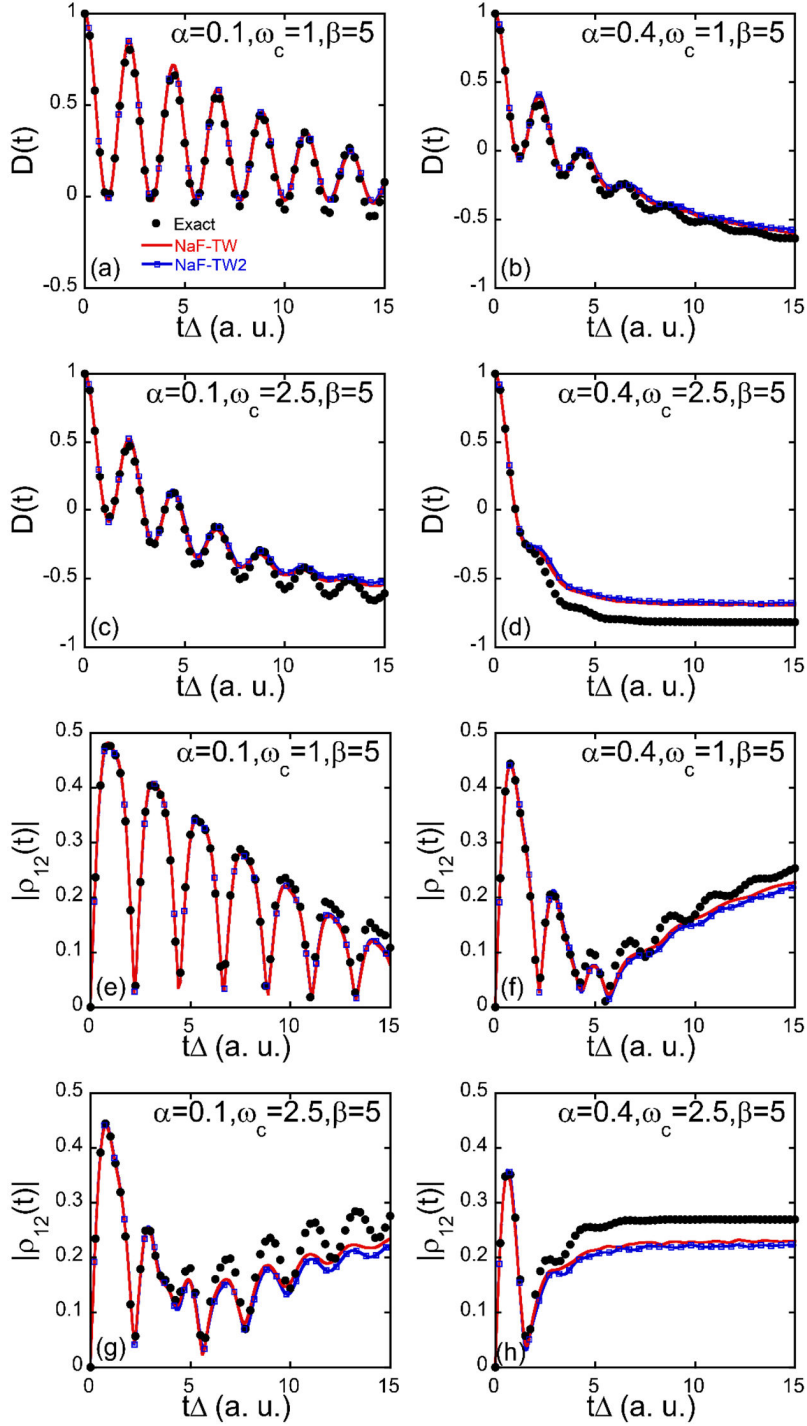

**Figure S8.** Comparison of the population dynamics (panels (a-d)) and coherence dynamics (panels (e-h)) of NaF-TW and NaF-TW2 for four spin-boson models discussed in the main text. The parameters for the models in panels (a-h) are consistent with those detailed in Figure 1 of the main

text. Red lines indicate: NaF-TW. Blue lines with hollow squares: NaF-TW2. Numerically exact results are denoted by black points.

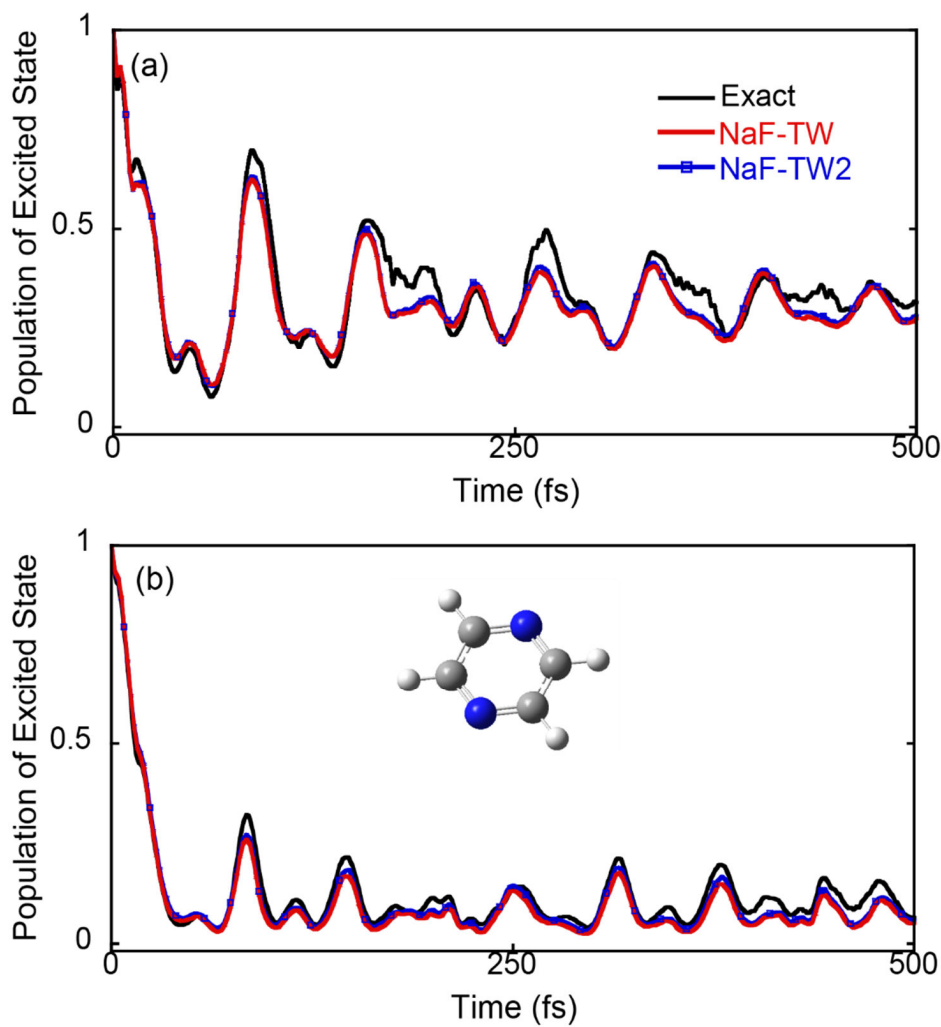

**Figure S9.** Comparisons of NaF-TW and NaF-TW2. Panels (a-b) depict the population dynamics of the second state of the 2-state LVCM with 3 modes for pyrazine [47] and that with 24 modes for the same molecule[48], respectively. Red lines: NaF-TW. Blue lines with hollow squares: NaF-TW2. Black lines: Numerically exact results produced by the MCTDH program[56].

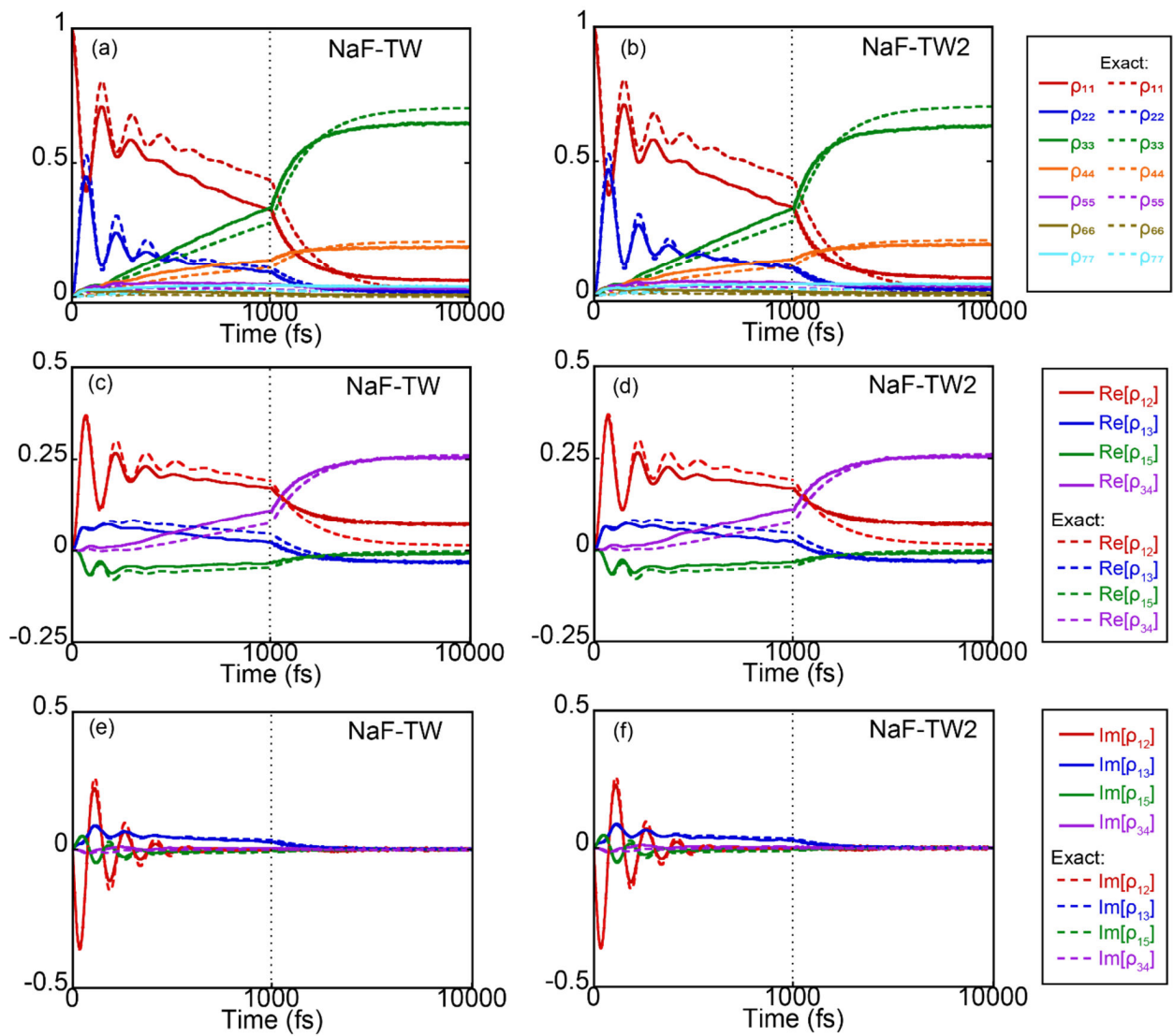

**Figure S10.** Comparisons of NaF-TW and NaF-TW2 results. Panels (a-b) depict the population dynamics of the FMO monomer as reproduced by the NaF-TW and NaF-TW2, respectively. Panels (c-f) show the corresponding coherence dynamics of the FMO monomer for the NaF-TW and NaF-TW2, respectively. The line colors in all panels match those used in Figure 2 of the main text.

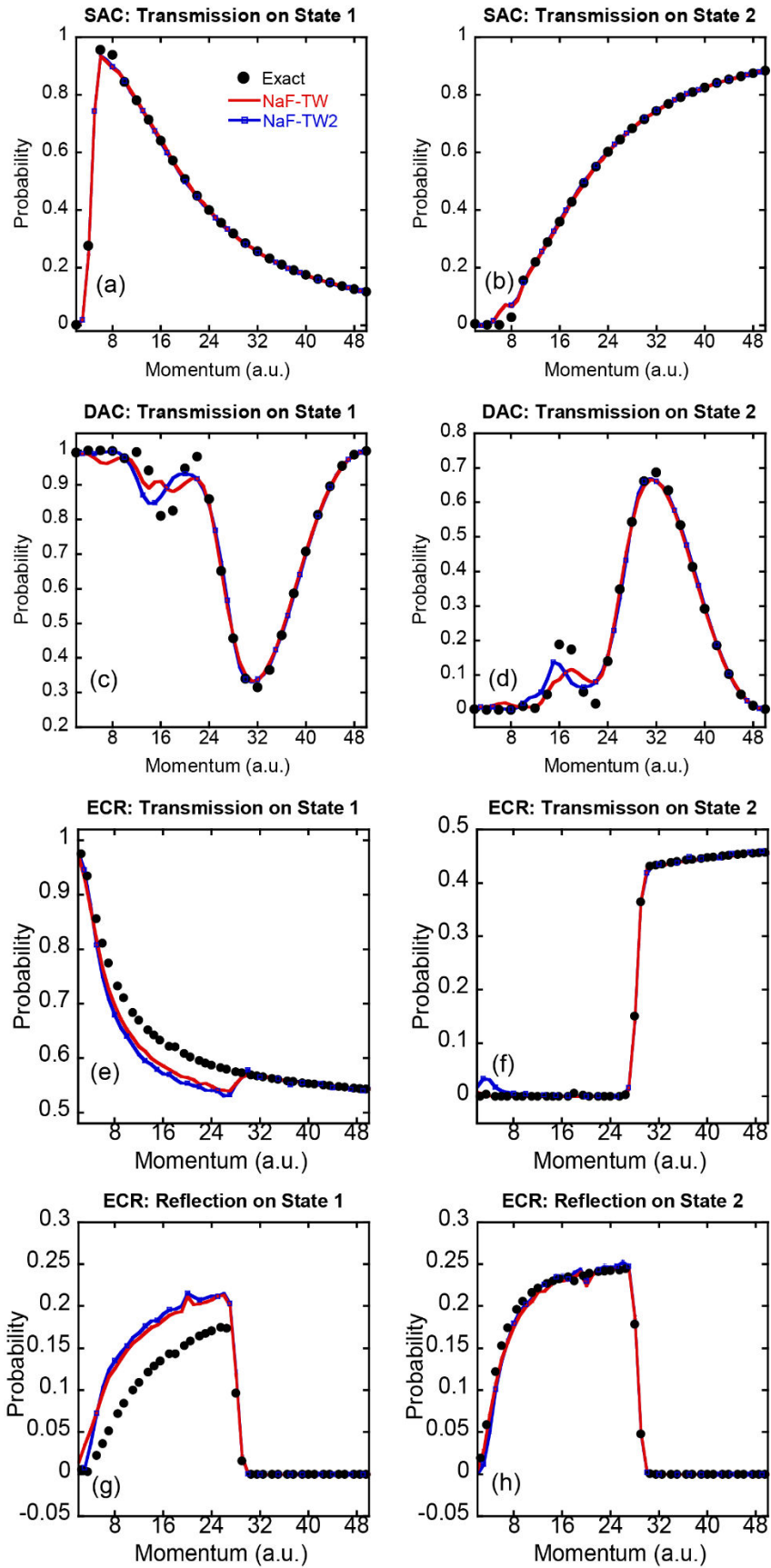

**Figure S11.** Comparisons of NaF-TW and NaF-TW2. Panels (a-b) depict the transmission probability on (adiabatic) State 1 and State 2 with respect to the center of momentum of the initial nuclear Gaussian wavepacket for Tully's SAC model, respectively. Panels (c-d) also show the transmission probability on (adiabatic) State 1 and State 2 but for Tully's DAC model. Panels (e-h) present the transmission probability on (adiabatic) State 1 and State 2, as well as the reflection probability on (adiabatic) State 1 and State 2 for Tully's ECR model, respectively. Red lines: NaF-TW. Blue lines with hollow squares: NaF-TW2. Black lines: Numerically exact results produced by DVR.

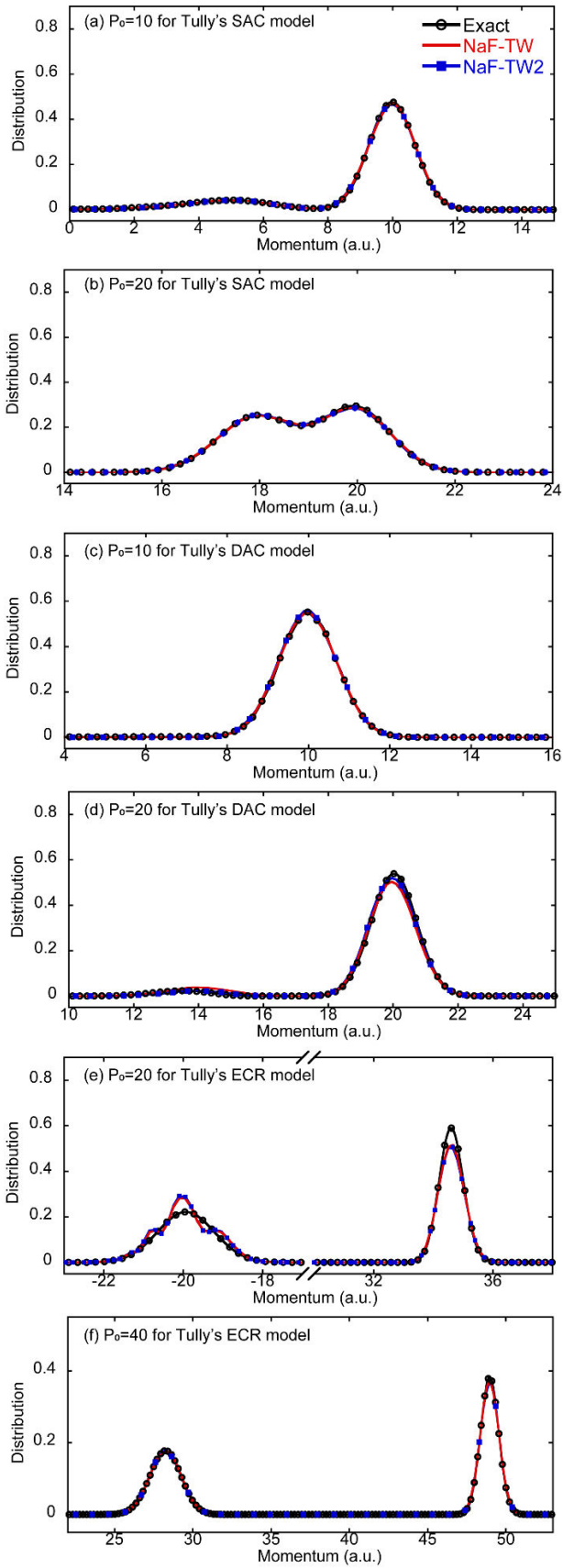

**Figure S12.** Comparisons of NaF-TW and NaF-TW2 results. Panels (a-b) depict the asymptotic nuclear momentum distribution for the initial momentum $P_0 = 10$ and $P_0 = 20$ of the center of the nuclear Gaussian wavepacket for Tully's SAC model, respectively. Panels (c-d) show the same distribution for the initial momentum $P_0 = 10$ and $P_0 = 20$ of the center of the nuclear Gaussian wavepacket for Tully's DAC model, respectively. Panels (e-f) present the same distribution for the initial momentum $P_0 = 20$ and $P_0 = 40$ of the center of the nuclear Gaussian wavepacket for Tully's ECR model, respectively. Red lines: NaF-TW. Blue lines with hollow squares: NaF-TW2. Black lines: Numerically exact results produced by DVR.

**S5: Mapping Triangle Window Functions onto Constraint Phase Space Leads to Novel Phase Space Formulations for the *F*-state Quantum Systems**

In this sub-section, we propose the phase space formulations on CPS which can be derived from the TW functions. Similar to Sub-section S1, consider an *F*-state discrete quantum system of electronic DOFs. Using action-angle variables, the expression of the population-population correlation function reads

$$\mathrm{Tr}_e\left[|n\rangle\langle n|e^{i\hat{H}t}|m\rangle\langle m|e^{-i\hat{H}t}\right] \mapsto \left(\overline{C}_{nn,mm}(t)\right)^{-1} p_{n\to m}^{\mathrm{SQC}}(t) , \tag{S59}$$

where

$$p_{n\to m}^{\mathrm{SQC}}(t) = \int \frac{2\mathrm{d}\mathbf{e}_0 \mathrm{d}\boldsymbol{\theta}_0}{(2\pi)^F \left(2-e_0^{(n)}\right)^{F-2}} K_{nn}^{\mathrm{SQC}}(\mathbf{e}_0, \boldsymbol{\theta}_0) K_{mm}^{\mathrm{bin}}(\mathbf{e}_t, \boldsymbol{\theta}_t) \tag{S60}$$

and

$$\overline{C}_{nn,mm}(t) = \sum_{k=1}^{F} p_{n\to k}^{\mathrm{SQC}}(t) . \tag{S61}$$

In eq (S60), the definition of $K_{nn}^{SQC}(\mathbf{e}_0, \boldsymbol{\theta}_0)$ is the same as eq (S3); substituting eq (S1) into eqs (16)-(17) of the main text (of the present letter) leads to

$$K_{mm}^{bin}(\mathbf{e}, \boldsymbol{\theta}) \equiv K_{mm}^{bin}(\mathbf{e}) = h(e^{(m)} - 1) \prod_{m' \neq m} h(1 - e^{(m')}) . \tag{S62}$$

We then transform the integral on the action-angle space, eq (S60), to the integral counterpart on a single CPS. Because a phase space point at any time $t$ always lies on a particular *iso-action* line $\sum_{n=1}^{F} e^{(n)} = \lambda$, the integral on the action space can be expressed as

$$\begin{aligned}
&\int_0^{+\infty} d\overline{e}^{(1)} \cdots d\overline{e}^{(F)} f(\overline{e}^{(1)}, \cdots, \overline{e}^{(F)}; \boldsymbol{\theta}) \\
&= \int_0^{+\infty} d\overline{\lambda} \int_0^{+\infty} d\overline{e}^{(1)} \cdots d\overline{e}^{(F)} f(\overline{e}^{(1)}, \cdots, \overline{e}^{(F)}; \boldsymbol{\theta}) \delta\left(\sum_{n=1}^{F} \overline{e}^{(n)} - \overline{\lambda}\right) \\
&\underset{\overline{e}^{(i)} = \lambda e^{(i)}}{\overset{\overline{\lambda} = (1+F\gamma)\lambda}{=}} (1+F\gamma) \int_0^{+\infty} \lambda^{F-1} d\lambda \int_0^{+\infty} de^{(1)} \cdots de^{(F)} f(\lambda e^{(1)}, \cdots, \lambda e^{(F)}; \boldsymbol{\theta}) \delta\left(\sum_{n=1}^{F} e^{(n)} - (1+F\gamma)\right)
\end{aligned} \tag{S63}$$

The transformation $\{\overline{e}_0^{(i)}\} \mapsto \{\lambda e_0^{(i)}\}$ leads to $\{\overline{e}_t^{(i)}\} \mapsto \{\lambda e_t^{(i)}\}$ at any time $t$. The integral of eq (S60) involved in the population-population correlation function can then be changed to

$$\begin{aligned}
p_{n \to m}^{SQC}(t) &= \int \frac{2d\mathbf{e}_0 d\boldsymbol{\theta}_0}{(2\pi)^F (2 - e_0^{(n)})^{F-2}} K_{nn}^{SQC}(\mathbf{e}_0) K_{mm}^{bin}(\mathbf{e}_t) \\
&= \frac{(1+F\gamma)}{(2\pi)^F} \int F d\mathbf{e}_0 d\boldsymbol{\theta}_0 \delta\left(\sum_{k=1}^{F} e_0^{(k)} - (1+F\gamma)\right) \times \\
&\quad \int_0^{+\infty} \frac{2\lambda^{F-1}}{F(2 - \lambda e_0^{(n)})^{F-2}} d\lambda h(\lambda - (e_0^{(n)})^{-1}) \prod_{n' \neq n} h\left(\frac{2}{e_0^{(n')} + e_0^{(n)}} - \lambda\right) \times \\
&\quad h(\lambda - (e_t^{(m)})^{-1}) \prod_{m' \neq m} h((e_t^{(m')})^{-1} - \lambda) \\
&= \frac{(1+F\gamma)}{(2\pi)^F} \int F d\mathbf{e}_0 d\boldsymbol{\theta}_0 \delta\left(\sum_{k=1}^{F} e_0^{(k)} - (1+F\gamma)\right) \overline{\Lambda}\left(\int_{h_1}^{h_2} \frac{2\lambda^{F-1}}{F(2 - \lambda e_0^{(n)})^{F-2}} d\lambda\right)
\end{aligned} \tag{S64}$$

where the time-dependent boundaries $h_1$ and $h_2$ rely on the indexes, $n$ and $m$,

$$h_1 = \max\left\{\left(e_0^{(n)}\right)^{-1}, \left(e_t^{(m)}\right)^{-1}\right\},$$

$$h_2 = \min\left\{\frac{2}{e_0^{(n')} + e_0^{(n)}}, \left(e_t^{(m')}\right)^{-1} \middle| n' \neq n, \, m' \neq m\right\}, \quad \text{(S65)}$$

and we define the function $\bar{\Lambda}(z) = \begin{cases} z, & \text{if } z > 0, \\ 0, & \text{otherwise} \end{cases}$. When $h_1 < h_2$ so that $p_{n \to m}^{\text{SQC}}(t) \neq 0$, $e_t^{(m)} > e_t^{(m')}$ for any $m' \neq m$. In addition, at time 0 the $n$-th state is occupied, thus $e_0^{(n)} > e_0^{(n')}$ for any $n' \neq n$.

Since $\dfrac{2\lambda^{F-1}}{\left(2 - \lambda e_0^{(n)}\right)^{F-2}}$ monotonically increases as the value of $\lambda$ does on the integration domain of $\lambda$, we obtain the analytical integral

$$\bar{\Lambda}\left(\int_{h_1}^{h_2} \frac{2\lambda^{F-1}}{F\left(2 - \lambda e_0^{(n)}\right)^{F-2}} d\lambda\right) = \frac{(F-1)!}{F} \prod_{n' \neq n} h\left(e_0^{(n)} - e_0^{(n')}\right) \prod_{m' \neq m} h\left(e_t^{(m)} - e_t^{(m')}\right) \times$$

$$\bar{\Lambda}\left(\phi\left(\min\left\{\frac{2}{e_0^{(n')} + e_0^{(n)}}, \frac{1}{e_t^{(m')}} \middle| n' \neq n, m' \neq m\right\}; e_0^{(n)}\right) - \phi\left(\max\left\{\frac{1}{e_0^{(n)}}, \frac{1}{e_t^{(m)}}\right\}; e_0^{(n)}\right)\right), \quad \text{(S66)}$$

where

$$\phi\left(\lambda; e^{(n)}\right) = \lambda^F \left(2 - \lambda e^{(n)}\right)^{1-F} \left[2(F-2)(F-1)\,_2\tilde{F}_1\left(1,1;F+1;\lambda e^{(n)}/2\right) + \frac{\left(6 - 2F - \lambda e^{(n)}\right)}{\Gamma(F)}\right]. \quad \text{(S67)}$$

Here, $_2\tilde{F}_1(a,b;c;z)$ is regularized hypergeometric function. Transforming eqs (S64)-(S67) back to coordinate-momentum phase space variables, the population-population correlation function reads

$$\text{Tr}\left[|n\rangle\langle n|e^{i\hat{H}t}|m\rangle\langle m|e^{-i\hat{H}t}\right] \mapsto \left(\bar{C}_{nn,mm}^{\text{sqz}}(t)\right)^{-1} \int_{\mathcal{S}(\mathbf{x},\mathbf{p};\gamma)} F d\mathbf{x}_0 d\mathbf{p}_0 \bar{w}_{nn,mm}^{\text{sqz}}\left(\mathbf{x}_0, \mathbf{p}_0; \mathbf{x}_t, \mathbf{p}_t\right) \\ \times K_{nn}^{\text{sqz}}\left(\mathbf{x}_0, \mathbf{p}_0\right) K_{mm}^{\text{sqz}}\left(\mathbf{x}_t, \mathbf{p}_t\right), \quad \text{(S68)}$$

where $K_{nn}^{\text{sqz}}(\mathbf{x}, \mathbf{p}) \equiv K_{nn}^{\text{sqz}}(\mathbf{e}, \boldsymbol{\theta}) = \prod_{n' \neq n} h\left(e^{(n)} - e^{(n')}\right)$. The time-dependent normalization factor is

S43

$$\overline{C}^{\text{sqz}}_{nn,mm}(t) = \sum_{k=1}^{F} \int_{\mathcal{S}(\mathbf{x},\mathbf{p};\gamma)} F d\mathbf{x}_0 d\mathbf{p}_0 \overline{w}^{\text{sqz}}_{nn,kk}(\mathbf{x}_0,\mathbf{p}_0;\mathbf{x}_t,\mathbf{p}_t) K^{\text{sqz}}_{nn}(\mathbf{x}_0,\mathbf{p}_0) K^{\text{sqz}}_{kk}(\mathbf{x}_t,\mathbf{p}_t) \quad \text{(S69)}$$

and the generalized weight function is

$$\overline{w}^{\text{sqz}}_{nn,mm}(\mathbf{x}_0,\mathbf{p}_0;\mathbf{x}_t,\mathbf{p}_t) \equiv \overline{w}^{\text{sqz}}_{nn,mm}(\mathbf{e}_0(\mathbf{x}_0,\mathbf{p}_0);\mathbf{e}_t(\mathbf{x}_t,\mathbf{p}_t)) = \frac{(1+F\gamma)^F}{F} \overline{\Lambda}(\tilde{z}_{nm}(\mathbf{e}_0,\mathbf{e}_t)) \quad \text{(S70)}$$

with

$$\tilde{z}_{nm}(\mathbf{e}_0,\mathbf{e}_t) = \phi\left(\min\left\{\frac{2}{e_0^{(n'\ne n)}+e_0^{(n)}},\frac{1}{e_t^{(m'\ne m)}}\bigg| n'\ne n, m'\ne m\right\};e_0^{(n)}\right) - \phi\left(\max\left\{\frac{1}{e_0^{(n)}},\frac{1}{e_t^{(m)}}\right\};e_0^{(n)}\right) \quad \text{(S71)}$$

for general $F$-state systems, which depends on both the values of $\mathbf{e}_0(\mathbf{x}_0,\mathbf{p}_0)$ and those of $\mathbf{e}_t(\mathbf{x}_t,\mathbf{p}_t)$. We denoted it as the squeezed (sqz) triangular window function approach. In this approach, the initial sampling is on the $U(F)/U(F-1)$ CPS, $\mathcal{S}(\mathbf{x},\mathbf{p};\gamma)$. The approach involves a time-dependent normalization factor for the population-population correlation function. The population-population correlation function of the sqz triangular window function approach is identical to that using TW functions for electronic DOFs of the $F$-state system.

Particularly, for the case of $F=2$, it is straightforward to derive $\phi(\lambda;e^{(n)}) = \lambda^2$. Correspondingly, $\overline{w}^{\text{sqz}}_{nn,mm}(\mathbf{x}_0,\mathbf{p}_0;\mathbf{x}_t,\mathbf{p}_t) \equiv \overline{w}^{\text{sqz}}_{nn,mm}(\mathbf{e}_0(\mathbf{x}_0,\mathbf{p}_0);\mathbf{e}_t(\mathbf{x}_t,\mathbf{p}_t)) = 2 - \frac{(1+F\gamma)^2}{2\min\{e_0^{(n)},e_t^{(m)}\}^2}$.

The $F=2$ case of the sqz formalism is used in the proof that the mapping formalism with TW functions offers an *exact* representation for the population dynamics for the pure two-state system (TSS), as demonstrated in our work of ref [1].

■ **AUTHOR INFORMATION**

**Corresponding Author**

*E-mail: jianliupku@pku.edu.cn

**ORCID**

Xin He: 0000-0002-5189-7204

Xiangsong Cheng: 0000-0001-8793-5092

Baihua Wu: 0000-0002-1256-6859

Jian Liu: 0000-0002-2906-5858

**Notes**

The authors declare no competing financial interests.

■ **ACKNOWLEDGMENT**

This work was supported by the National Science Fund for Distinguished Young Scholars Grant No. 22225304. We acknowledge the High-performance Computing Platform of Peking University, Beijing PARATERA Tech Co., Ltd., and Guangzhou supercomputer center for providing computational resources. We thank Youhao Shang, Haocheng Lu, and Bingqi Li for useful discussions. We also thank Bill Miller for having encouraged us to investigate the window function approach.